\documentclass[twocolumn]{aastex63}

\usepackage{times}

\usepackage{graphicx}

\usepackage{tikz}

\usepackage[T1]{fontenc}
\usepackage{aecompl}
\usepackage{calligra}
\usepackage{soul}
\DeclareMathAlphabet{\mathcalligra}{T1}{calligra}{m}{n}
\DeclareFontShape{T1}{calligra}{m}{n}{<->s*[2.2]callig15}{}

\usepackage{hyperref}
\hypersetup{
    pdfnewwindow=true,      % links in new window
    colorlinks=true,       % false: boxed links; true: colored links
    linkcolor=orange,      % color of internal links
    citecolor=cyan,        % color of links to bibliography
    filecolor=orange,      % color of file links
    urlcolor=orange           % color of external links
}

\def\apjl{ApJL}               % Astrophysical Journal, Letters
\def\apjs{ApJS}               % Astrophysical Journal, Supplement
\def\aap{A\&A}                % Astronomy and Astrophysics
\usepackage{color}

\def\orb{\mathrm{orb}}

\def\min{\mathrm{min}}
\def\max{\mathrm{max}}
\def\Edd{\mathrm{Edd}}
\def\cav{\mathrm{cav}}
\def\rep{\mathrm{rep}}
\def\acc{\mathrm{acc}}
\def\dec{\mathrm{dec}}
\def\core{\mathrm{c}}

\def\CBDstm{\mathrm{stm}}
\def\ring{\mathrm{rng}}
\def\bin{\mathrm{bin}}
\def\MS{\mathrm{MS}}

\def\Ncav{\mathrm{N_\cav}}
\def\rcav{\mathrm{r_\cav}}

\def\GW{\mathrm{GW}}
\def\Msun{{M_{\odot}}}
\def\Rsun{{R_{\odot}}}
\def\rhosun{{\rho_{\odot}}}
\def\rhoavg{{\bar{\rho}_*}}

\def\Mdot{\dot{M}}
\def\MEdd{\dot{M}_{\mathrm{Edd}}}
\def\Machstm{\mathcal{M}_{\mathrm{d}}}

\def\crit{\mathrm{crit}}
\def\partial{\mathrm{prtl}}

\def\Heat{\mathrm{Heat}}

\def\Bfacmax{\mathcal{B}_{\max}}
\def\Bfact{\mathcal{B}}
\def\tdecay{\mathcal{T}}

\def\AGW{\beta_{\GW}}

\def\eg{{\em e.g.}}
\def\ie{{\em i.e.}}

\def\BHcirc{\tikz\draw[black,fill=black] (0,0) circle (.25ex);}

\def\mbhb{\BHcirc \BHcirc}

\usepackage{units}
\newcommand{\Order}[1]{\mathcal{O}({#1})}
\newcommand{\gsn}[0]{GSN 069 }
\newcommand{\rxj}[0]{RX J1301.9+2747 }

\newcommand{\eroi}[0]{eRO-QPE1 }
\newcommand{\eroii}[0]{eRO-QPE2 }

\usepackage{lineno}
\usepackage{units}

\begin{document}

\shorttitle{CBD TDE}
\shortauthors{D'Orazio et al.}

\title{Stellar Stripping and Disruption in Disks around Supermassive Black Hole Binaries: \\ Repeating nuclear transients prior to LISA events}

\author[0000-0002-1271-6247]{Daniel J. D'Orazio}
\affiliation{Space Telescope Science Institute, 3700 San Martin Drive, Baltimore, MD 21218, USA}
\email{dorazio@stsci.edu}
\affiliation{Department of Physics and Astronomy, Johns
Hopkins University, 3400 North Charles Street, Baltimore,
Maryland 21218, USA}
\affiliation{Niels Bohr International Academy, Niels Bohr Institute, Blegdamsvej 17, 2100 Copenhagen, Denmark}

\author[0000-0000-0000-0000]{Christopher Tiede}
\affiliation{Niels Bohr International Academy, Niels Bohr Institute, Blegdamsvej 17, 2100 Copenhagen, Denmark}

\author[0000-0000-0000-0000]{Lorenz Zwick}
\affiliation{Niels Bohr International Academy, Niels Bohr Institute, Blegdamsvej 17, 2100 Copenhagen, Denmark}

\author[0000-0003-4799-1895]{Kimitake Hayasaki}
\affiliation{Department of Astronomy and Space Science, Chungbuk National University, Cheongju 361-763, Republic of Korea}
\affiliation{Department of Physical Sciences, Aoyama Gakuin University, Sagamihara 252-5258, Japan}

\author[0000-0000-0000-0000]{Lucio Mayer}
\affiliation{Department of Astrophysics, University of Zurich, Winterthurerstrasse 190, 8057 Zurich, Switzerland}

%%%%%%%%%%%%%%%%%%%%%%%%%%%%%%%%%%%%%%%%%%%%%%%%%%%%%%%%%%%%%%%%
%%%%%%%%%%%%%%%%%%%%%%%%%%%%%%%%%%%%%%%%%%%%%%%%%%%%%%%%%%%%%%%%

\begin{abstract} 
If supermassive black hole binaries (SMBHBs) are driven together by gas disks in galactic nuclei, then a surrounding nuclear star cluster or in-situ star-formation should deliver stars to the disk plane. Migration through the circumbinary disk will quickly bring stars to the edge of a low-density cavity cleared by the binary, where the stellar orbit becomes trapped and locked with the binary decay. Here we explore the scenario where the trapped stellar orbit decays with the binary until the binary tidally strips the star in a runaway process. For Sun-like stars, this occurs preferentially for $10^4-10^6 \Msun$ SMBHBs, as the SMBHB enters the LISA band. We estimate that the runaway stripping process will generate Eddington-level X-ray flares repeating on hours-to-days timescales and lasting for decades. The flaring timescales and energetics of these circumbinary-disk tidal-disruption events (CBD-TDEs) match well with the recently discovered Quasi-Periodic Eruptions. However, the inferred rates of the two phenomena are in tension, unless low-mass SMBHB mergers are more common than expected. For less-dense stars, stripping begins earlier in the SMBHB inspiral, has longer repetition times, lasts longer, is dimmer, and can occur for more massive SMBHBs. Whether CBD-TDEs are a known or a yet-undiscovered class of repeating nuclear transients, they could provide a new probe of the elusive SMBH mergers in low mass / dwarf galaxies, which lie in the sweet-spot of the LISA sensitivity.
\end{abstract}


\section{Introduction}
When galaxies harboring supermassive black holes (SMBHs) collide, the SMBHs can form a binary at the center of the nascent galaxy \citep{Begel:Blan:Rees:1980}. If the SMBH binary (SMBHB) can efficiently interact with its environment, to remove orbital angular momentum and energy, it can reach separations where gravitational radiation efficiently drives the binary towards merger \citep{MerrittMilos:2005:LRR}. The late inspiral and merger would be detectable with low and mid-frequency gravitational wave detectors such as the Pulsar Timing Arrays \citep[PTAs, \eg,][]{FosterBacker:1990, NANOG-GWB:2023, EPTA_GWB_2023}, TianQin \citep{TianQin+2021} and LISA \citep{LISA_Astro+2023, LISA_redbook:2024}. 

Environmental interactions include those with gas surrounding the binary in a circumbinary disk \citep[CBD, \eg, ][]{ArmNat:2002:ApJL} and with stars from a surrounding nuclear star cluster \citep[\eg,][]{Quinlan:1996}. 
Both types of interactions can lead to observational identifiers of a SMBHB. Gas interactions with the binary can lead to periodically variable photometric and spectroscopic signatures due to accretion rate variability and relativistic effects \citep[\eg,][]{binlite+2024, DOrazioCharisi:2023}. Stellar interactions can lead to tidal disruption events (TDEs).

A TDE is a luminous transient event that typically occurs around a SMBH at the center of an inactive galaxy, and is characterized by flares lasting for several months to years \citep[see][for a review]{Gezari2021}. The standard theoretical picture is that stars approach the SMBH on nearly parabolic orbits, are torn apart by its tidal forces \citep{Hills1975}, and release a fraction of their gravitational energy as the stellar debris accretes onto the SMBH, producing bright flares \citep{Rees1988}. 
Over the past fifteen years, the number of TDEs or TDE candidates has increased significantly, but is still only a hundred or more.
However, with the upcoming operation of the Vera C. Rubin Observatory \citep[VRO;][]{Andreoni+2022}, the number of observed TDEs is expected to increase by an order of magnitude or more within the next decade. One of the scientific motivations for studying TDEs is that they provide opportunities to find fingerprints of intermediate-mass black holes (IMBHs), SMBHBs, and isolated, recoiling SMBHs \citep{Komossa2015}.

For SMBHBs, TDEs present unique phenomena that provide insights into both black hole dynamics and accretion processes. When a star is disrupted in such a system, its debris interacts not only with the tidal field of a SMBH, but also with the gravitational potential of the binary, leading to complex accretion dynamics \citep[\eg,][]{Ricarte+2016}. \citet{LiuChen:2009,Liu+2014} demonstrated that TDEs in SMBHBs could produce periodic but intermittent flares due to debris streams interacting with the secondary SMBH, offering a potential observational signature of binary systems. \citet{HayasakiLoeb:2016,Coughlin+2017} investigated the hydrodynamics of such events around close SMBHBs, and showed that a chaotic motion of the debris stream significantly modifies the accretion behavior, leading to unique electromagnetic signals. In particular, for merging SMBHs, an accretion disk formed around the secondary SMBH generates a periodic lightcurve due to special relativistic boosting, potentially providing a method to diagnose gravitational wave emission from the merging SMBHs \citep{HayasakiLoeb:2016}.
These studies highlight the potential of TDEs to serve as a unique diagnostic tool for detecting SMBHBs or merging SMBHs and understanding the interplay between stellar disruption, accretion physics, and binary evolution.

In reality, both gas disks accreting onto the black holes and surrounding stars (in a nuclear star cluster or formed in the disk) should exist in tandem, providing opportunities for disk and TDE-induced binary signatures, as well as possible combinations, which we investigate here. The interaction between AGN disks and surrounding stars has been studied with recently renewed interest in the context of the pairing and merging of stellar-mass BHs through the ``AGN-channel'' \citep{McKernan+2012, Stone_AGNchan+2017, Bartos_AGNchan+2017}. 
One outcome of this interaction is the eventual grind-down of stellar orbits into the plane of the disk \citep[\eg, ][]{Bartos_AGNchan+2017, Fabj+2020, MacLeodLin:2020}, and their rapid migration towards the central accretor \citep[\eg, ][]{Secunda+2019}.  

Here we argue that the same process of grind-down and migration occurs for disks around SMBHBs, that is until the migrating star approaches the inner part of the CBD, where the binary clears a low density cavity \citep[\eg][]{AL94, D'Orazio:CBDTrans:2016}. At this cavity a natural and robust migration trap \citep{Masset_traps+2006, ThunKley:2018, Kley_cavtraprad:2019} locks the stellar orbit to that of the decaying SMBHB until either the trap breaks, the SMBHB decouples from the disk, or the star begins to be tidally stripped by the binary. 

We explore the latter case here, arguing that the stellar stripping events can runaway and fully disrupt the star, significantly altering the feeding of the binary by the CBD and resulting in Eddington-level X-ray events repeating on hours to days timescales and lasting for a few to tens of years. Furthermore, for Solar-like stars, these events, which we call circumbinary disk tidal disruption events (CBD-TDEs), would accompany inspiraling LISA SMBHBs or their progenitors. There are also intriguing similarities to the recently discovered Quasi-Periodic Eruptions \citep[QPEs, \eg, ][and references therein]{Nicholl+2024}.

Our study proceeds as follows. In \S\ref{S:Setup} we motivate the trapping of stars at the edge of circumbinary cavities around SMBHBs and the consequences for disruption of these stars while the SMBHB is approaching, or in, the LISA band. In \S\ref{S:Disruption} we build a model for partial to full disruption of such a star by the SMBHB. In \S\ref{S:AccretonEmission} we investigate the consequences of stripped stellar debris falling onto the binary and generating EM emission. In \S\ref{S:Discussion} we compare predicted signatures of this new multi-messenger event to known EM transients, the QPEs, present related post-SMBHB merger signatures, and discuss required future work on this topic and opportunities for (gravitational-wave) astrophysics. In \S\ref{S:Conclusion} we conclude.

\section{Problem Setup and Characteristic Scales}
\label{S:Setup}

\begin{figure}
\begin{center} 
\includegraphics[scale=0.3]{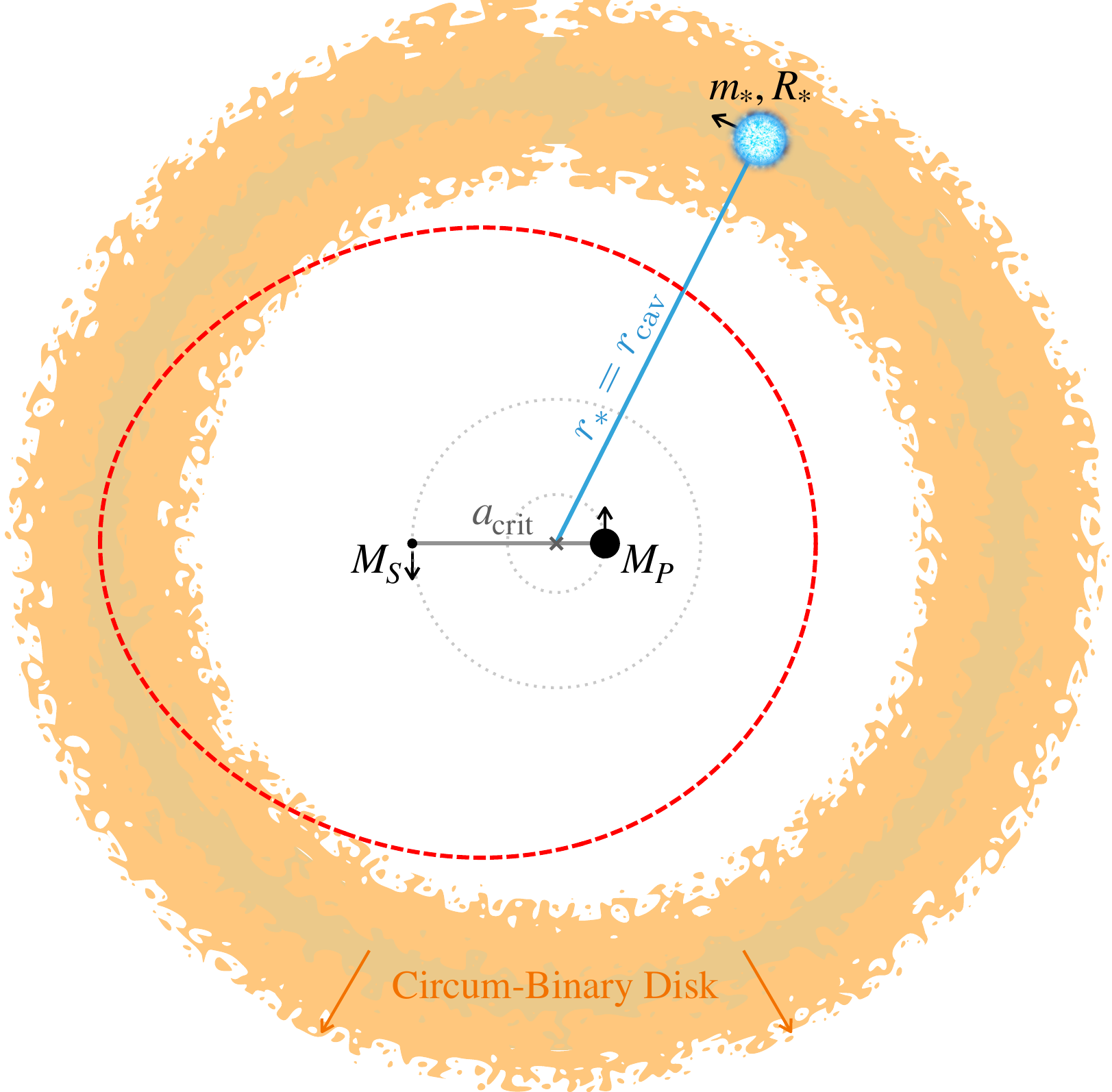} 
\end{center}
\vspace{-15pt}
\caption{
Schematic of the circumbinary disk (CBD) plus star around a 3:1 mass ratio supermassive black hole binary with components $M_P$ and $M_S$ at the critical separation $a_{\crit}$ (Eq. \ref{Eq:acrit}), where the maximum extent of the tidal radius of the binary (red-dashed line) is equal to the distance to the edge of the circumbinary cavity. At this separation the star begins to be tidally stripped once per conjunction of the secondary black hole and star. Each binary component may have its own accretion disk (minidisk) not shown.
 }
\label{Fig:schematic}
\end{figure}

Figure \ref{Fig:schematic} illustrates the problem setup. Throughout, we consider a SMBHB orbiting in the plane of a CBD, on a circular orbit with separation $a$ and component masses $M_P$ and $M_S$ such that the total binary mass is $M=M_P+M_S$ and the binary mass ratio is $q \equiv M_S/M_P \leq 1$. We consider binaries with near unity mass ratio such that a low density cavity is formed at the inner edge of the CBD with an average radial location $\rcav \equiv \Ncav a \approx 2a-4a$ \citep[\eg,][]{AL94, D'Orazio:CBDTrans:2016}. A star of mass $m_*$ and radius $R_*$ orbits in the plane of the CBD at a distance $r_*$ from the binary center of mass.

%%%%%%%%%%%%%%%%%%%%%%%%%%%%%%%%%%%%%%%%%%%%%%%%
%%% Fig 1 schematic %%%
%%%%%%%%%%%%%%%%%%%%%%%%%%%%%%%%%%%%%%%%%%%%%%%%
%%%%%%%%%%%%%%%%%%%%%%%%%%%%%%%%%%%%%%%%%%%%%%%%

\subsection{Stars stuck at the edges of circumbinary cavities}

%\textcolor{red}{comparison with CB planets here?}

We expect such a system to arise in the centers of galactic nuclei that harbor accreting SMBHBs and their associated nuclear star clusters. Stars in the nuclear cluster will quickly be dragged into the plane of the AGN disk \citep{Bartos_AGNchan+2017}, or be formed at the disk edge \citep{Stone_AGNchan+2017}, and once in the disk midplane will begin to migrate through the disk, towards the binary.
If the migrating star is not permanently halted by encounters with other bodies \citep[\eg,][]{Batygin_MMRrev:2015}, or migration traps \citep[\eg,][]{Bellovary+2016,Grishin_Therm+2023}, it will continue to migrate inwards until it reaches the edge of the low-density cavity carved by the binary. Multiple studies have shown that negative Linblad and positive co-rotation torques cancel at the cavity edge and so halt migration of the star there \citep{Masset_traps+2006, ThunKley:2018, Kley_cavtraprad:2019}. This robust trapping mechanism arises because the co-rotation torque depends on the steepness of the disk density profile at the position of the star. When the star reaches the edge of the cavity, the density profile is truncated very steeply generating a large positive co-rotation torque that balances the negative outer Linblad torque \citep{Masset_traps+2006}. Hence, the star will halt its migration at the cavity edge.

For this situation to be realized, the capture and migration timescale of the star must be short compared to the lifetime of the CBD, the binary, and the star. Assuming that the CBD ultimately drives the binary to merger, we approximate the CBD lifetime, and so the binary lifetime, by the typical AGN lifetime of $\sim 10^7-10^8$ years \citep{PMartini:2004}. We then consider two cases for delivery of stars to the disk and to the inner CBD cavity.

\paragraph{Formation in the outer disk}
In the first scenario, stars are formed in a gravitationally unstable outer region of the accretion disk, and so begin migrating via type I migration upon forming a dense core, in the protostar stage \citep{DerdzinskiMayer:2023}. For disks that power bright AGN, the young star will very likely not open a gap in the disk, and it will migrate inwards via type I migration. The type I migration timescale for an object of mass $m_* \ll M$ orbiting at angular frequency $\Omega=\sqrt{GM/r^3_*}$ around central mass $M$, at location $r_*$ in a disk with local surface density $\Sigma$ and aspect ratio $h\equiv H/r_*$ (for scale height $H$), is approximately \citep[\eg,][]{Ward:1997},
\begin{equation}
    t_I  \approx \frac{1}{2 f_{\rm{mig}}} \left( \frac{M}{m_*} \right) \left( \frac{M}{\Sigma r^2_*} \right) h^2 \Omega^{-1},
    \label{Eq:typeI}
\end{equation}
where $f_{\rm{mig}}$ is a numerically calibrated factor which we take to be $2$ \citep[\eg,][based on results of \cite{Kanagawa+2018}]{TagawaAGN+2020}.
To evaluate the disk dependent quantities, we use steady-state $\alpha$-disk models \citep{SS73}. Assuming that the star is in the outer part of the disk where gas pressure and free-free opacity dominate \citep[\eg,][]{HKM09}, and that the outer edge of the disk is set by the self-gravitating radius $r_{\rm{SG,ff}}$, allows us to write $\Sigma$, $h$, and the migration timescale as functions of the total binary mass, stellar mass, viscous $\alpha$ parameter, and accretion rate parameterized by the Eddington fraction $f_{\Edd} \equiv \Mdot/\MEdd$. This gives a characteristic timescale for stars to migrate to the inner CBD cavity edge from the self-gravitating outer edge, 
\begin{eqnarray}
    t_I  &\approx& 1.45 \times 10^5 \ \mathrm{yr} \ \left( \frac{m_*}{\Msun} \right)^{-1} \left( \frac{M}{10^5 \Msun} \right)^{1/45} \times \nonumber \\ 
    &\times& \left( \frac{\alpha}{0.3} \right)^{41/45} \left( \frac{f_{\Edd}}{0.1} \right)^{-29/45}.
\end{eqnarray}
This is much shorter than main-sequence stellar lifetimes \citep[of order $10^7-10^{10}$ years][]{FarrMandel_comment:2018} and typical AGN lifetimes. Hence, stars formed at the edge of an AGN disk may migrate to the cavity edge before the AGN turns off and before the star evolves off of the Main Sequence.

\paragraph{Grind down from surrounding star cluster}
In a second scenario, the star initially exists on an orbit that is misaligned with the CBD but loses orbital angular momentum in the direction perpendicular to the disk through multiple disk encounters. 
Alignment times are most extreme for highly inclined orbits, taking in excess of $\mathcal{O}(10^9)$ years, but are much shorter ($\lesssim 10^6$ years) for less extreme ($\lesssim35^{\circ}$ inclination) systems \citep[\eg][]{Fabj+2020}. 
\citet{Bartos_AGNchan+2017} estimate that for $\sim 10^5 \Msun$ central SMBHs, of order a few percent of orbiting objects can be dragged into the disk within $10^6$ years while more than half can be dragged in over $10^8$ years \citep[see also][in the context of the stellar disruption loss cone]{MacLeodLin:2020}. Hence, the alignment plus ensuing migration timescale can be of order $10^6$ years for some objects, shorter than stellar evolution and CBD depletion timescales.

\paragraph{Stellar fate}
If the AGN lifetime can be associated with a CBD accretion stage that drives the SMBHB to merger, we conclude that stars must be brought to the cavity edge well before the binary merges, the disk is depleted, or the star significantly evolves (see also \S\ref{Ss:caveats}). As the SMBHB orbit decays, star(s) will be trapped at the cavity edge of the CBD, and the CBD cavity size will shrink with the decaying binary.

The stellar orbit will track the shrinking cavity and SMBHB orbit until the first of the following events occur:
\begin{itemize}
    \item \textbf{Disk-decoupling:} the binary inspiral outpaces the viscous inflow of the disk (\S\ref{Ss:decouple}),
    \item \textbf{Migrator-decoupling:} the binary inspiral outpaces the inward migration of the star (\S\ref{Ss:decouple}), or,
    \item \textbf{Tidal stripping\footnote{We use the term tidal stripping interchangeably with tidal peeling \citep{ChengCheng_PTDE+23}, partial disruption, or Roche-lobe overflow.} :} the binary shrinks to a critical separation where tidal effects from the SMBHs in the central binary begin partially disrupting the star (\S\ref{Ss:crit_scales}) in a circumbinary disk tidal disruption event ({\bf CBD-TDE}).
\end{itemize}
If either disk- or migrator-decoupling occur first, the star and the cavity stay behind and intact (though not necessarily at the same radius) as the binary races to merger. This scenario could result in a post-SMBHB-merger disruption of the star that we address further in \S\ref{S:post_merger_dsiruption}.
If instead the star first enters the tidal radius of the binary, it will begin to be tidally stripped while orbiting in the disk. We next estimate binary separations where each of these events occur. We then focus on the binary-star-disk parameter space where CBD-TDEs ensue and explore the multi-messenger signatures of such events.

\subsection{Disk- and migrator-decoupling}
\label{Ss:decouple}

\paragraph{Disk-decoupling}
At compact separations, the binary will begin to decay more quickly than the viscous inflow of the disk. At this point, the assumption of a fixed cavity-edge position in units of binary semi-major axis breaks down as the cavity moves inwards more slowly than the binary \citep[\eg, ][]{Farris:2015:GW, Krauth:2023, Franchini_decouple+2024, Dittmann_decouple+2023}, leaving the star behind with the cavity edge.

To derive the decoupling semi-major axis we equate the viscous inflow velocity at the cavity edge with the binary GW inspiral rate and solve for the semi-major axis at decoupling \citep[\eg,][]{Dittmann_decouple+2023}. 
The viscous inflow velocity depends on the kinematic co-efficient of viscosity $\nu$, which we parameterize with the $\alpha$ prescription $\nu = \alpha h^2 (r^2 \Omega)$, for constant viscous parameter $\alpha$, (dimensionless) disk aspect ratio $h$, and disk Keplerian angular frequency $\Omega$, evaluated at CBD cavity radius $r_{\cav} = N_{\cav}a$. 
Because the solution depends on the disk model (radial dependence of $h$), we use multiple disk models (\citealt{SS73}, \citealt{Sirko_Goodman:2003}) below to solve for the disk-decoupling separation (See Appendix \ref{A:disk_models}). 
We gain insight, however, from first considering disks with constant $h$. In this case, disk decoupling occurs when the binary reaches separations smaller than,
\begin{eqnarray}
    \label{Eq:adec_tII}
    a_{\dec D} &=& \left[\frac{128}{15} \frac{q}{(1+q)^2} \frac{\sqrt{N_{\cav}}}{\alpha h^2} \right]^{2/5} r_G \\ \nonumber
    &\approx& 431 r_G \left( \frac{N_{\cav}}{2} \right)^{1/5} \left( \frac{\alpha}{0.005}\right)^{-2/5} \left( \frac{h}{10^{-2}} \right)^{-4/5}.
\end{eqnarray}
where the second line is evaluated for $q=0.33$, $N_{\cav}=2$, and values for $\alpha$ and $h$ that we motivate further below.

\paragraph{Migrator-decoupling}
At compact separations, the binary may also outpace the inward migration of the star. If this happens before disk decoupling, the star will lag behind the cavity as it shrinks. We solve for the migrator-decoupling separation analogously to the above calculation, by replacing the inward viscous radial velocity of the fluid with the inward type I migration velocity of the star, which we approximate by $(N_{\cav}a)/t_I$, for $t_I$ given by Eq. (\ref{Eq:typeI}). In this case the solution depends on the radial dependence of both the disk surface density $\Sigma(r)$ and disk aspect ratio $h(r)$. As for disk decoupling, we use \cite{SS73} and \citet{Sirko_Goodman:2003} disk models to estimate typical decoupling semi-major axes (Appendix \ref{A:disk_models}).

The latest possible decoupling occurs when the disk- and migrator-decouplings occur at the same separation. This is because independent of disk model $a_{\rm dec, D} \propto h^{-4/5}$, while for a disk with constant $h$ and $\Sigma$ with radius,
\begin{eqnarray}
\label{Eq:adec_TI}
    a_{\dec M} &=& \left[\frac{32}{5 f_{\mathrm{mig}} }\frac{q}{(1+q)^2} \left( \frac{M}{m_*}\right) \left( \frac{M}{\Sigma r^2_G}\right) \right]^{2/9} \frac{h^{4/9} }{N^{1/3}_{\cav}} r_G \\ \nonumber
    &\approx& 432 r_G \left( \frac{h}{10^{-2}}\right)^{4/9} \left( \frac{M/m_*}{10^{5}}\right)^{2/9} \left( \frac{M/(\Sigma r^2_G)}{10^{11}}\right)^{2/9},
\end{eqnarray}
where $\Sigma$ is consistent with $\alpha$-disk models with near unity $f_{\rm Edd}$ and low values of $\alpha$ (as in Eq.~\ref{Eq:adec_tII}). Hence, $a_{\rm dec M} \propto h^{4/9}$ and the two decoupling radii scale oppositely with $h$.
The minimum decoupling separation, where $a_{\rm dec M} = a_{\rm dec D}$, then depends on the cavity size, binary mass ratio, star-binary mass ratio and the specific disk model, which sets the radial dependencies (and the binary-to-disk mass ratio). This matching criterion motivated the choice of $\alpha$ in Eq. (\ref{Eq:adec_tII}). However, as we explain below and in Appendix \ref{A:disk_models}, the maximum decoupling scale does not deviate greatly from this estimate when considering more viscous ($\alpha\rightarrow1$), steady-state disks.

Finally, for type I migration to continue as detailed above, the star must have enough disk material continually delivered to its orbit to carry away the angular momentum of the stellar orbit, and the star should not open a gap in the disk during its journey. The first condition requires that either the local disk mass be at least of order the stellar mass, or that the angular momentum advected through the orbit of the star $\dot{M}l$, integrated over the migration timescale, be of order the angular momentum of the orbit ($m_*l$) --- for specific orbital angular momentum $l$ and steady state disk accretion rate $\dot{M}$. This condition is, approximately, that either of
\begin{equation}
     2 \pi \int^{r_{\cav}}_0\Sigma(r) r dr \gtrsim m_*  \quad \mathrm{or}\quad f_{\Edd} \dot{M}_{\Edd} t_I \gtrsim m_*,
     \label{Eq:mstr-mdisk-conditions}
\end{equation}
hold\footnote{While we make heuristic arguments here, \citet{Rafikov:2013, Rafikov:2016} has shown rigorously that the local disk mass can be much less than the perturber mass and still efficiently decay its orbit towards merger.}, for disk surface density $\Sigma$, steady accretion rate through the disk $\dot{M}=f_{\Edd} \dot{M}_{\Edd}$, and type I migration timescale given by Eq. (\ref{Eq:typeI}). This condition for efficient type I migration is satisfied throughout the evolution for high Eddington ratios (even for low $\alpha$ parameters, see \S\ref{S:AccretonEmission}).

The second condition can be estimated from gap opening criteria of \eg, \cite{Crida+2006, DuffellMac:2013:smallqGapOpen} but does not become an issue for the fiducial disk parameters and stellar-mass stars considered above. In addition, if the stars form in the gravitationally unstable, outer-AGN disk, the gap opening timescale is longer than the migration timescale in such a disk \citep[see, \eg,][]{MalikMMM:2015}, so a gap should not form regardless of stellar mass. Furthermore, the opening of a gap could still facilitate the stars migration with the binary as it merges, because it would be locked into the viscous inflow of the disk, though the cavity trap may not operate. We leave this part of parameter space, for gap opening systems, to future work and continue by determining the implication of near-Solar-mass stars closely approaching the SMBHB as it inspirals.

\subsection{ Critical binary separation for onset of tidal stripping}
\label{Ss:crit_scales}

The star will begin to be tidally stripped when the combined tidal fields from each SMBH in the binary exceed the surface gravity of the star. We follow \citet{CoughlinNixon:2022} but include the (radial) tidal field from both orbiting SMBHs with the star on a circular orbit around the binary center of mass. Stellar-surface force balance requires,
\begin{eqnarray}
        \frac{GM_P}{(d_P(\xi) - R_*)^2} &-& \frac{GM_P}{(d_P(\xi) + R_*)^2} + \nonumber \\ 
    \frac{GM_S}{(d_S(\xi) - R_*)^2} &-& \frac{GM_S}{(d_S(\xi) + R_*)^2} 
    = \frac{Gm_*}{R^2_*},
    \label{Eq:bin_tid_eq1}
\end{eqnarray}
where $d_P(\xi)$ and $d_S(\xi)$ are the time-dependent distances from primary and secondary SMBHs to the star,
\begin{eqnarray}
\label{Eq:dPdS}
    d_P(\xi) &=& a\left[ N^2_{\cav} + \left(\frac{q}{1+q}\right)^2 - \frac{2N_{\cav} q}{1+q}\cos{\xi}\right]^{1/2} \\ \nonumber
    d_S(\xi) &=& a\left[ N^2_{\cav} + \left(\frac{1}{1+q}\right)^2 - \frac{2N_{\cav}}{1+q}\cos{\left(\xi - \pi\right)} \right]^{1/2},
\end{eqnarray}
and where $\xi$ is the angle subtending the lines between primary, binary barycenter, and the star.

First observe that the tidal force on the LHS of Eq. (\ref{Eq:bin_tid_eq1}) is maximized when $\xi=\pi$, when the star, primary and secondary are all aligned, and the secondary is closest to the star (this can be seen by the location of the maximal extent of the dashed red line in Figure \ref{Fig:schematic}).
We then observe, that for the stellar and SMBH masses considered here, the stellar radius is always small compared to the tidal radius for any of the single, component SMBHs, $(m_*/M)^{1/3} \lesssim 10^{-1} \rightarrow 10^{-2}$, which allows use of the tidal approximation for the LHS of Eq. (\ref{Eq:bin_tid_eq1}). 
Hence, the critical binary semi-major axis for which the star begins to be disrupted in the tidal limit is given by solving Eq. (\ref{Eq:bin_tid_eq1}) with $\xi=\pi$ and 
$d_P(\pi), d_S(\pi) \gg R_*$,
\begin{eqnarray}
\label{Eq:acrit}
    a_{\crit} &=& 4^{1/3}  \left[ \Bfacmax(N_{\cav}, q)  \left(\frac{M}{m_*}\right) \right]^{1/3} R_* \\ \nonumber 
    \Bfacmax(N_{\cav}, q) &\equiv&  \frac{(1+q)^2}{\left[(1+q)N_{\cav} + q\right]^3} + \frac{q(1+q)^2}{\left[(1+q)N_{\cav} - 1 \right]^3},
\end{eqnarray}
where $\Bfacmax$ is a binary correction factor that depends only on the size of the cavity and the binary mass ratio and is less than unity for all mass ratios when $N_{\cav}\gtrsim1.43$.
In the limit that $q\rightarrow0$, $ \Bfacmax=N^{-3}_{\cav}$, so that the distance from the star to the primary at the center of mass is $N_{\cav} a_{\crit} = r_T = 4^{1/3}(M/m_*)^{1/3} R_*$. This is the single-BH tidal radius $r_T$ but with a factor of $4^{1/3}$ specific to the onset of tidal stripping at the stellar surface. This factor is in agreement with numerical simulations for single SMBHs (see \citealt{CoughlinNixon:2022} and \S\ref{S:Disruption} for further discussion).

For $N_{\cav}=2$, $q=0.33$, and stellar parameters $m_*=2\Msun$, $R_*=2\Rsun(m_*/\Msun) \equiv 2 R_{MS}$, 
\begin{eqnarray}
    a_{\crit} &\approx&  635 r_G \left(\frac{  \Bfacmax(\Ncav,q) }{\Bfacmax(2,0.33)}\right)^{1/3} \times \nonumber \\
     &\times&  \left(\frac{M}{10^5\Msun}\right)^{-2/3} \left( \frac{\rhoavg}{\rhosun/2^{5}}\right)^{-1/3} ,
    \label{Eq:acrit_eval}
\end{eqnarray}
where $r_G\equiv GM/c^2$, $\rhoavg\equiv 3m_*/(4 \pi R^3_*)$ is the average stellar density, $\rhosun$ is the Solar value, and 
$\Bfacmax(2,0.33)\approx0.1938$. 
%$\Bfacmax(2,1)=608/3375\approx0.18$. 
Here we chose a star with twice the expected radius because tidal heating is likely to expand a Sun-like star to this size (see further motivation below and in Appendix \ref{A:Tidal_Heating}). 
Hence, for our chosen, less-dense-than-Solar star, the critical binary semi-major axis for stellar disruption is larger than the decoupling separations (Eqs. \ref{Eq:adec_tII} and \ref{Eq:adec_TI}). For a Sun-like star this value is $2^{5/3}$ smaller, $\sim 195 r_G$.

%%%%%%%%%%%%%%%%%%%%%%%%%%%%%%%%%%%%%%%%%%%%%%%%
%%% Fig 2 acrit scales %%%
%%%%%%%%%%%%%%%%%%%%%%%%%%%%%%%%%%%%%%%%%%%%%%%%
\begin{figure*}
\begin{center} $
\begin{array}{cc}
\hspace{-5pt}
\includegraphics[scale=0.252]{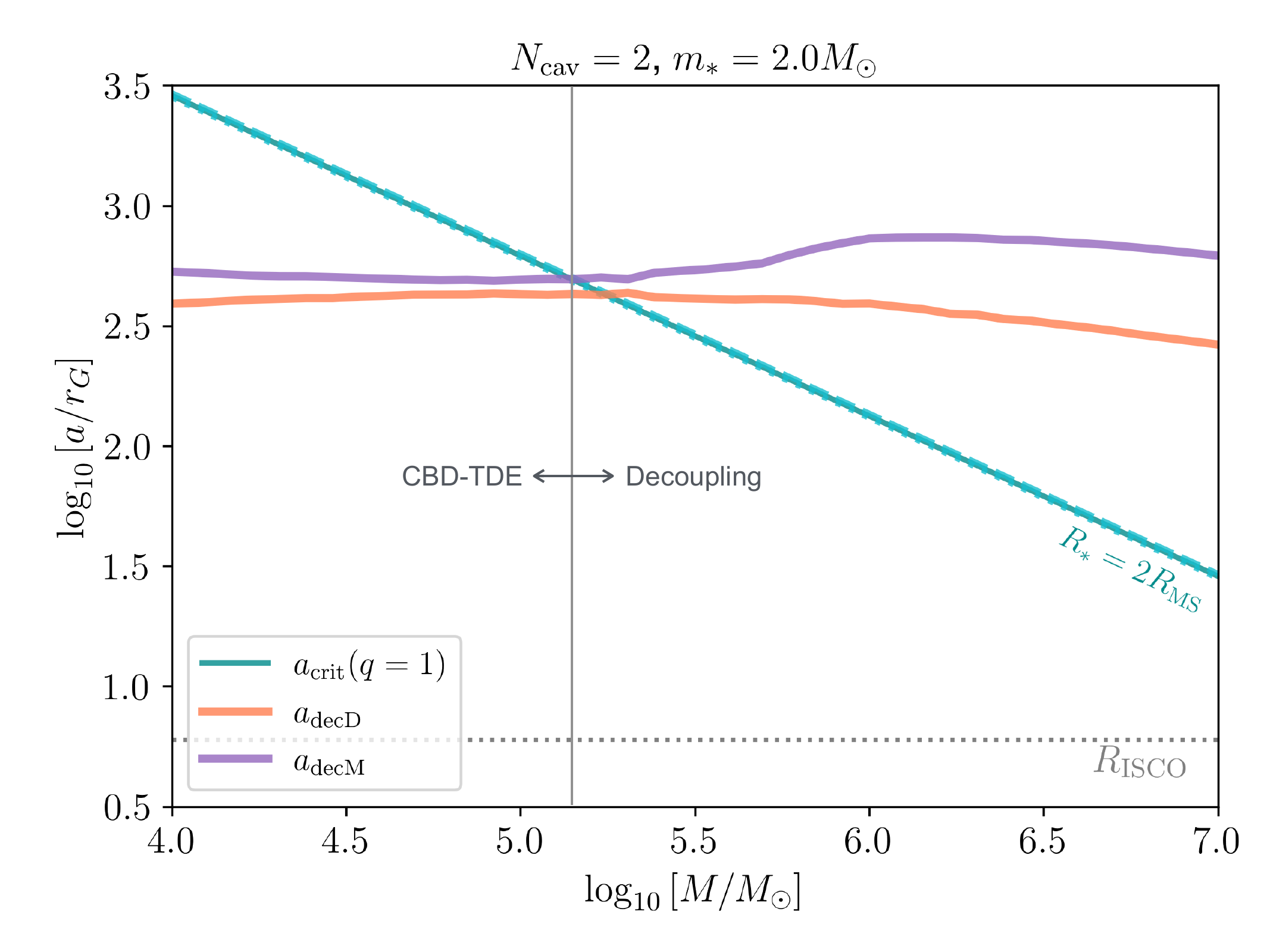} &
\hspace{-10pt}
\includegraphics[scale=0.56]{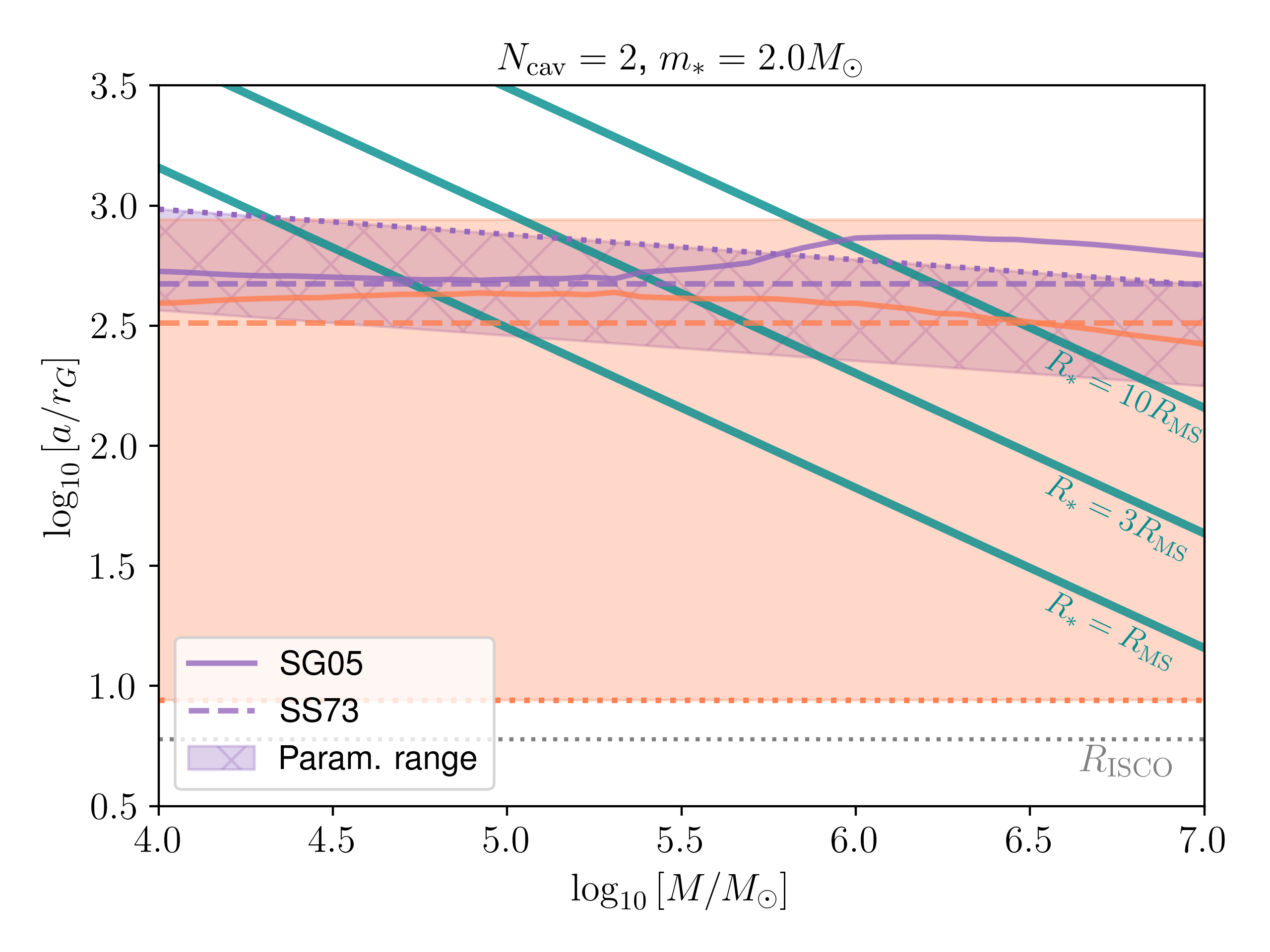} 
\end{array}$
\end{center}
\vspace{-15pt}
\caption{
\textbf{Left, Fiducial:} 
Critical semi-major axes for disruption ($a_{\crit}$, cyan line) compared to those for binary-disk decoupling ($a_{\dec D}$, orange) and binary-stellar migration decoupling ($a_{\dec M}$, purple), for fiducial values and the \citealt{Sirko_Goodman:2003} disk models. $a_{\crit}$ is also drawn for different mass ratios ($q=0.1$, dotted cyan and $q=0.39$ dashed cyan), but the variation is within the width of the $q=1$ line. {\em Disruption is allowed when the cyan line falls above the purple and orange lines.}
\textbf{Right, Parameter Dependence:} 
Critical disruption separations drawn for three stellar densities parameterized by the stellar radius.
The shaded regions indicate ranges of possible decoupling separations given a range of  disk parameters. The solid (dashed) lines show decoupling radii for fiducial disk parameters for the \citealt{Sirko_Goodman:2003} (\citealt{SS73}) disk model. Dotted orange and purple lines denote disk models with the same parameters in the extremes of the shaded regions, highlighting the inverse dependence of the two types of decoupling on disk parameters.
}
\label{Fig:acrit}
\end{figure*}
%%%%%%%%%%%%%%%%%%%%%%%%%%%%%%%%%%%%%%%%%%%%%%%%

\subsection{Decoupling vs. tidal stripping}
\label{Ss:2.4}

Figure \ref{Fig:acrit} plots critical and decoupling semi-major axes to illustrate which stellar densities, SMBHB masses, and disk parameters result in tidal stripping of the star vs. stellar decoupling from the binary. The left panel shows the fiducial case which we motivate below (see Table \ref{Table:Fid}). We parameterize the stellar density by the stellar radius at fixed mass. We choose the radius to be a multiple of a convenient, and approximately Main-Sequence relation, $R_* \propto R_{\MS}$, with $R_{\MS} \equiv \Rsun \left(m_*/\Msun\right)$. The cyan lines show the critical radii for the onset of tidal-stripping $a_{\crit}$ for a $2 \Msun$ star, with radius $R_*=2 R_{\MS} = 4 \Rsun$, and different mass ratios ($q=1$, solid; $q=0.39$, dashed; $q=0.1$, dotted). These lines can hardly be distinguished from one another, showing the weak dependence on binary mass ratio. The purple and orange lines show the migrator- and disk-decoupling semi-major-axes computed from the \cite{Sirko_Goodman:2003} model. Disk parameter values, $\alpha=5.0\times 10^{-3}$, $f_{\Edd}=0.5$, and binary mass ratio $q=0.33$ are chosen such that the two decoupling separations are close to each-other, as motivated in the previous section.
Comparison of the cyan line and the larger of the purple and orange lines shows that tidal stripping of the star in the CBD will occur for smaller SMBHBs, $M\lesssim 10^5 \Msun$. For larger SMBHBs, the star will decouple before it can be tidally stripped and no CBD-TDE is expected.

The right panel of Figure \ref{Fig:acrit} demonstrates that this result is most sensitive to the stellar density and is less sensitive to the disk model. In the right panel we drawn cyan lines for three different stellar densities, parameterized by the stellar radius at fixed stellar mass. In addition to the decoupling separations computed from the \citet{Sirko_Goodman:2003} model, we compute decoupling separations for a \cite{SS73} disk, with the same disk parameters, and represent a range of possible decoupling separations by the shaded regions.

The orange shaded region shows a range of binary separations where disk-decoupling could occur, $a_{\dec D}$ (using Eq. \ref{Eq:adec_tII} for a range of disk scale heights $h\in [0.01, 0.1]$ and viscous parameter values $\alpha \in [0.001,1]$).
The purple shaded region shows a range of binary separations where migrator-decoupling could occur, $a_{\dec M}$ (using Eq. \ref{Eq:adec_TI}). Because the migrator-decoupling separations depend on the disk scale height and density as a function of radius, we assume a \citet{SS73} (middle-region, see Appendix \ref{A:Acc_model}) disk model with a range of  $\alpha \in [0.001,1]$ and accretion rates parameterized by the Eddington ratio  $f_{\Edd} \in [0.1,1]$.

A dotted line is drawn at the extremes of each shaded region, which here corresponds to the thickest disks (largest $h$). This illustrates what was found in \S\ref{Ss:decouple}, that the two decoupling separations scale inversely with disk thickness. As argued above, this causes the smallest stellar decoupling separations to occur when $a_{\dec M} = a_{\dec D}$. Importantly, this minimum decoupling radius is rather insensitive to the value of $\alpha$, for all plotted models.
Higher values of $\alpha$ significantly decrease the disk-decoupling (orange) separation, and only slightly increase the migrator-decoupling (purple) separation. This results in the larger-of-the-two decoupling separations being nearly the same as those depicted in the left panel of Figure \ref{Fig:acrit}, even for $\alpha\rightarrow 1$ (Appendix \ref{A:disk_models}).

In summary, the maximum decoupling separation is approximately constant with binary mass, disk model, and disk parameters. Hence, stellar disruption is allowed for combinations of stellar density and binary mass which allow the stellar stripping for SMBHB separations above a value of $a_{\dec} \sim 10^{2.75} r_G$ (Figure \ref{Fig:acrit} and Eqs. \ref{Eq:adec_tII} and \ref{Eq:adec_TI}).
The right panel of Figure \ref{Fig:acrit} shows that, as expected from Eq. (\ref{Eq:acrit_eval}), the less dense stars are disrupted earlier in the inspiral and so can be disrupted by more massive SMBHBs, up to $10^6 \Msun$ for the least dense stars plotted. The denser-than-fiducial $R_*=R_{\MS}$ stars disrupt for lighter, $\lesssim 10^{4.75} \Msun$ SMBHBs.

Based on the above arguments, we adopt fiducial parameters summarized in Table \ref{Table:Fid} that lead to successful disruption before decoupling in an $\alpha$ disk. These result in an average stellar density of $\rhosun/2^5$, as used in evaluating Eq. (\ref{Eq:acrit_eval}) and throughout.
Such a puffed up star is expected from tidal heating of a Sun-like star as it moves inwards with the SMBHB on its nearly-circular orbit. This is detailed further in Appendix \ref{A:Tidal_Heating}. Accretion and formation in the disk could also alter the stellar structure away from Main-Sequence expectations (\eg, \citealt{DerdzinskiMayer:2023, DittmannCantiello:2024} and see further discussion in \S\ref{S:Discussion}). For now we simply take the above as our fiducial case to explore the disruption process. Alternatively, $\sim10^4 \Msun$ SMBHBs, would disrupt a Sun-like star before star or binary decouples from the disk.

\begin{table}
 \begin{center}
 \title{Summary of Fiducial Parameters}
    \begin{tabular}{c  c  c  c  c  c  c}
    \hline \hline
    $M/\Msun$ & $q$ & $N_{\cav}$ & $m_{*0}/\Msun$ & $R_{*0}/R_{\MS}$ & $f_{\Edd}/\eta$ & $\alpha$ \\
    \hline 
    $10^{5}$ & $0.33$ & $2$ & $2$ & $2$ & $4.5$ & $0.1$ \\
    \hline 
    \end{tabular}
    \end{center}
    \vspace{-10pt}
    \caption{\textbf{Summary of fiducial parameters.} From left to right: total binary mass, binary mass ratio, central cavity radius in units of binary separation, (pre-stripping) initial stellar mass, initial stellar radius in units of $R_{\MS} \equiv \Rsun (m_*/\Msun$), disk Eddington ratio over accretion efficiency, and viscous parameter.}
\label{Table:Fid}
\end{table}

%%%%%%%%%%%%%%%%%%%%%%%%%%%%%%%%%%%%%%%%%%%%%%%%
%%% Fig 3 GW scales %%%
%%%%%%%%%%%%%%%%%%%%%%%%%%%%%%%%%%%%%%%%%%%%%%%%
\begin{figure*}
% \vspace{-10pt}
\begin{center} $
\begin{array}{cc}
\hspace{-5pt}
\includegraphics[scale=0.56]{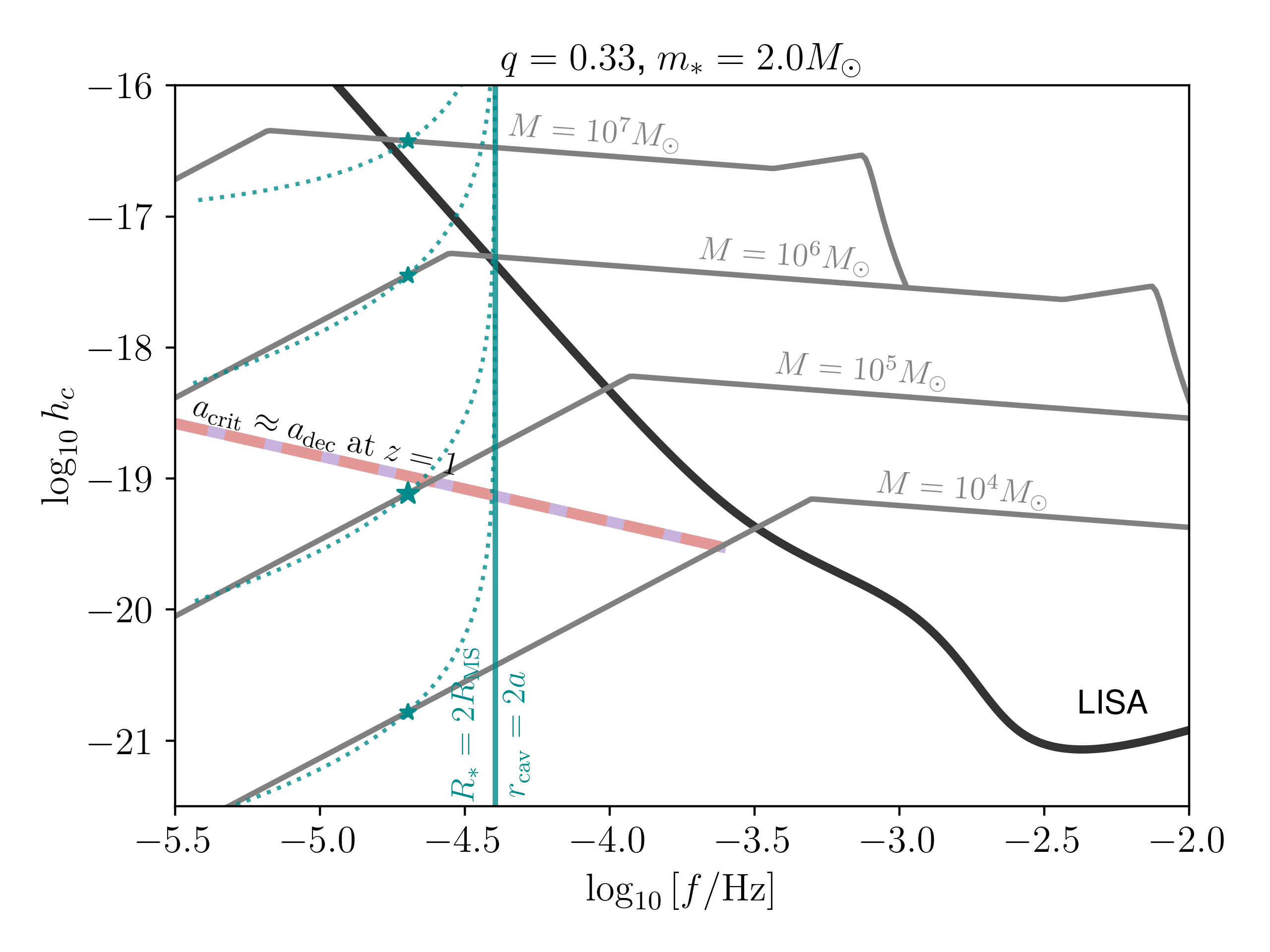} &
\hspace{-15pt}
\includegraphics[scale=0.56]{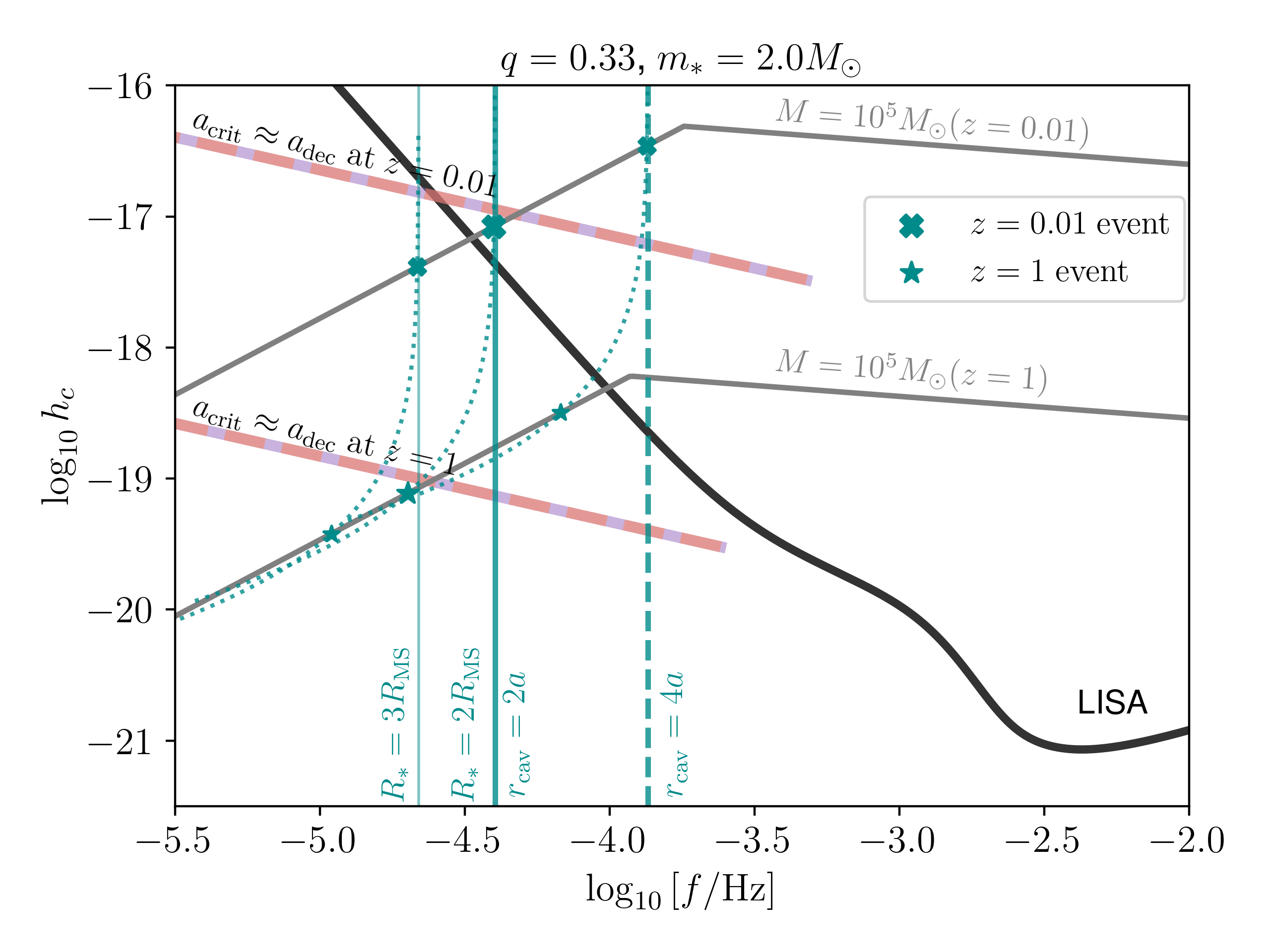} 
\end{array}$
\end{center}
\vspace{-15pt}
\caption{
\textbf{Left, Fiducial:} Gravitational wave properties of the SMBHB when stellar-stripping begins. The solid black line represents the sensitivity of LISA while the gray lines are SMBHB inspiral tracks for the labeled binary masses, mass ratio, and redshift. The vertical cyan line is the critical rest-frame frequency where the fiducial star at $\rcav=2a$ ($N_{\cav}=2$) will begin to be stripped, for all SMBH masses (corresponding to the cyan lines in Fig. \ref{Fig:acrit}). The dotted cyan lines show how this moves down and to the left with increasing redshift. The cyan stars show where disruption would begin for the different binary masses at $z=1$ (the largest symbol representing the fiducial case). Above the purple-orange dashed line, disk-or migrator-decoupling may occur before the disruption begins (Fig. \ref{Fig:acrit}).
\textbf{Right, Parameter Dependence:}   
The same as the left panel but demonstrating dependence on source redshift, lower stellar density (thin cyan $R=3 R_{\mathrm{MS}}$ line), or increased CBD cavity size (dashed $\rcav=4a$ vertical line).
}
\label{Fig:LISA}
\end{figure*}
%%%%%%%%%%%%%%%%%%%%%%%%%%%%%%%%%%%%%%%%%%%%%%%%

\subsection{Coincidence with gravitational wave emission}
The SMBHB orbital frequency at the critical binary separation is dependent only on the average stellar density, the binary mass ratio, and the CBD cavity radius. For the fiducial stellar density,
\begin{eqnarray}
    \label{Eq:fcrit}
    f_{\crit} &=& \frac{1}{2} \sqrt{ \frac{G \rhoavg }{3 \pi \Bfacmax(N_{\cav},q)}  } \\ \nonumber 
    &\approx& 2.0\times10^{-5} \mathrm{Hz} \ \left(\frac{  \Bfacmax(\Ncav,q) }{\Bfacmax(2,0.33)}\right)^{-1/2} \left(\frac{\rhoavg}{\rho_{\odot}/2^5
}\right)^{1/2},
\end{eqnarray}
where $\Bfacmax(N_{\cav},q)$ is given in Eq. (\ref{Eq:acrit}). For fixed stellar properties, this frequency varies very little with mass ratio, as evident from the very weak dependence of $a_{\crit}$ on $q$ in the left panel of Figure \ref{Fig:acrit}.
This SMBHB orbital frequency at the onset of tidal stripping corresponds to an orbital period of $13.8$ hours ($2.45$ hours for a Solar-density star). For a circular binary orbit, the observed GW frequency at the onset of disruption is $f=(1+z) 2 f_{\crit} \approx 4.0 \times10^{-5} (1+z)$~Hz ($2.3\times10^{-4} (1+z)$~Hz for a Solar-density star), which is at the low-frequency-edge of the LISA band for low redshifts. The GW characteristic strain $h_c$ can then be computed assuming a chirp mass and source redshift. 

Figure \ref{Fig:LISA} shows where stripping events fall in the observed $h_c$-$f$ space, with relation to the LISA band (black sensitivity curve from \citealt{RobsonCornsih+2019}), for various system parameter choices. 
The gray lines show PhenomA \citep{Ajith_Phenom+2007} characteristic strain tracks of SMBHB mergers with the labeled total mass and a mass ratio of $q=0.33$. Vertical cyan lines show the rest-frame GW frequency emitted by the binary when it begins to disrupt the star, and the cyan markers indicate where stellar stripping begins at the redshift indicated in the legend. The dotted-cyan curves show how the onset of disruption changes with redshift, going to lower frequency and lower strain-amplitude for sources at higher redshift. Finally, we draw an orange-purple dashed line at constant $a = 10^{2.75}r_G$ in $h_c$-$f$ space. This approximates the dividing line between systems which will decouple before stellar disruption begins (above the line) and systems which will strip the star as the SMBHB inspirals, generating a CBD-TDE (below the line; see Figure \ref{Fig:acrit}).

In the left panel of Figure \ref{Fig:LISA} we focus on the fiducial stellar density and cavity radius at $z=1$, for a range of SMBHB masses. The systems that result in stellar stripping (cyan stars below the orange-purple dashed line), occur for the less massive SMBHBs, and so just outside of the LISA band at this redshift and fiducial system parameters.

In the right panel of Figure \ref{Fig:LISA} we fix the SMBHB mass to $10^5\Msun$ and show how the onset of stellar stripping moves in $h_c$-$f$ space with redshift, CBD cavity size, and stellar density. For lower redshifts the source appears with higher GW amplitude and frequency causing the SMBHB to begin stripping the star while emitting in the LISA band (large cyan cross). A larger cavity (dashed vertical line with $r_{\cav}=4a$) causes the disruption to begin even later in inspiral and so further into the LISA band, however, such a system would likely decouple from the disk before disruption (as it lies above the $z=0.01$ orange-purple dashed line).
On the other hand, a less dense star (thin vertical cyan line with $R_*=3R_{\MS}$), causes the disruption to occur earlier during the inspiral, moving the event out of the LISA band, but safely before decoupling. Each of the above behaviors are monotonic.

The time from the onset of stellar stripping to the GW-driven SMBHB merger, assuming circular orbits, is,
\begin{eqnarray}
    \frac{T_{\GW}}{(1+z)} &\approx& 267.6 \ \mathrm{yr}\  \left(\frac{(1+q)^2/q}{5.3603}\right) \left(\frac{\Bfacmax(N_{\cav}, q)}{\Bfacmax(2,0.33)}\right)^{4/3} \nonumber \\ 
    &\times&  \left( \frac{M}{10^5 \Msun} \right)^{-5/3} \left( \frac{\rhoavg}{\rho_{\odot}/2^5} \right)^{-4/3}, 
    \label{Eq:TGW}
\end{eqnarray}
which scales strongly with stellar density (2.6 years for a Sun-like star, which disrupts in the LISA band). Hence, such CBD-TDEs occur around SMBHBs which will soon merge in the LISA band. While this is a small portion of the SMBHB lifetime, such an event does not depend on the coincidence of a star arriving in this window, rather the star can be loaded up at the cavity edge long before due to the cavity trap mechanism described above.

Figures \ref{Fig:acrit} and \ref{Fig:LISA} together suggests that stellar stripping events mediated by a disk surrounding an inspiralling SMBHB would occur for less-than-Solar density stars $\mathcal{O}(10^2)$ years before SMBHBs with masses in the range $10^4-10^6 \Msun$ enter the LISA band. Such events could occur while the SMBHB is emitting in the LISA band provided the event is at low redshift, or that CBDs mediating the inspiral are more dense and thicker than the steady-state $\alpha$-disks considered here (delaying decoupling), and tidal heating does not accelerate the tidal stripping process more than estimated here (Appendix \ref{A:Tidal_Heating}).
We next model the implied CBD-TDEs and consider their observational implications.

\section{Tidal Stripping Process}
\label{S:Disruption}

We now consider the case where the star enters the tidal radius of the SMBHB before decoupling. The stellar orbit shrinks slowly on a near-circular orbit as the binary decays until the star reaches a distance $N_{\cav}a_{\crit}$ from the binary center of mass and the outer parts of the star begin to be stripped once per encounter time with the smaller SMBH in the SMBHB.  Before investigating a detailed model of repeated stellar stripping, and whether this results in a final disruption, we first present important timescales for stellar stripping, and repetition of encounters.

Mass will be stripped from the surface of the star at approximately the free-fall time of the star,
\begin{eqnarray}
    \label{Eq:taustr}
    \tau_* \equiv \sqrt{\frac{3\pi}{32}} \frac{1}{\sqrt{G \rhoavg}} =  \frac{P_*}{2^{5/2}}  \approx 2.77 \ \mathrm{hrs} \left(\frac{\rhoavg}{\rhosun/2^5}\right)^{-1/2},
\end{eqnarray}
which we have written in terms of the stellar density and also the orbital period at the stellar surface $P_* = 2 \pi \Omega^{-1}_* \equiv 2 \pi R^{3/2}_* /\sqrt{Gm_*}$.

The time between stripping events at the onset of tidal stripping is found from the difference in angular frequencies of the stellar orbit at the position of the cavity edge and the binary orbit,
\begin{eqnarray}
    \delta \Omega &\equiv& \Omega_{b} - \Omega_{\rm{cav}}  = \sqrt{\frac{GM}{a^3} }\left(1 - \frac{1}{\sqrt{N^3_{\cav}}}. \right), 
    \label{Eq:DOm}
\end{eqnarray}
Evaluated at the critical separation, the time between stripping events at the onset of stellar stripping is,
\begin{eqnarray}
\label{Eq:Trep}
    T_{\rm{rep}} &\equiv& \frac{2 \pi}{\delta \Omega_{\crit}} =  \frac{2 \Bfacmax^{1/2} N^{3/2}_{\cav}}{N^{3/2}_{\cav}-1} P_* \\ \nonumber
    &\approx& 21.4 \mathrm{hrs} \left( \frac{\rhoavg}{\rhosun/2^5}\right)^{-1/2},
\end{eqnarray}
where we evaluated for $N_{\cav}=2$, $q=0.33$ (near the maximal value for $\Bfacmax$ at $q\approx0.39$), and fiducial stellar properties. The time between strippings depends only on $q$, $N_{\cav}$ and $\rhoavg$, because $a^3_{\crit}\propto M$.

\begin{figure}
% \vspace{-10pt}
\begin{center} $
\begin{array}{c}
\hspace{-10pt}
\includegraphics[scale=0.55]{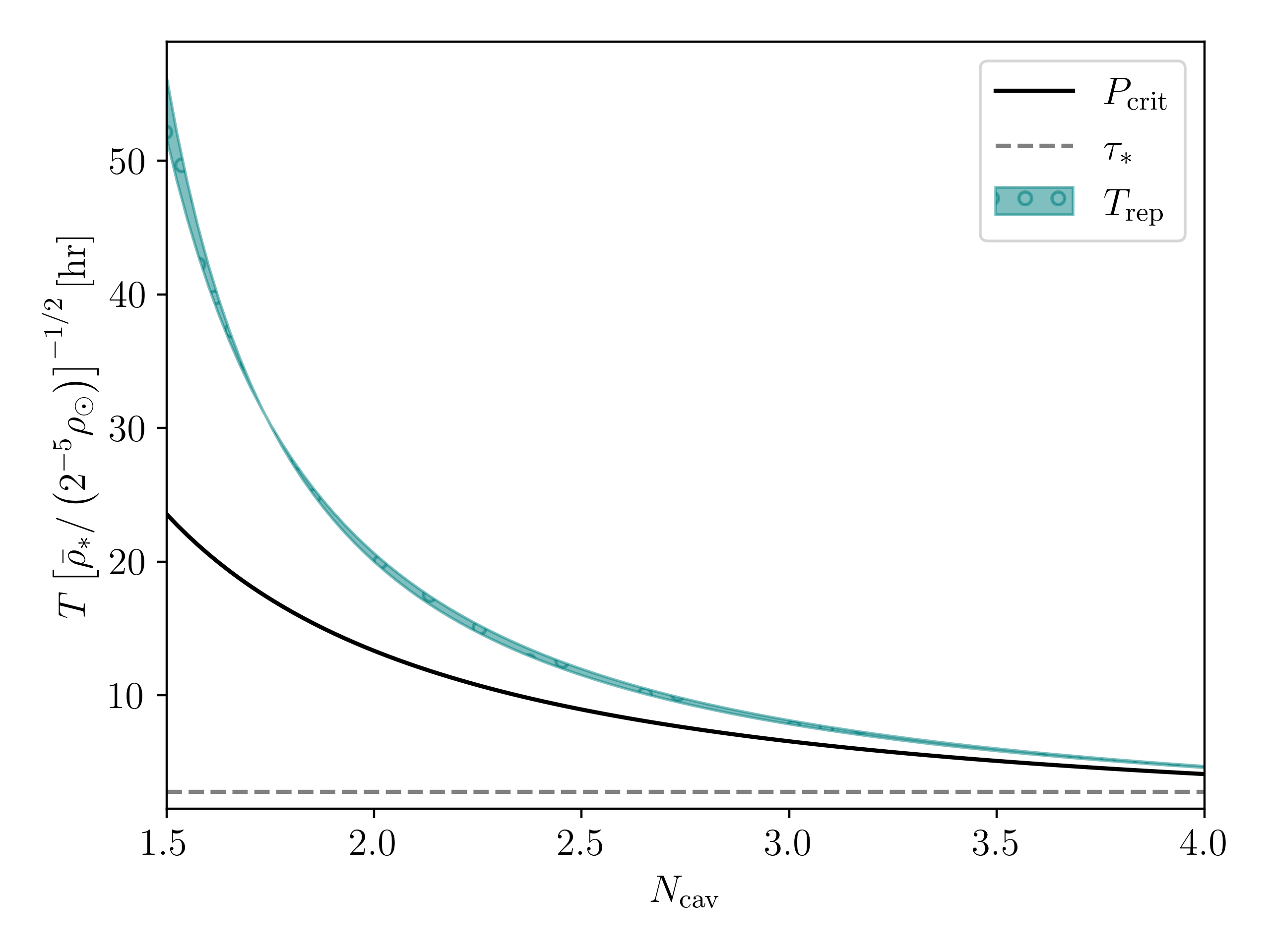} 
\end{array}$
\end{center}
\vspace{-15pt}
\caption{
Timescale to strip material from the star $\tau_*$, Time between stripping events $T_{\rep}$, and the SMBH binary orbital period at the onset of stripping $P_{\crit}$ vs. cavity size. Each of the plotted quantities is independent of SMBHB mass. The shaded blue range represents the weak binary mass ratio ($q \in [0.1,1]$) dependence of the time between stripping events.
}
\label{Fig:Trep_tstrip}
\end{figure}

Figure \ref{Fig:Trep_tstrip} shows the relationship between these timescales and the binary orbital period at the critical separation $P_{\crit}$, as a function of the relevant system parameters. 
Because the time between strippings, the stripping time, and the critical orbital period have the same dependence on $\rhoavg$, and are independent of binary mass, Figure \ref{Fig:Trep_tstrip} represents the relationship between the three timescales for all systems parameters. The time to strip the star is always shorter than the time until the next stripping.

%%%%%%%%%%%%%%%%%%%%%%%%%%%%%%%%%%%%%%%%%%%%%%%%
%%% Fig 4 timescales %%%
%%%%%%%%%%%%%%%%%%%%%%%%%%%%%%%%%%%%%%%%%%%%%%%%
%%%%%%%%%%%%%%%%%%%%%%%%%%%%%%%%%%%%%%%%%%%%%%%%

\subsection{Repeated stellar stripping model}
\label{Ss:DisruptModel}

To model the tidal stripping process we assume a polytropic star with $P=K_{\gamma}\rho^{\gamma}$. Because the stellar stripping time is always shorter than the time between strippings (Figure \ref{Fig:Trep_tstrip}), we assume that after each stripping event the star re-settles to the same polytropic relation but with a smaller mass and new radius set by the polytropic mass-radius relation. We further assume that no excess energy, \eg, from fallback or accretion, is deposited in the outer layers of the star between strippings (see \S\ref{S:Discussion} for further discussion). As a proof of principle example, we assume that the outer portion of the star can be modeled with a $\gamma=5/3$ polytrope appropriate for stars with convective envelopes. We further assume that the star orbits the binary center of mass on a circular orbit, which decays with the SMBHB orbit due to emission of gravitational radiation by the SMBHB.\footnote{The GW decay rate of the stellar orbit will be much slower, by factor $\sim (4/N^3_{\cav})(m_*/M)$.}

We build a differential equation to model the time-dependent stellar mass and mass stripping rate, $m(t)$ and $\dot{m}(t)$, (hereafter dropping the $*$ subscript on $m_*$ for brevity of notation) as the binary decays and the star is stripped by tidal forces. As shown in \citet{ChengCheng_PTDE+23}, the analytical models of \citet{Zalamea:2010} capture well the tidal stripping events simulated with full hydrodynamical calculations. Hence, we adapt these models to the case at hand. We define a small parameter,
\begin{equation}
    \Delta(t; m) \equiv R_* - R_{T} ,
\end{equation}
where $R_*$ is the time-dependent stellar radius, which expands as $R_* \propto m^{-1/3}$ when the star is stripped, for a $\gamma=5/3$ polytrope. $R_{T}$ is the time-dependent tidal radius, defined in the frame of the star (inversely from Eq. \ref{Eq:acrit}),
\begin{equation}
    R_{T} = a(t)\left[4 \Bfact(t) \left(\frac{M}{m}\right) \right]^{-1/3},
    \label{Eq:RstrTid} 
\end{equation}
where $\Bfact(t)$ is the time dependent version of $\Bfacmax$ in Eq. (\ref{Eq:acrit}), and encodes the time changing distances from each SMBH and the star,
\begin{equation}
    \Bfact(t) = \frac{q/(1+q)}{d^3_P(\delta\Omega t)} + \frac{1/(1+q)}{d^3_S(\delta\Omega t)},
   \label{Eq:rstrOFt}
\end{equation}
where $d_P$ and $d_S$ are defined in Eqs. (\ref{Eq:dPdS}), and $\delta\Omega$ is given by Eq. (\ref{Eq:DOm}) and represents an oscillating component due to stellar and SMBHB orbital motion. 
A slowly changing component of $d_P$ and $d_S$ arises from the binary semi-major axis $a(t)$ decaying due to GWs \citep{Peters64},  $a(t) = \left[a^4_{\crit} - 4\beta_{\GW} t \right]^{1/4}$,
for $\beta_{\GW} \equiv (64/5) (G^3/c^5)M^3 q/(1+q)^2$.

We then define an instantaneous accretion rate in the limit that $\Delta$ remains small throughout the partial disruption phase by computing the mass that can be liberated in the shells between $R_T$ and $R_*$, and dividing by the stellar stripping time (Eq. \ref{Eq:taustr}),
\begin{equation}
    \dot{m} \approx 4 \sqrt{\frac{32\pi}{3}} \sqrt{G\rhoavg}  R^2_* \int^{\Delta}_{0}{\rho(z) dz},
\end{equation}
where $\rhoavg$ is allowed to evolve in time with the changing stellar properties, and $\rho(z)$ is derived from hydrostatic equilibrium in the plane-atmosphere limit,
\begin{equation}
    \rho(z) = \left(\frac{\gamma-1}{\gamma}\frac{Gm}{K_{\gamma} R_*} \right)^{\frac{1}{\gamma-1}}\left(\frac{z}{R_*} \right)^{\frac{1}{\gamma-1}},
\end{equation}
which recovers the expression in \cite{Zalamea:2010} for $\gamma=5/3$.

Integrating, the stellar mass loss rate\footnote{This may be an overestimate, as we do not account for the time for the material to flow out of the L1 and L2 nozzles \citep[\eg,][]{LinialSari:2017}. This will ultimately delay the runaway disruption process below, but not change the qualitative picture.} is related to stellar mass as,
\begin{widetext}
\begin{equation}
    \dot{m} = 8 \sqrt{2} \left(\frac{\gamma-1}{\gamma}\right)^{\frac{\gamma}{\gamma-1}} \left( K_{\gamma}\right)^{-\frac{1}{\gamma-1}} \left( Gm \right)^{\frac{\gamma+1}{2(\gamma-1)}}  \left(R_*\right)^{\frac{3\gamma-5}{2(\gamma-1)}} \left( \frac{\Delta}{R_*} \right)^{\frac{\gamma}{\gamma-1}} ,
    \label{Eq:mDEQ}
\end{equation}
\end{widetext}
which is a differential equation for the stellar mass as a function of time, when considering stripping by the time changing potential of the orbiting and decaying SMBHB potential.

The adiabatic constant for our fiducial case is given by the standard polytropic relation, $K_{5/3}\approx0.424 G m^{1/3} R_*$, and is fixed by a choice of initial stellar mass $m_0$ and radius $R_{*0}$. We numerically solve this differential equation for $\dot{m}(t)$ and $m(t)$ with initial condition $m(0) = m_0$. We integrate until $R_T = R_{\mathrm{c}} =\kappa_c R_*(t)$, with $\kappa_c \approx 0.51$, which is the location of maximal self gravity for the $\gamma=5/3$ polytrope, where full disruption ensues \citep{CoughlinNixon:2022}.

%%%%%%%%%%%%%%%%%%%%%%%%%%%%%%%%%%%%%%%%%%%%%%%%
%%% Fig 5 mdot PDE %%%
%%%%%%%%%%%%%%%%%%%%%%%%%%%%%%%%%%%%%%%%%%%%%%%%
\begin{figure*}
% \vspace{-10pt}
\begin{center} $
\begin{array}{cc}
\includegraphics[scale=0.65]{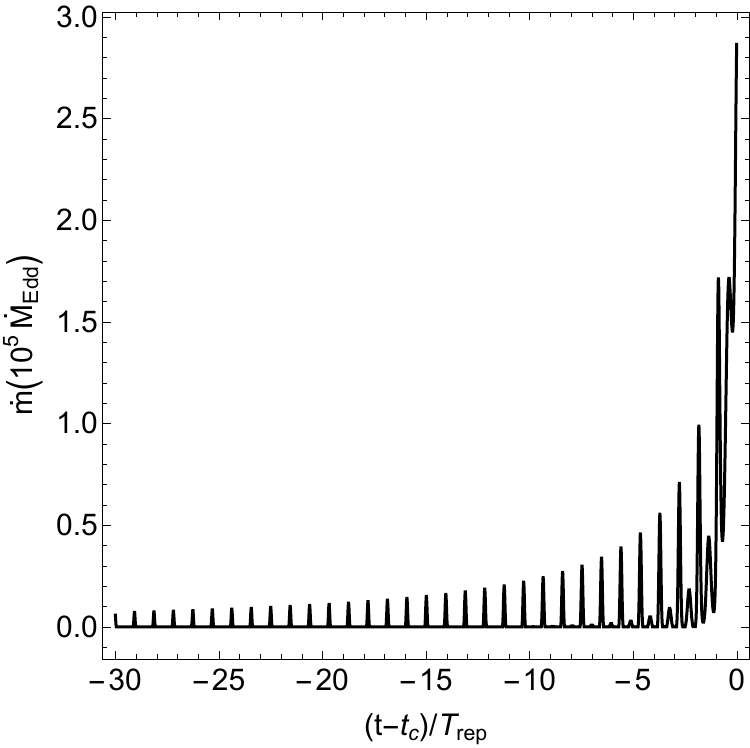} &
\includegraphics[scale=0.65]{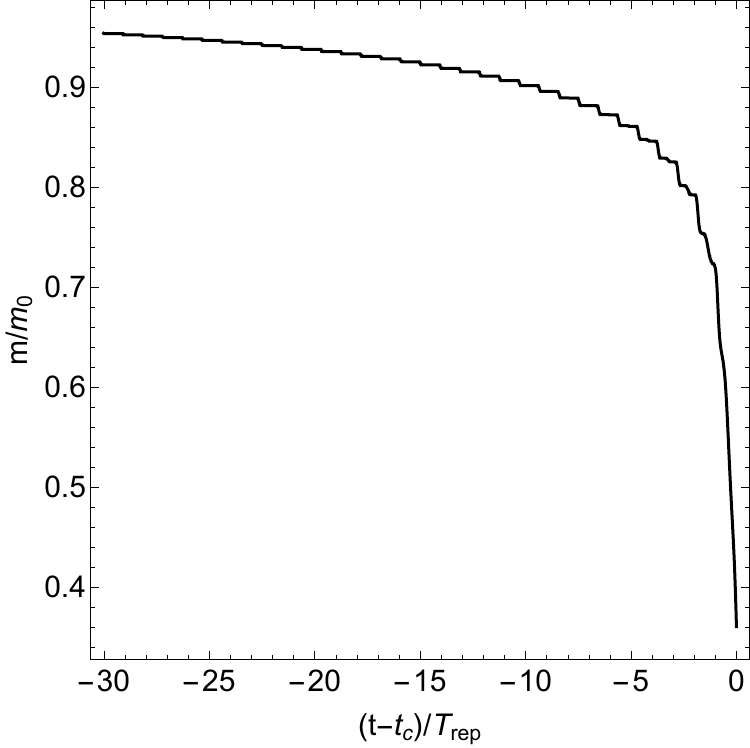}
\end{array}$
\end{center}
\vspace{-15pt}
\caption{
Result of solving Eq. (\ref{Eq:mDEQ}) including tidal stripping from both the primary and secondary SMBHs, up to the point of full disruption, for a $\gamma=5/3$ polytropic star and the fiducial system (Table \ref{Table:Fid}). This shows the last $32$ encounters (stripping events) representing the final $\approx 27$ days before full disruption, during which time the majority of the star is stripped. The onset of tidal stripping begins at $t=0$, $\approx21$ years before full disruption at time $t_c$. The SMBHB will merge $\approx246$ years after the full disruption. Times are in the source frame. Results are representative of solutions with $N_{\cav}=2$ and intermediate binary mass ratios (see \S\ref{Ss:scalings}).
}
\label{Fig:disrupt}
\end{figure*}
%%%%%%%%%%%%%%%%%%%%%%%%%%%%%%%%%%%%%%%%%%%%%%%%

Before solving the equation, we first observe that the stripping process will runaway (\ie, unstable mass-transfer) when $\Delta \geq 0$ and $\dot{\Delta} > 0$. The first condition is satisfied for binary separations below the critical separation. The latter condition will always be true as long as the radius of the star goes as $m$ to a negative exponent ($-1/3$ for the case at hand) and as long as $\dot{m}, \dot{a} < 0$. This is true in our case, where the star is losing mass for $\Delta \geq 0$ and $a$ is decreasing. Hence, the stripping events will penetrate deeper and deeper into the star liberating more and more mass until time $t_{\core}$ defined by $R_T(t_\core) = R_{\core}(t_{\core})$, when the entire star is disrupted. Using Eq. (\ref{Eq:RstrTid}) and $R_c(t) = \kappa_c R_{*0} (m(t)/m_0)^{-1/3}$ This condition can be simplified to,
\begin{equation}
    \frac{m(t_\core)}{m_0} = \kappa^{3/2}_c \left( \frac{P_{\crit}}{P(t_\core)}\right) \approx \kappa^{3/2}_c,
    \label{Eq:mcore_est}
\end{equation}
where $P$ is the binary orbital period and the approximation follows because, as we will see from solving Eq. (\ref{Eq:mDEQ}), the binary evolution timescale is slow compared to the runaway stripping of the star, and so $P_{\crit} \approx P(t_c)$. For the $\gamma=5/3$ polytrope, this implies a disrupted core mass of $m(t_\core) \approx (0.51)^{3/2}m_0 \approx 0.36 m_0$.

Figure \ref{Fig:disrupt} shows that this is indeed the case with an example solution to Eq. (\ref{Eq:mDEQ}) for the fiducial system 
(Table \ref{Table:Fid}). The left panel shows the instantaneous stripping rate of the star in units of $10^5 \MEdd(M)$ (in units of Eddington for the total binary mass) versus time before full disruption, in units of the time between encounters at the critical radius, $T_{\rep}$ (Eq. \ref{Eq:Trep}). The right panel shows the corresponding mass of the star over time, normalized to its initial mass.  

Stripping of the star occurs in sharp spikes lasting $\approx 0.1 T_{\rep}$ ($\approx 2$ hrs) each and separated by $T_{\rep} \approx 21.4$ hrs. The symmetric spikes correspond to the portion of the stellar and binary orbit where the tidal scale defined in Eq. (\ref{Eq:RstrTid}) passes within the stellar radius, peaking at closest approach to the secondary black hole. Tidal stripping is indeed a runaway process that intensifies over time as the SMBHB decays in its orbit via GW emission and as the star's radius grows with each stripping ($R_* \propto m^{-1/3}$). While this is not the rate that matter falls onto the binary, we have scaled it in terms of the Eddington mass accretion rate onto the binary for reference. Near the end of the repeated stripping process, the stripping rates reach a few times $10^5 \MEdd$. This can be understood simply from dividing the mass lost in a stripping event, which approaches $0.1\Msun$ in the final passages, to the stripping time of $\sim 3$~hours (Eq. \ref{Eq:taustr}), resulting in rates of $\approx 300 \Msun$ per year. Compared to the Eddington rate for a $10^5 \Msun$ SMBH of $10^{-3} \Msun$ per year, this gives $\approx 3\times 10^5 \MEdd$.

During the final $5-10$ disruption events, weaker and wider stripping spikes appear between the main spikes. These occur at the second peak of the time-dependent binary tidal field occurring at the closest approach of the primary to the star. The feature is of course dependent on the binary mass ratio, such that for an equal mass SMBHB, the period of spikes is doubled from the disparate mass-ratio case. Over the final two-to-three passes, $\dot{m}$ never reaches zero over an orbit and the star is continuously stripped.\footnote{The $\Delta \ll R_*$ approximation assumed to derive Eq. (\ref{Eq:mDEQ}) only begins to break down in these final orbits before full disruption.}  In agreement with our estimate above, approximately two-thirds of the star is stripped before the tidal radius of the binary finally surpasses the radius of maximum stellar gravity resulting in a complete disruption of the remaining core of mass $m_c\approx0.36 m_{0}$. 
This final disruption (not plotted in Fig.~\ref{Fig:disrupt}) would unleash a third of a solar mass into the CBD in a few hours, at $\sim 10^6 \MEdd$. We examine the fate of this gas released into the disk in \S\ref{S:AccretonEmission}.

While Figure \ref{Fig:disrupt} shows the final 32 encounters ($\approx27$ days in the source frame), 
tidal stripping begins at $t=0$, $\sim 8600$ stripping times before final disruption, or $\approx 21$ years in the binary source frame. 
The orbital frequency of the SMBHB changes by $3\%$ during this time, not appreciably deviating from the starting position marked in Figure \ref{Fig:LISA} (large cyan star), and validating the approximation used to estimate the remaining core mass in Eq. (\ref{Eq:mcore_est}). After final stellar disruption, the SMBHB will merge in $\approx 246(1+z)$ years. 

Due to modeling uncertainties\footnote{Numerically, this number converges with integrator working precision but requires tracking small changes in mass over many strippings. Furthermore, the result may be sensitive to modeling approximations such as multidimensionality and the details of overflow through the Lagrange points \citep[\eg][]{LinialSari:2017}.}, the inferred time from the beginning of the stripping event to the full disruption is less certain than the shape of the mass loss curves and the final core mass.  We do not, however, expect the duration of this stage to differ drastically from the estimate here -- it likely depends more strongly on stellar parameters than modeling uncertainties.
Regardless, the most important and observationally relevant stage of evolution is the inevitable final tens of strippings, shown in Figure \ref{Fig:disrupt}.

\subsection{Scaling with system parameters}
\label{Ss:scalings}

The shape of the time evolution of the stripping event depicted in Figure \ref{Fig:disrupt} is largely unchanged for different binary masses and stellar densities. The stripping rate in units of Eddington scales nearly linearly with the binary mass $M$, while $\rhoavg$ scales the x-axis by dictating at what binary separation tidal stripping begins, the stripping rate, and the time between stripping events (Figure \ref{Fig:Trep_tstrip}).  For example, for a fixed stellar mass with a factor of $\mathcal{X}_*$ change in stellar radius, the time between stripping events and the stripping time will change as $\mathcal{X}^{3/2}_*$, and the event will occur with a factor of $(M/10^5\Msun)^{5/3}(\mathcal{X}_*)^{4}$ change in the time until SMBHB merger.

Changing $q$ primarily dictates when secondary spikes in the stripping rate occur due to the primary SMBH. While smaller $q$ does not appreciably change when stripping begins, it does slow down the SMBHB inspiral, resulting in a slower disruption process.

Changing $N_{\cav}$ can change the shape of the disruption mass-loss curve in Figure \ref{Fig:disrupt} and accelerate the stellar stripping over time. This is because larger $N_{\cav}$ results in the stripping occurring later in the inspiral, when the binary orbit is shrinking more quickly, and the star is more quickly overflowing its Roche-lobe. Note, however, that cases with larger $N_{\cav}$ are more likely to result in decoupling before disruption (\S\ref{S:Setup}).

All of the above parameter changes only have a small effect on the final core mass, weakly changing the balance of Eq. \ref{Eq:mcore_est}). The system always strips unstably with a final core mass before total disruption of $(0.3\rightarrow0.5) m_0$, with variations due to the amount of binary evolution during stripping, and to where in the stripping cycle the core (complete disruption) radius is reached.

\section{Fallback, Accretion and EM emission}
\label{S:AccretonEmission}

We next consider the fallback debris stream. Stripped mass from the star, with specific orbital energy $\epsilon_{\rm d}$, will fall back to the star on the Keplerian orbital timescale \citep{EvansKochanek1989},
\begin{equation}
    t_{\mathrm{fb}} = \frac{\pi}{\sqrt{2}}\frac{GM}{\epsilon^{3/2}_{d}}.
    \label{Eq:tkep}
\end{equation}
The specific orbital energy of debris stripped from the star, $\epsilon_{\rm d}$, is distributed by a spread in specific energies, $\delta \epsilon$, caused by the tidal force: 
\begin{eqnarray} \epsilon_{\rm d} = \epsilon_{\ast} \pm \delta \epsilon, \label{Eq:eps_str} 
\end{eqnarray}
where $\epsilon_\ast$ is the binding energy of the star and is negative by definition. For a standard single-SMBH TDE, where a star plunges on a nearly parabolic orbit, the stellar orbital energy is approximately zero, resulting in $\epsilon_{\rm d} = \pm \delta \epsilon$ \citep{Rees1988,EvansKochanek1989}. In contrast, $\epsilon_{\rm d} = \epsilon_\ast \pm \delta \epsilon$
for eccentric TDEs
or $\epsilon_{\rm d}=-\epsilon_\ast\pm\delta\epsilon$ for hyperbolic TDEs  \citep{Hayasaki+2013,Hayasaki+2018,Park2020}.

For a CBD-TDE with sufficiently small orbital eccentricity $e_*$, the specific orbital energy of the star is given by 
\begin{eqnarray}
\epsilon_\ast 
= -\frac{GM}{2r_{\rm cav}},
    \label{eq:binde}
\end{eqnarray}
where $r_{\rm cav}=N_\cav{a}$ represents the semimajor axis of the stellar orbit around the binary barycenter. Using the framework of single-SMBH TDEs, the energy imparted by binary-SMBH tidal forces can be approximated by the first-order term in the Taylor series of the binary tidal potential, $\phi(d)$, as 
\begin{eqnarray} 
\delta \epsilon 
&=& 
\phi(d_{P}+\Delta d, d_{S}+\Delta d) - \phi(d_{P}, d_{S}) 
\nonumber \\
&\approx& 
\left[\frac{GM_P}{d_P^2} + \frac{GM_S}{d_S^2}\right]R_{*},
\label{eq:binarytidalpot}
\end{eqnarray} 
where $\Delta d/d\ll1$ and $\Delta d = R_{*}$ is adopted in our case. Substituting Eqs.~(\ref{eq:binde}) and (\ref{eq:binarytidalpot})
%, at the stellar orbital phase of maximal tidal stripping ($\xi=\pi$ in Eqs. \ref{Eq:dPdS}), 
into Eq.~(\ref{Eq:eps_str}), gives the orbital energy of the stellar debris in CBD-TDEs:
\begin{eqnarray}
\epsilon_{\rm d}
&=&
-\frac{GM}{2r_{\rm cav}}
\pm
\left[\frac{GM_P}{d_P^2} + \frac{GM_S}{d_S^2}\right]R_{*}
\label{eq:cbdtde-debris-energy}
\end{eqnarray}

Eq.~(\ref{eq:binarytidalpot}) indicates that the spread in orbital energy of the debris is small compared to the orbital energy of the star. This means that the fractional change, $\delta r_{\rm d}$, in orbital radii of the stellar debris stream, for a circular orbit, must also be small. Using $\delta r_{\rm d}$, the binding energy of the stellar debris can be expressed by $\epsilon_{\rm d}=-(1/2)GM/(r_{\rm cav}-\delta r_{\rm d})$.
Equating it with Eq.~(\ref{eq:cbdtde-debris-energy}) yields
\begin{eqnarray}
    \label{Eq:stream_spread}
     \frac{\delta r_{\mathrm{d}}}{r_{\cav}} &=& \left|-\frac{\delta \epsilon}{\epsilon_*}\right|
%     \frac{a_{\crit}}{r_{\cav}} 
     = 2 \mathcal{C} N_{\rm cav}
     \left(\frac{R_*}{a_{\crit}} \right),
     \\ \nonumber
     \mathcal{C} &\equiv&   \frac{(1+q)}{\left[ (1+q)N_{\cav} + q\right]^2} + \frac{q(1+q)}{\left[ (1+q)N_{\cav} - 1\right]^2}, 
\end{eqnarray}
where Eq.~(\ref{Eq:dPdS}) with $\xi=\pi$ and $a=a_{\rm crit}$ was used for the derivation.
Because $2 N_{\rm cav}\mathcal{C} \sim 1$ and $R_*/a_{\crit} \sim (m_*/M)^{1/3} \ll 1$, this quantity is always small\footnote{In agreement with the tidal approximation used to derive Eq. (\ref{Eq:acrit}).}, and the stream orbits with very little deviation from the stellar orbit. Indeed, in units of stellar radius, $\delta r_{\mathrm{d}}/R_* = 2 \mathcal{C} N_{\cav}^2 \sim 2$\footnote{The same conclusion can be found from comparing the escape velocity of debris from the star and the stellar orbital velocity.}.
Compared to a scale height of the disk, $\delta r_{\mathrm{d}} / H = (\delta r_{\mathrm{d}} / r_{\cav})(r_{\rm cav}/H)$, which is of order unity for a typical disk aspect ratio, $H/r_{\rm cav}\sim10^{-2}$. 
Therefore, material stripped from the star will follow the orbit of the star with a radial extent of approximately one stellar diameter and fill the co-orbital region of the stellar orbit (with approximate size $H$).

In this case, the minimum and maximum times for stripped debris to fall back to the star along its circular orbit are 
\begin{eqnarray}
    \label{Eq:tminmax_bary}
      t_{\min, \max} &=& \frac{2 \Bfacmax^{1/2} N^{3/2}_{\cav} }{\left[ 1 \pm 2 N_{\cav} \mathcal{C} \left(4 \Bfacmax \frac{M}{m_*} \right)^{-1/3} \right]^{3/2} } P_*. 
\end{eqnarray}
Compared to the stellar orbital time around the binary barycenter, $P_{\cav} \equiv 2 \pi r^{3/2}_{\cav} / \sqrt{G (M+m_*)}$, we see, as expected, that some of the stripped material leads the star in its orbit while some trails behind. In the rotating frame of the stellar orbit, the trailing and leading debris streams meet each other on the opposite side of the cavity, after time,
\begin{equation}
    t_{\mathrm{col}} = \frac{1}{2}\frac{t_{\min} P_{\cav}}{P_{\cav} - t_{\min}},
    \label{Eq:tcol}
\end{equation}
which is approximately $17$ stripping events, or $26$ binary orbital periods, for fiducial parameters.

Because the stream orbits are displaced in radius by the same amount as their widths $\sim R_*$, they may impact each-other. However, the relative velocity of the streams when they collide will be low $\delta v_{\mathrm{d}}/v_* = 2 (R_*/r_{\cav}) \sim 10^{-2}$ compared to the Keplerian orbital velocity of the star $v_*$. The Mach number of the stream collision (assuming background disk sound speed $c_s$) will be 
$\Machstm = \delta v_{\mathrm{d}}/c_s \approx 2 R_*/H$. Hence, for $R_* \gtrsim H$, the stream collision may result in a shock that will dissipate energy. However, given the small shock velocity $\delta v_{\mathrm{d}}$ compared to Keplerian, and that the minimum post-shock velocity for a strong non-radiative shock is $(1/4) \delta v_{\mathrm{d}}$, the shock-induced change in stream orbital energy, and so orbital radius, will be only a $(\delta v / v_*)^2 \sim 10^{-4}$ level effect.
Shocks will also provide a negligible heat source given that the specific energy converted into heating is $1/2 (\delta v)^2 \sim 10^{-8}c^2$ for our fiducial system. Then even for mass delivered to the shock at the maximum stripping rate of $10^5\MEdd$, the heating rate is $\lesssim 10^{-3} L_{\Edd}$. Stream-stream shock heating will only weakly affect disk dynamics and will not generate bright, Eddington-level, flares.

\subsection{Model for accretion of disrupted material}
\label{S:Mdotbin_Model}
We first first compute the accretion rate of disk and stream material onto the binary components before considering observational features in the next subsections.

While the stellar debris streams are not immediately pulled onto the binary components, they can be accreted via gas streamers from the CBD cavity edge.
Due to the oscillating, quadrupolar nature of the binary gravitational potential, gas streams can be pulled from the edge of the CBD cavity through the binary L2 and L3 Lagrange points and fall ballistically towards the perturbing SMBH in the binary. This process of stream generation was first described by \citet[][]{ArtyLubow:1996, LubowArty:1997, Hayasaki:2007}, explored further by \cite[\eg,][]{ShiKrolik:2012:ApJ, DHM:2013:MNRAS, D'Orazio:CBDTrans:2016}, and observed in many recent studies.
As material fills the co-orbital region of the star, at the cavity edge, the binary will pull in streams, which we refer to as CBD streams hereafter, from the accumulating ring of disk and stellar material, once-to-twice per orbital period (depending on the binary mass ratio).

In the simplest picture, part of the CBD-stream that is pulled into the cavity impacts the accretion disk around the individual SMBH(s), mini-disk(s) hereafter, and deposits a fraction $f_{\acc}$ of its mass to the minidisk, which is then accreted. The remaining CBD-stream mass is ejected outwards, where it impacts and re-joins the edge of the CBD cavity. Numerical hydrodynamical simulations measure $f_{\acc}$ to be of order $1-10\%$ \citep{DHM:2013:MNRAS, Tiede:2022}\footnote{
By assuming simply that part of the stream is delivered to the SMBHs and part is flung out into the disc, we can recover the basics of the circular orbit binary problem while ignoring more complicated processes by which matter from the stream can transfer between SMBHs via the binary L1 point \citep[\eg,][]{Bowen_slosh+2017}, or intersect the other SMBH through more complicated dynamics within the cavity \citep[\eg,][]{Dunhill+2015}.}. Hence, even though material pulled from the cavity edge has higher specific angular momentum than the binary, it can still be accreted by being pulled into the cavity and dissipating excess angular momentum there. Furthermore, the formation of minidisks themselves occurs through a similar process \citep[see, \eg, Fig.~3 of][]{SBCC+2024}, suggesting that existing minidisks are not required for the CBD streams to deliver material to the SMBHs.

The majority $(1-f_{\acc})$ of a CBD stream, with mass $M_{\CBDstm}$, that is pulled into the cavity is expelled back into the CBD, adding mass to the cavity edge. Without stellar debris also feeding material to the cavity edge, the process of CBD-stream generation and expulsion mediates orbit-averaged steady-state inflow in the disk, \ie, over the course of a binary orbital period $P$, the mass flux through a ring at the edge of the cavity is $\dot{M}_{0} - M_{\CBDstm}/P + (1-f_{\acc}) M_{\CBDstm}/P = 0$, where $\dot M_0$ is the accretion rate at infinity -- in steady-state $\dot{M}_{0}= f_{\acc} M_{\CBDstm}/P$.

When we include the stripped stellar debris, extra mass is added to the ring at each stripping, at an average rate of $\left< \dot{m} \right>_T \approx \left< \dot{m} (\tau_* / T_{\rep}) \right>$, which is the average instantaneous rate spread out over the time between stripping events\footnote{The time for the material to spread in azimuth (Eq. \ref{Eq:tcol}) represents a delay time rather than a rate decrease.}. This causes mass to build up at rate $\dot{M}_{\ring}$, from which more massive CBD streams are pulled, resulting in higher CBD-stream accretion rates and higher mass deposition rates at the cavity edge. 

Because the viscous spreading time of the accumulating ring is longer than the time between stripping events that fill the disk with new material, we model $\dot{M}_{\ring}$, and the resulting enhanced accretion onto the binary, with a conservation equation for the mass in the co-orbital ring of the star.
To do so we assume that the mass pulled from the CBD ring into the CBD-stream is proportional to the mass in the ring ($M_{\mathrm{CBDstm}} \propto M_{\ring}$), and account for rates at which mass enters and exits (see Appendix \ref{A:Acc_model} for more details),
\begin{eqnarray}
    \dot{M}_{\ring} &=& \dot{M}_0\left( 1 - \frac{M_{\ring}(t)}{M_{\ring,0}} \right) +  \left<\dot{m}_*\right>_T + m_c\delta(t-t_c), \nonumber \\
    \label{Eq:Mring}
\end{eqnarray}
where, $M_{\ring,0}$ is the initial, steady-state local disk mass before disruption, and $m_c$ is the remaining stellar core mass released in the final disruption at time $t_c$.
Terms on the right-hand-side represent the inflow of mass to the cavity edge from infinity, the net rate at which mass is delivered via CBD streams to the minidisks, the rate at which tidal stripping adds material to the disk, and a step function addition of stellar core mass $m_c$ from the final disruption.

We solve this equation using the solution for $\dot{m}(t)$ (\S\ref{Ss:DisruptModel})\footnote{In practice we compute $\left<\dot{m}_*\right>_T$ by solving Eq.~(\ref{Eq:mDEQ}), with the time-independent $\mathcal{B}_{\max}$ replacing $\mathcal{B}(t)$, and re-scaling the time coordinate so that the entire disruption process has the same duration as in the full, oscillating solution.} and initial condition set by steady-state solutions for the disk surface density $\Sigma$ \citep[inner region with $b=1$ in][]{HKM09}, $M_{\ring}(0) = M_{\ring,0} = 2 \pi \left[\Sigma r \right]_{r_{\crit}} \Delta R$, which is evaluated at the cavity, at the critical separation. For the ring width we take the width of the region over which stellar debris is deposited into the disk, $\Delta r = 2 R_*$.

The solution for $M_{\ring}(t)$ allows us to compute the accretion rate onto the binary, which simplifies to $\dot{M}_{\mathrm{bin}} = M_{\ring}(t)/\tdecay$ with $\tdecay \equiv M_{\ring,0}/\dot{M}_{0}$ (see Appendix \ref{A:Acc_model}). The left panel of Figure \ref{Fig:Acc_Mdodel_Ex} shows this computed accretion rate onto the binary in orange while the black line shows the input, averaged stripping rate $\left<\dot{m}_*\right>_T$, for the same fiducial system as in Figure \ref{Fig:disrupt} (Table \ref{Table:Fid}).

%%%%%%%%%%%%%%%%%%%%%%%%%%%%%%%%%%%%%%%%%%%%%%%%
%%% Fig 6 Accretion %%%
%%%%%%%%%%%%%%%%%%%%%%%%%%%%%%%%%%%%%%%%%%%%%%%%
\begin{figure*}
\begin{center} 
$
\begin{array}{cc}
\includegraphics[scale=0.7]{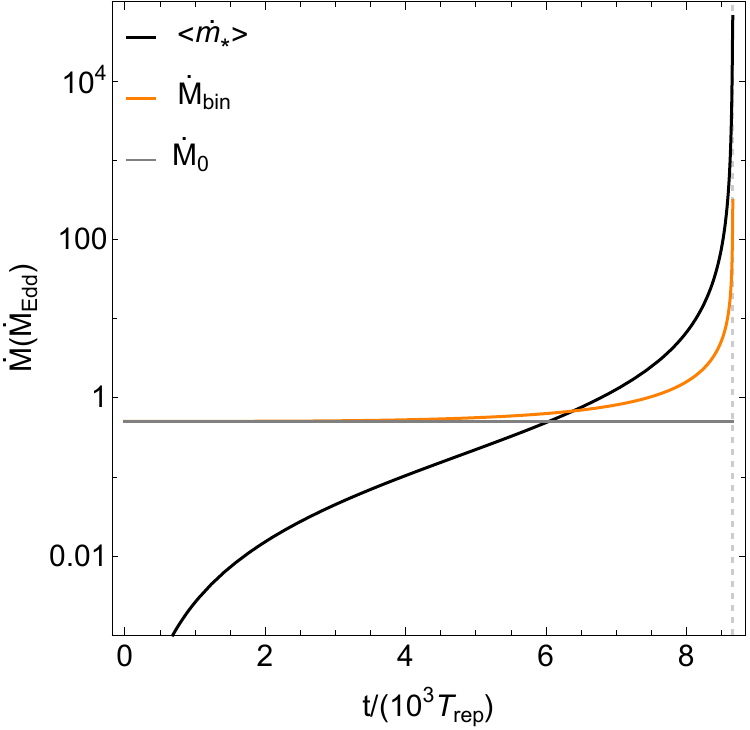} & 
\includegraphics[scale=0.7]{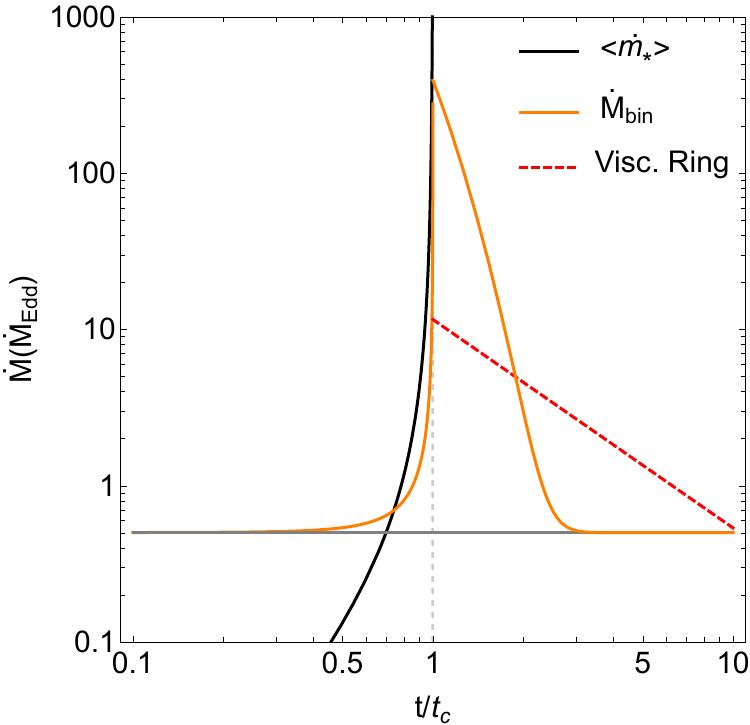} 
\end{array} $
\end{center}
\vspace{-15pt}
\caption{
\textbf{Left:} The average stellar-stripping rate (black line) and solution to Eq. (\ref{Eq:Mring}) for the amplitude of the accretion rate onto the binary (orange), computed from the beginning of the disruption process to full stellar disruption (grey-dashed line). \textbf{Right:} Continuation of the left panel after full stellar disruption and the final decay. We use fiducial system parameters (Table \ref{Table:Fid}) corresponding to the event modeled in Fig. \ref{Fig:disrupt}.
}
\label{Fig:Acc_Mdodel_Ex}
\end{figure*}
%%%%%%%%%%%%%%%%%%%%%%%%%%%%%%%%%%%%%%%%%%%%%%%%

We can learn key features of the solution and implications for binary accretion by considering the post-disruption accretion rate onto the binary. Solving Eq. (\ref{Eq:Mring}) for $t>t_c$, we find,
 \begin{equation}
    \dot{M}_{\mathrm{bin}} = \dot{M}_0  \left[ \left( \frac{M_{\ring, c} + m_c}{M_{\ring,0}} - 1\right) e^{-(t-t_c)/\tdecay} + 1  \right] ,
\end{equation}
where $M_{\ring,c}>M_{\ring,0}$ is the solution to Eq. (\ref{Eq:Mring}) just before core mass $m_c$ is added to the disk in the final disruption. 

First notice that the peak accretion rate compared to the quiescent accretion rate $\dot{M}_0$ is given by the ratio of stripped stellar mass retained in the disk at full stellar disruption ($M_{\ring,c} + m_c$) to the initial local disk mass $M_{\ring,0}$. Because the stripping is a runaway process that ends in a final disruption of $(0.3-0.5) m_0$, this means that the approximate maximum binary accretion rate enhancement is 
\begin{equation}
    \dot{M}^{\mathrm{peak}}_{\mathrm{bin}} \approx \left( \frac{m_0}{M_{\ring,0}} \right)\Mdot_0.
    \label{Eq:Mdotbin_peak}
\end{equation}
The binary accretion rate will approach this peak soon after the final core disruption of the star and then decay exponentially back to the steady-state disk value on characteristic timescale $\tdecay$, which is the viscous time required for gas to diffuse across the ring with width $\Delta r = 2 R_*$ (Appendix \ref{A:Acc_model}). While we did not explicitly model viscous inflow, this timescale appears from the assumption of a steady state accretion rate $\dot{M}_0$ before disruption. It is associated with spacial scale $\Delta r$ because the binary pulls streams directly from the ring. For comparison, if the ring was initialized around a single black hole, it would take a viscous time over distance $r$, not $\Delta r$, to reach the black hole and accrete (such a scenario is plotted for comparison with the red-dashed line in Fig.~\ref{Fig:Acc_Mdodel_Ex}).

The right panel of Figures \ref{Fig:Acc_Mdodel_Ex} shows the full evolution of the binary accretion rate in orange, for the fiducial system. For reference, we also plot the quiescent disk accretion rate $\Mdot_0$ (horizontal gray line), the averaged stripping rate (black), and the disruption time $t_c$ (vertical dashed gray line). 
In the fiducial case, the binary accretion rate peaks at $\sim400\times$ Eddington (enhancement of $\sim800$) and decays in time $\tdecay \approx t_c$, which is $\approx 10^4 T_{\rep}$, or $\approx 21$ years for the fiducial rest frame value of $T_{\rep} \approx 21$ hours. This peak accretion rate and timescale depend strongly on the system parameters, as we show below.

The above solution shows that the peak accretion rate and its decay time are set by the disk mass at the critical binary separation (Eq. \ref{Eq:acrit}) and the viscous time there. The former is constrained by the requirements laid out in \S\ref{S:Setup}. Specifically,
the second of Eqs. (\ref{Eq:mstr-mdisk-conditions}), the angular momentum flux condition, sets the required local disk mass needed to keep the star migrating at the type I rate, and so sets the maximum accretion rate enhancement,
\begin{equation}
    %\frac{m_0}{m_{d,\crit}}
     \frac{ \dot{M}^{\mathrm{peak}}_{\mathrm{bin}} }{\Mdot_0} \leq \sqrt{\frac{1}{8\pi}} \left(\frac{M}{M_{\ring,0}} \right)  \left(\frac{\Mdot_0 P_{\crit}}{M_{\ring,0}} \right)^{1/2} h.
      \label{Eq:Mdotbin_peak_condition}
\end{equation}
Because $M_{\ring,0}$ decreases at smaller radii, the peak accretion enhancement will decrease for less dense stars,
which are disrupted at larger separations. Note, however, that this will depend on the disk model through $\Sigma(r)$ and $h(r)$ and on possible relations between the stellar mass and density, \eg, if less dense stars that disrupt further out are much more massive than those that disrupt closer in.

\begin{figure}
\begin{center} 
$
\begin{array}{c}
\includegraphics[scale=0.26]{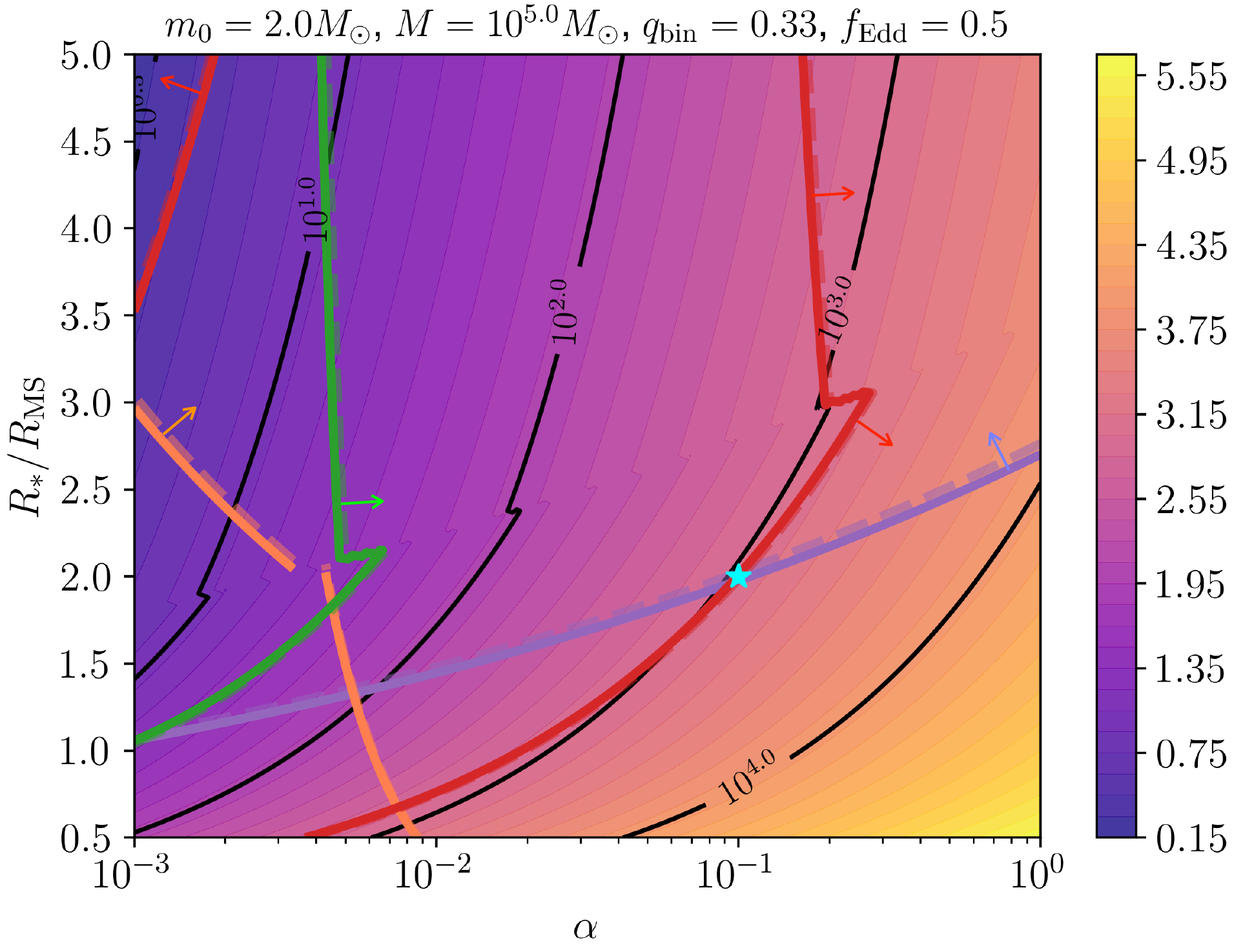}
\end{array} $
\end{center}
\vspace{-15pt}
\caption{
Purple-yellow colored contours (with highlighted values in black) of log$_{10}$ of the maximum accretion rate enhancement for a steady-state $\alpha$-disk and the system parameters labeled in the figure. Thick solid lines show the limits of where disk properties allow consistent migration of the star to the critical disruption radius. Red lines delineate the space via condition (\ref{Eq:mstr-mdisk-conditions}), green is the gap opening condition of \citep{Crida+2006}, orange is for disk-decoupling and purple is for migrator-decoupling (see \S\ref{S:Setup}). Arrows point in the direction where consistent solutions exist. 
The upper-right region with 100s-1000s$\times$ enhancements in the accretion rate is allowed for the parameters chosen here. The cyan star denotes the fiducial system, for which Figures \ref{Fig:disrupt} and \ref{Fig:Acc_Mdodel_Ex} show example evolutions.
}
\label{Fig:Max_Acc}
\end{figure}

Figure \ref{Fig:Max_Acc} draws contours of the log$_{10}$ maximum enhancement factors (colorbar and selected black contours) for the fiducial $2 \Msun$ star over a range of stellar radii and disk viscosities, at the labeled fiducial parameters. 
The enhancement factor ranges from unity to $10^5$ for the most compact stars. However, these very large enhancement factors occur only in regions of parameter space where we expect the star to have already decoupled from the cavity before disruption begins (\S\ref{S:Setup}). 
This figure is unchanged for constant ratio of Eddington factor to accretion efficiency, $f_{\Edd}/\eta$.

To make this clear, we also draw solid contours delineating the space where disruption is expected. Green delineates gap opening criteria, red delineates angular momentum supply criteria (Eq. \ref{Eq:mstr-mdisk-conditions}), orange, disk-decoupling, and purple, migrator-decoupling. This shows that, for the considered parameters, the upper right part of this plot is viable, and in this allowed regime, we expect enhancement factors of up to 100's-1000's. The cyan star shows the fiducial system considered throughout. The maximum enhancement factor at the location of the cyan star is agreement with the slightly lower enhancement inferred from the orange line in Figure \ref{Fig:Acc_Mdodel_Ex}.

Our fiducial system is at the edge of the acceptable parameter space delineated by the constraints of \S \ref{S:Setup}. Note that these are approximate constraints based on simple disk models but they suggest that this process is consistent for the fiducial system and those with larger stellar radius and higher disk viscosity. We discuss this point further in \S \ref{S:Discussion}.

%%%%%%%%%%%%%%%%%%%%%%%%%%%%%%%%%%%%%%%%%%%%%%%%
%%% Fig 7 Max Accretion Param Space%%%
%%%%%%%%%%%%%%%%%%%%%%%%%%%%%%%%%%%%%%%%%%%%%%%%
%%%%%%%%%%%%%%%%%%%%%%%%%%%%%%%%%%%%%%%%%%%%%%%%

The decay (viscous) time from this peak accretion rate scales with the radius of the cavity at the critical separation as $\tdecay \propto r_{\crit}\Delta r/\nu(r_{\crit}) \propto r^{1/2}_{\crit} R_*$, for a standard $\alpha$ disk. Hence, the decay time of the accretion enhancement scales with stellar properties as $\tdecay \propto R^{3/2}_* \propto \rho^{-1/2}_*$. Hence, for less dense stars, the accretion rate enhancement decreases, but lasts for a longer time.

\subsection{Bright X-ray flares from enhanced binary accretion}
\label{Ss:fallback_flares}

We now consider observable features of the enhanced binary accretion rate. As modeled above, binary accretion is mediated by CBD-stream impacts onto the minidisks surrounding the SMBHs.
In this subsection we follow \citet{Roedig_SEDsigs+2014} to estimate the emission from stream-mini-disk impacts. These impacts imply shocks\footnote{A lower temperature shock may arise from the stream's later impact with the CBD, but we leave this likely sub-dominate emission mechanism for future investigation.} with maximum post-shock temperatures of $T_s \approx 10^{10} \unit{K} (a/a_{\crit})^{-1}$, using Eq. (12) of \citet{Roedig_SEDsigs+2014} evaluated at the critical separation ($a_{\mathrm{crit}} \approx 500 GM/c^2$) for our fiducial system (Table \ref{Table:Fid}). However, pair-production will act as a thermostat to limit the post-shock temperatures to $10^9\unit{K}$ before such high temperature are reached\footnote{If $T_s \gtrsim 10^{10} \unit{K}$ is reached, pair-annihilation neutrinos would become effective at cooling the flow.}. Hence, we cap the post-shock temperature at $T_s \sim 10^9\unit{K}$ below, and so the following arguments will hold even for $10\times$ more widely separated binaries (equivalently $10^3 \times$ less dense stars).

At these high temperatures Bremsstrahlung and Inverse Compton (IC) scattering of hot electrons off of thermal photons generated by accretion onto the SMBHs likely dominate the radiative shock cooling. In the standard binary accretion scenario, IC is the dominant mechanism, with IC emissivity being $\mathcal{O}(10^4)$ times larger. 
In the CBD-TDE case, the mass-enhanced streams feeding the binary could diminish this ratio as the ambient photon energy density scales the IC emissivity while the ion number density scales the Bremsstrahlung emissivity. However, even if the photon energy density stays the same from the un-enhanced case (though it will likely increase due to the increased accretion rate), then the ion density will increase by approximately the ring mass enhancement of the previous subsection Eq. (\ref{Eq:Mdotbin_peak}). Hence, at most, Bremsstrahlung emissivity could be boosted by a factor of this ratio, which approaches $\lesssim10^4$ for the highest values of $\alpha$ in the allowed region of Figure \ref{Fig:Max_Acc}. In the fiducial case, this factor is an order of magnitude smaller and we consider IC to be the dominant cooling mechanism.

In the case of IC cooling, the hot-spot generated by the shock impact will release its thermal energy in a small fraction of a binary orbital time (Eq. 17 of \citealt{Roedig_SEDsigs+2014}). 
Because the IC cooling time is at least as fast as the time for stream material to be delivered to the minidisk (of order the binary dynamical time), the luminosity associated with IC cooling of the debris-stream shock at the minidisk will be, conservatively, the thermal energy per average rest frame particle mass times the binary mass accretion rate, $\Mdot_{\bin}$, computed in the previous section,
\begin{eqnarray}
        \frac{L_X}{L_{\Edd}} &\approx& \frac{3}{2} \frac{k T_s}{\eta m_p c^2} \frac{\dot{M}_{\bin}} {\dot{M}_{\Edd}} \approx  \\ \nonumber
        &\approx&   1 \left(\frac{\eta}{0.1}\right)^{-1} \left(\frac{T_s}{10^9 \mathrm{K}}\right) \left(\frac{\Mdot_{\bin}}{10^3 \dot{M}_{\Edd}}\right), 
        \label{Eq:LX}
\end{eqnarray}
where we approximate the average ion mass by the proton mass $m_p$, $k$ is Boltzmann's constant and $\eta$ is the accretion efficiency. The implied low efficiency for radiation production derives from the pair-creation thermostat discussed above.
The radiation spectrum, in this highly Comptonized limit will be a Wien spectrum with average photon energy given by the (maximum) shock temperature estimate above, $\left< h\nu \right> \lesssim 250~\mathrm{KeV} (T_s/10^{9}\mathrm{K})$. 

These first approximations suggest repeating Eddington-level X-ray transients from merging SMBHBs. Mass-loaded CBD-stream impacts with the minidisks generate high-energy X-ray flares which last for a fraction of the binary orbital time. These flares repeat on the binary orbital time and increase in luminosity until reaching Eddington levels soon after stellar disruption, thereafter slowly decaying in peak amplitude over decades (for the fiducial system), as described in the previous subsection.

\subsection{Super-Eddington outflow}
\label{Ss:outflow}

The analysis of \S\ref{S:Mdotbin_Model} implies super-Eddington binary accretion rates (Eq. \ref{Eq:Mdotbin_peak_condition} and Figure \ref{Fig:Max_Acc}). For the case of accretion onto a single black hole, such super-Eddington accretion has been argued to result in radiation driven outflows \citep{Jiang+2019}. Here we estimate the luminosity released when the radiation escapes from such outflows, following analysis for an analogous scenario for a single black hole \citep{Linial_circTDE:2024}. We assume that a quasi-spherical outflow is launched at velocity $v_w = \beta_w c$ for $\beta_w \approx 0.1$, from the sonic radius, $r_s \sim r_G\beta^{-2}_w$, of each SMBH in the binary. 
The outflow then expands until radiation can escape beyond the trapping radius, when the photon diffusion time becomes shorter than the wind expansion time, or equivalently when the
optical depth (to electron scattering) becomes smaller than $\beta^{-1}_w$. If we assume that a fraction $f_w\lesssim 1$ of the mass accreted onto the binary is driven into the outflow, $\Mdot_w = f_w \Mdot_{\bin}$, then \citet{Linial_circTDE:2024} show that the escaping radiation has luminosity,
\begin{equation}
    L \approx L_{\Edd} \left(\frac{\Mdot_{\bin}}{10^2 \Mdot_{\Edd} } \right)^{1/3} \left( \frac{f_w}{\eta} \right)^{1/3} \left( \frac{\beta_w}{0.1} \right)^{2/3},
\end{equation}
for each SMBH (assuming equal accretion and outflow rates), and for accretion efficiency $\eta$.

The energy of escaping emission will be,
\begin{eqnarray}
        k T &=& 55 \mathrm{eV} \left( \frac{M}{10^5 \Msun} \right)^{-1/4}  \left(\frac{\Mdot_{\bin}}{10^2 \Mdot_{\Edd} } \right)^{5/12} \\ \nonumber
        &\times& \left( \frac{f_w}{\eta} \right)^{5/12} \left( \frac{\beta_w}{0.1} \right)^{-1/12},
\end{eqnarray}
for each black hole, again assuming they are identical.

Hence, the super-Eddington rate of accretion onto the binary could power an outflow that would emit as a blackbody emitting in the far UV to soft X-rays. 
This emission may be strongly modulated or even peaked in conjunction with the stream accretion events once per orbit.
The wind expansion time to the trapping radius is of order minutes, much shorter than the time between stream accretion events. Hence, if a hypothetical buffering timescale between the stream impacts and the driving of an outflow is also short, this suggests outflow emission may be modulated at the binary orbital period during the build up to final disruption and for the decades duration of the Super-Eddington decay in the right panel of Figure \ref{Fig:Acc_Mdodel_Ex}.

Note that these calculations are simplified extensions of the picture for accretion onto a single black hole. While we would still expect radiation driven outflows to form from accretion onto the individual black holes, a number of differences may arise in the binary case. For example, a strong asymmetry could be imparted to the outflow, from the binary angular momentum and the surrounding circumbinary disk. Also note that the wind launching radius $r_s \sim 100 r_G$ is usually a factor of a few smaller than the critical separation (Eq. \ref{Eq:acrit}) for tidal stripping to begin, hence radiation driven outflows could be launched from each SMBH in the binary and interact with each-other and the CBD, creating strong shocks and liberating even more---and higher energy---radiation than estimated here.

We summarize this section by stating that, for both emission scenarios, we predict Eddington level X-ray flares repeating on the binary orbital period of $\mathcal{O}(10)$ hours. The flares will rapidly increase in magnitude up to the point of disruption and then decay over $\mathcal{O}(10)$ years, for the fiducial system. During this time the binary will decrease its orbital period by a few percent due to GW emission.

\section{Discussion}
\label{S:Discussion}

We have argued that a generic feature of gas-driven, low-mass SMBHB mergers %in nuclear star clusters 
will be the migration to and trapping of stars at the edges of CBDs (\S\ref{S:Setup}). A selection of such stars will track the SMBHB to late inspiral where they can be tidally stripped by the binary in a runaway process which fully disrupts the star (\S\ref{S:Disruption}). For the Solar-like stars considered here, this process results in Eddington-level, repeating X-ray flares (\S\ref{S:AccretonEmission}) occurring for SMBHBs at the low-frequency end of, or just outside of, the LISA band.

\subsection{Comparison with known nuclear transients}
\label{Ss:QPEs}

While the above emission models are relatively simple, they do capture the important timescales and energetics of the problem. Common to both models (\S\ref{Ss:fallback_flares} and \S\ref{Ss:outflow}) is the prediction of repeating X-ray flares with periods $\Order{10\unit{hrs}-\mathrm{days}}$ occurring around relatively low central masses $\Order{10^5 - 10^6 \, \Msun}$, with Eddington-level peak X-ray luminosity, $L_X\sim10^{43}~M/(10^5\Msun)$ erg/s.

Such basic properties are consistent with a recently discovered class of repeating nuclear transient, QPEs. QPEs are characterized as $\Order{< \unit{hrs}}$ long X-ray flares that recur on timescales of $\Order{10\unit{hrs}}$. The set of known QPEs also appear coincident with galactic nuclei of comparatively low-mass (post-starburst) host galaxies with implied central SMBH masses $\Order{10^5 - 10^{6.5}\,\Msun}$, although this is typically inferred from the low-mass-end of SMBH-galaxy relations that suffer from large uncertainties. Their host spectra in quiescence are consistent with radiatively efficient disk-accretion with Eddington ratios of $\Order{0.1}$, but lack evidence for extended flows like typical type-I Active Galactic Nuclei. We have not explicitly computed the X-ray flare duty cycle, but if we assume it does not deviate greatly from the $10\%$ stripping duty cycle of our disruption model, then we find further agreement with QPE observations. Table \ref{Table:QPE} in Appendix \ref{A:QPE} summarizes QPE and CBD-TDE properties.

The flare structure of QPEs is characterized by a fast rise and slightly more gradual decline with higher energy X-rays peaking earlier and evolving faster than softer X-rays \citep{Miniutti:QPE-Nature:2019, Giustini:QPE-RXJ1301.9+2747:2020, Arcodia:eRO-QPE1:2022}.
While this is consistent with the propagation of perturbations in a radiatively efficient accretion flow, the bolometric-luminosity--temperature relationships do not follow the usual scalings for blackbody emission possibly suggesting a shock origin for the emission \citep{Miniutti:QPE-Nature:2019}, consistent with our model for X-ray shock emission in \S\ref{Ss:fallback_flares}.

The observed set of QPEs have also been circumstantially linked with occurrences of TDEs. Beyond the overlap in host galaxy properties with those preferred for TDEs (low-mass, post-starburst galaxies \citealt{French:QPE-galaxies:2020, Wevers:QPE-galaxies:2022}), a few sources show signs of current or previous TDE activity. One such system, \gsn, has even possibly experienced two TDEs separated by approximately 10 years \citep{Muniutti:GSN069review:2023}.
Intriguingly, the final disruption of the star in our model, or the associated release of the CBD cavity-mass build-up described in \S\ref{S:Mdotbin_Model}, may masquerade as a TDE in the same system. Hence, one possible strength of this model is that stellar-disruptions are naturally predicted as a feature of QPE flaring. As we motivate below, multiple CBD-TDE events could occur in succession if stellar orbits of migrating stars are packed tightly in the CBD \citep{Kratter_CBDplanetstability:2014}, \eg, in mean motion resonant chains \citep[\eg,][]{Batygin_MMRrev:2015}.

Beyond these basic properties, QPEs exhibit a rich range of timing variations. Two sources, \gsn \citep{Muniutti:GSN069review:2023} and \eroii \citep{Arcodia:Nature-QPE-eRO-12:2021}), posses a remarkably consistent burst periodicity characterized by a two-burst alternating pattern of a strong flare followed by a ``long'' interval and a weaker flare followed by a ``short'' interval. 
Two other sources exhibit notably less regular timing behavior between flares (\rxj, \citealt{Giustini:QPE-RXJ1301.9+2747:2020}) and possibly even present with overlapping burst behavior (\eroi, \citealt{Arcodia:Nature-QPE-eRO-12:2021}). 
Such variations could be realized for a mildly eccentric stellar orbit or an eccentric binary. 
For a circular binary orbit and a mildly eccentric stellar orbit, we find that the disruption timings of the star can follow the above long-short and overlapping timing variations as well as a number of other distinct variations. While intriguing, we leave the eccentric problem, and its possible explanation of QPE timing variations to future work.

Finally, we note that the presence of an inner-binary could provide a natural explanation for the periodic quiescence emission observed in \gsn and \rxj. In the SMBHB model the shape and amplitude of this quiescent variability would depend on binary parameters via, \eg, Doppler boost or accretion variability models \citep[\eg,][]{PG1302Nature:2015b, binlite+2024}, providing an independent constraint on system parameters that generate flares, and so an additional test for the model proposed here.

%\textbf{
Another possibly unique aspect of this model is the time evolution of the flares. Figure \ref{Fig:Acc_Mdodel_Ex} shows that the accretion flares are brightest starting very shortly before total stellar disruption and within time $t_c$ after disruption. This predicts two unique features: 1) the flare magnitudes decay over a decade timescale (for fiducial parameters) and 2) a small fraction of events may show growing flare magnitudes followed by decay on longer timescales. Given observed flare repetition periods and magnitudes, one could estimate upper limits on $t_c$, thereby linking the flare repetition timescale and the lifetime of the entire event, predicting the timescale for which the flares should disappear within the CBD-TDE model.
%}

\paragraph{Comparative abundances}

To gauge consistency of associating some fraction of the so-far discovered QPEs with CBD-TDEs, we compare the inferred number density of QPEs with an estimated number density of low-mass SMBHBs at the critical separation.
\cite{Arcodia+2024} measure the abundance of QPEs, with 0.5-2~keV X-ray luminosity above $10^{41.5}$ erg s$^{-1}$, to be $\mathcal{N}_{\mathrm{QPE}} \approx 600^{+4730}_{-430} $Gpc$^{-3}$. A maximum number density of CBD-TDEs is given by the number density of expected SMBHBs, in the $\sim 10^4 - 10^6 \Msun$ mass range, below the critical separation.

To estimate the latter, we model the SMBHB population with a steady-state continuity equation, decaying due to GW-driven, circular orbit inspiral (a good approximation near merger). Then the number density of SMBHBs is given by the volumetric merger rate, $\mathcal{R}_{\mbhb}$, times the GW decay time at the critical separation \citep[\eg][]{ChristianLoeb:2017}. A volumetric merger rate upper limit can be estimated as follows.

The census of low mass SMBHs that would pair into $\sim10^5 \Msun$ SMBHBs is very uncertain and so too is the predicted merger rate $\mathcal{R}_{\mbhb}$ for such low mass SMBHBs.  \citet{Husko+2022} present the most exhaustive study of galaxy merger rates to date that extends to the mass scale of dwarf galaxies relevant in this context. The study is based on a compilation of results from all major cosmological volume simulations as well as semi-analytical models of galaxy formation. In the relevant galaxy mass range, between $10^8 \Msun$ and $10^{10} \Msun$, they obtain an integrated merger rate of $\mathcal{R}_{\mbhb} \sim 0.01$ Gpc$^{-3}$yr$^{-1}$ at $z < 0.1$ (nearly the same for major and minor mergers). The mapping between galaxy merger rates and SMBHB merger rates is highly uncertain, especially at this mass scale, as the occupation fraction of low mass SMBHBs in dwarf galaxies is likely less than unity and several astrophysical perturbations might hamper orbital decay at larger than pc-scale separations \citep[the so called "last kpc problem", see][]{LISA_Astro+2023}. Nevertheless, the galaxy merger rate is an upper limit on the SMBHB merger rate.

From the merger rate we estimate the number density of low-mass SMBHBs below the critical separation (corresponding to orbital period $P_{\crit}$) from the solution to the steady-state continuity equation for SMBHBs undergoing GW decay on circular orbits. For $q=0.33$, this is \citep{ChristianLoeb:2017},
\begin{eqnarray}
    \mathcal{N}_{\mathrm{\mbhb}} \approx 3 \left(\frac{P_{\crit}}{13.8 \mathrm{hrs}} \right)^{8/3} \left(\frac{M}{10^5 \Msun} \right)^{-5/3} \mathrm{Gpc}^{-3},
    \label{Eq:Nabundance}
\end{eqnarray}
which is 100 times smaller than the lower limit for QPEs from \citealt{Arcodia+2024}.

Clearly the estimated rate depends strongly on the critical orbital period for the onset of stellar disruption, $P_{\crit}$. This increases with lower stellar densities as $P_{\crit} \propto \rhoavg^{-1/2}$, so that $\mathcal{N}_{\mathrm{\mbhb}}  \propto \rhoavg^{-4/3}$. The more probable, lower-stellar-density events will have longer repetition periods $T_{\rep} \propto \rhoavg^{-1/2}$, last longer $\mathcal{T} \propto \rhoavg^{-1/2}$, and be dimmer (though scaling weakly with $\rhoavg$, as described in \S\ref{S:Mdotbin_Model}). Such low-density stellar disruptions should also generate interesting repeating nuclear transients on day or longer repetition timescales. To match the QPE abundances, $\rhoavg$ would need to be $10^{3/2} \times$ less dense than fiducial, resulting in $5.6 \times$ longer flare repetition times (few-day repetition timescales). 

Hence, while unlikely that the QPE rate can be explained by our fiducial CBD-TDE model given current expectations for LISA merger rates, it may not be out of the question, especially for CBD-TDEs with longer repetition times. 
It may also be worth considering further as a connection between QPEs and CBD-TDEs would imply high rates of merging low mass SMBHBs in LISA and offer a way to study the poorly understood population of low mass SMBHBs.

\subsection{Post-decoupling stellar disruption}
\label{S:post_merger_dsiruption}
In the case that disk- or migrator-decoupling occurs before the tidal stripping can begin (\S\ref{S:Setup}), the star and disk are left behind as the SMBHB merges. Even in these cases, the star can be brought within a few tidal radii of the remnant SMBH. Multiple mechanisms could result in the disruption of the star by the remnant SMBH. In order of time after SMBHB merger:
\begin{enumerate}
    \item The GW recoil kick could put the star on an eccentric orbit, bringing its new pericenter within the tidal disruption radius of the remnant SMBH. For a star that decouples at $n>1$ critical semi-major axes distance from the SMBHB barycenter, an eventual disruption will occur on the stellar orbital time, $2 \pi n^{3/2}\Omega_{\cav}$, which is $\sim 1.6 n^{3/2}$ days after SMBHB merger, for fiducial values.
    \item The disk will diffuse inwards towards the remnant SMBH bringing the star close enough for tidal stripping or complete disruption on a low eccentricity orbit. This will occur on the cavity viscous time of $\sim 10-100$ years (depending on the disk thickness), and may observationally resemble near-circular orbit stellar disruptions studied in \cite{Linial_circTDE:2024} or undergo moderately eccentric-orbit disruptions \citep{Hayasaki+2013}.
    \item
    The star eventually evolves off of the main sequence and is tidally disrupted as it enters a giant phase, over the course of its main-sequence lifetime ($\gtrsim 10^7$ yr). In this case, the resulting flares from tidal disruption of giant stars are expected to last for much longer than in the case of a Main-Sequence disruption \citep{MacLeod+2012}.
\end{enumerate}  
These scenarios could each source bound-TDEs, with the first uniquely occurring days-to-months after a low mass SMBHB merger, the second decades to a century after, and the last having the properties of an evolved star disrupted on a bound orbit. Interestingly, the first type of event would precede, and the second type would be coincident with, EM signatures indicative of a decoupling circumbinary disk \citep[\eg,][]{Krauth:2023, Zrake+2024}. 
We plan to study such scenarios in future work, as they represent the complement of the stripped stars studied here, and may signify post-SMBHB-merger systems.

\subsection{Caveats and future directions}
\label{Ss:caveats}

Finally, we consider possible impacts of approximations made in this work and the efforts that would be required to improve the accuracy of our predictions.

\paragraph{Migration and cavity trap}
The trapping condition from \cite{Masset_traps+2006} is quite robust, requiring that the width of the cavity edge be $\lambda \lesssim 8/3 h r_{\cav} \approx 0.5 a$, which is easily satisfied in CBD simulations with $h=0.1$ at the cavity edge, as well as for the more sharply truncated thinner disks \citep[\eg,][]{Tiede:2020}. However, other complications could prevent the star from reaching the trap, or from staying there until SMBHB merger; for example, 
non-cavity migration traps \citep[\eg][]{Bellovary+2016, Grishin_Therm+2023},
mean motion resonances with the binary \citep{ThunKley:2018}\footnote{During the preparation of this work, \cite{RevedStone_StarBinResCap:2024} showed how an eccentricity driving 2:1 resonant capture could generate highly eccentric tidal disruption events during SMBHB inspiral, in the absence of gas.} or with other migrators in the CBD \citep{Batygin_MMRrev:2015, Secunda+2019}. \citet{ThunKley:2018} find that migrators can be trapped in a 11:1 resonance with the binary for 10's of thousands of binary orbits. However, the 11:1 resonance exists at $11^{2/3} \approx 5a$, which would change our model by increasing the value of $N_{\cav}$. This further motivates the mechanisms described here as such resonances could hold even if the CBD dissipates. Furthermore, it would delay the disruption until later in the inspiral, pushing the CBD-TDE further towards or into the LISA band.

%\textbf{
Mean motion resonances between multiple stars in the disk could generate resonant chains that queue up stars to be subsequently disrupted by the binary. 
The time between disruptions is limited by the shorter of times for the star further out in a resonant chain to migrate to the cavity once its predecessor has been destroyed, or the time for the tidal radius to reach out to the orbit of the second star due to the GW decay of the SMBHB. 
These timescales depend on the distance between stars packed into the inner CBD. The minimum such distance is set by stability arguments. For two stars orbiting a single SMBH the minium packing separation is $\Delta r_* \gtrsim 3.5 (m_*/3M)^{1/3} r_* \approx 0.07 N_{\cav} a$ \citep{Gladman:1993}, and is a few times larger for the case of a binary \citep{Kratter_CBDplanetstability:2014}.
This implies a minimum time for queued-up stars to refill the cavity trap under type I migration of, $t_{\mathrm{refill}} \gtrsim t_{I}\Delta r_*/\rcav \approx 250$ years for $t_{I}$ (Eq. \ref{Eq:typeI}) evaluated at $\rcav$ in the fiducial model. This is of order the GW residence time of the SMBHB (Eq. \ref{Eq:TGW}), and so for fiducial parameters type I migration is too slow to cause refilling of stars to the binary tidal radius. Such queued up stars would allow refilling of the cavity-trapped star in the event that one is knocked out earlier in the SMBHB inspiral (see below). 
For less-dense stars that are disrupted earlier in the SMBHB evolution, the lifetime of the binary is much longer and many such events could become possible in one SMBHB system. This would further enhance the CBD-TDE probability of Eq. (\ref{Eq:Nabundance}).
%}

%\textbf{
While the stellar refilling timescale through migration is slow compared to the SMBHB lifetime, the time needed for the second, queued-up star to enter the disruption radius of the binary is much shorter: $T_{\GW} \Delta r_*/\rcav \sim 20$ years. This is approximately $t_c$, the time to fully disrupt the first star. Hence, if the second star does not decouple, it will begin disruption just after the first finishes. This is worth further consideration as it suggests multiple disruption events in succession.
%}

Throughout we assumed that torques from tidal stripping of the star do not immediately displace the stellar orbit from the cavity trap. We estimate the validity of this assumption as follows. The cavity trap torque must be of order the type I migration torque, as it is generated from the balance of an outer Linblad torque and co-rotation torque. The timescale associated with type I migration torques $t_I$ is given by Eq. (\ref{Eq:typeI}). Timescales for change in the stellar orbit can be approximated from
\begin{equation}
    \tau^{-1}_{\orb*} = \frac{\dot{a}_*}{a_*} \approx \frac{\dot{m}_*}{M+m_*} + \frac{\dot{l}_*}{l} \approx X \frac{\dot{m}_*}{M}
\end{equation}
for specific torque on the stellar orbit $\dot{l}_*$, with specific angular momentum $l$, and where we have approximated the relative contribution of stripping and gravitational torques with the of-order-unity prefactor $X$.

We require $t_I/\tau_{\orb*} \lesssim 1$ for the trap to hold, implying that the maximum amount of mass stripped from the star per stellar orbit (with period $T =  N^{3/2}_{\cav}P$, for binary orbital period $P$) is,
  \begin{equation}
    \dot{m}_* T \lesssim \frac{8\pi}{\chi} \left(\frac{m_d}{M}\right) h^{-2} m_*.
    \label{Eq:mstrip_lim}
\end{equation}
For a typical $\alpha$-disk at this separation around the fiducial SMBHB, ($h\approx 0.006$, $m_d/M \approx 10^{-7}$) the trap will hold for stripped debris streams that are $\lesssim 0.1 m_*$. This condition for thin disks is satisfied until the final stripping events leading to total disruption. However, by this time in the evolution the disk mass has also increased by two orders of magnitude (following \S\ref{S:Mdotbin_Model}), and the trap can hold for a range of disk properties. From the $h^{-2}$ dependence in Eq. (\ref{Eq:mstrip_lim}) it is apparent that if the disk is thickened at the cavity edge, without significantly increasing its density, then the minimum allowed stripping mass drops to where even the reinforced density from the stripping may not be enough to hold the star in the cavity-edge orbit. This requires a consistent evolution of the CBD and stellar orbit to understand further.

If the trap does eventually break due to the stripping process, the star may be pulled into the cavity to be disrupted or ejected back into the disk.
In the former case, our picture will become one of an initial tidal stripping stage followed more quickly by full disruption of a larger core mass. If instead the trap breaks earlier in the inspiral, due to other possible  perturbations, \eg, strong turbulence or vortices at the cavity edge \citep[\eg,][]{CimmermanRafi:2024}, then tidal disruption may still occur as the star enters the CBD cavity and interacts with the binary components.

If the disk can hold denser stars than considered here, without their premature decoupling before disruption (\S\ref{S:Setup}), then extremely high enhancement factors are expected. This is because the star would be much more massive than the local disk (lower region of Figure \ref{Fig:Max_Acc}). Consideration of different disk models, \eg, thicker but less dense magnetically dominated disks \citep[c.f.][]{Hopkins_MADisk+2024}, or those which are altered by the binary \citep{Rafikov:2013, Tiede+2024}, will affect the stellar types expected to survive the binary inspiral towards disruption.

\paragraph{Stellar structure} 
Parts of our model depend strongly on the stellar density and structure. Stars that survive migration to the cavity trap have not followed a standard, single-star evolutionary pathway. If their orbits were dragged into the disk, then they experienced stripping and ablation before a stage of evolution in the dense medium of the disk. If they were formed in the disk, then they may be representative of a non-standard initial mass function.

Recent studies of in-situ star formation in the outer regions of AGN disks show that the mass function of stars is skewed to large masses in the range from a few to a few tens of solar masses \citep{DerdzinskiMayer:2023, ChenJiang+2023}, and that the interaction with the surrounding disk induces prompt inward migration on timescales even faster than Type I \citep{DerdzinskiMayer:2023}, in line with our scenario. As these stars may suffer tidal mass loss already during their protostellar stage, the ensuing structure might differ from a standard polytrope, and numerical simulations are under way to investigate this.

Stars accreting in AGN disks may evolve very differently from those in isolation and have larger masses and radii \citep[\eg][]{DittmannCantiello:2024}, resulting in less-than-fiducial density stars which disrupt much earlier in the SMBHB inspiral. This could result in transients with lower energy and longer repetition timescales than considered here, but with higher occurrence rates given that SMBHBs spend a longer time at wider separations (Eq. \ref{Eq:Nabundance}). 

Interestingly, because $a_{\crit} \propto (M/\rhoavg)^{1/3}$, low density stars formed and evolved in the AGN disk allow not only disruptions at larger radii, but alternatively disruption by more massive SMBHBs, with disruption and repetition timescales growing as $\rhoavg^{-1/2}$ (Fig. \ref{Fig:Trep_tstrip}). We save such possible longer events around more massive SMBHBs for a future investigation. 

If a star become too massive, it could open a gap in the disk and possibly break the cavity trap\footnote{However, this may almost never occur even for massive stars, if they migrate quickly \citep{MalikMMM:2015}.}. If many stars are migrating through the CBD with different evolution and growth rates \citep[\eg,][]{Gaia_AGNstars+2024}, then the migration process may act as a filter on stellar mass and density, retaining for disruption only those that will not decouple or clear a gap.

Furthermore, runaway disruption of the star depends on how the stellar envelope is modeled \citep{SobermanPhinney+1997}. Here we assumed a $\gamma=5/3$ polytrope, consistent with a convective envelope. However, different equations of state, which cause the star to contract upon losing mass, could prevent the runaway stripping process.

%\textbf{
Finally, the mass loss rate computed from our polytropic stripping model (\S\ref{Ss:DisruptModel}) assumed that accretion or returning fallback streams (\S\ref{S:AccretonEmission}) do not deposit significant energy into the stellar envelope. If cooling in the envelope does not exceed energy deposition before the next stripping event, then the extra heating could lead to an altered, likely faster mass loss rate to be investigated in future analyses.
%}

\paragraph{Emission modeling}
Basic scales of the problem have lead us to predict multiple avenues for Eddington level, high energy emission modulated at the binary orbital period, but a number of more complex signatures may arise from a more complete treatment of the disruption, fallback, magnetic field, and radiation. Numerical hydrodynamical and radiative transfer calculations are likely required to gain further traction on the accretion and emission processes in this gaseous three-body problem.

\section{Conclusion}
\label{S:Conclusion}
We have argued that the existence of stars around accreting SMBHBs necessitates their delivery into a circumbinary disk where they will migrate down to and be trapped at the circumbinary cavity, tracking the binary as it inspirals. For standard $\alpha$-disk models and $\lesssim 10^5 \Msun$ SMBHBs, stars that are less dense than Solar will track the binary down to where the tidal radius of the binary reaches into the disk and begins to periodically strip the outer layers of the star. This occurs just before the SMBHB enters the LISA band and results in a runaway disruption (unstable Roche-lobe overflow) of the star. The stripped stellar material builds up mass at the edge of the circumbinary cavity from which the binary pulls streams and accretes at higher and higher rates that can be super Eddington years before, and decades after, the final disruption of the star, depending on the stellar density. 

We consider the non-thermal shock-induced emission from accretion streamers impacting the minidisk around each SMBH in the binary and also outflows from the modulated Super-Eddington accretion. For the sub-Solar-density stars considered here, both emission mechanisms suggest Eddington-level X-ray flaring with 10's of hours repetition times. 
These timescales and energetics, along with the preference for low-mass SMBHBs resemble characteristics of the QPEs, though if all QPEs were caused by this mechanism, the LISA rates for $\sim10^5\Msun$ SMBHBs would be a factor of 100 higher than currently expected. Rates are in less contention for days-long flares resulting from the disruption of less-dense stars.

Whether or not a subset of the QPEs are associated with disrupting stars in circumbinary disks, or instead represent a yet-undiscovered repeating nuclear transient, the mechanism explored here could offer an EM tracer of gas-driven, low-mass SMBHB inspirals, just before or as they enter the LISA band. In the event that the disruption process can be tied with GWs from the SMBHB, associated GW de-phasing along with the predicted EM emission could be used as a multi-messenger probe of black hole binary accretion physics and the SMBHB merger process. Such an opportunity requires further investigation into this possible, bright EM signature of inspiralling SMBHBs.

\acknowledgments
This project was conceived of at the RESCEU-NBIA Meeting on Gravitational Wave Science at the University of Tokyo, and has benefited from discussions with Eric Coughlin, Johan Samsing, Cl\'ement Bonnerot, Alejandro Vigna-G\'omez, Matt Nicholl, and Itai Linial. DJD also thanks participants of the Lorentz Center Workshop on ``Multi-messenger observations of supermassive black hole binaries'' for their input, especially Mark Avara, and Massimo Dotti, and for an enlightening run through Leiden with Zolt\'an Haiman.
D.J.D. and C.T. acknowledge support from the Danish Independent Research Fund through Sapere Aude Starting Grant No. 121587. C.T. also received support through the European Union’s Horizon research and innovation program under Marie Sklodowska-Curie grant agreement No. 101148364. 
L.Z. was supported by ERC Starting Grant no. 101043143.
K.H. was supported by the National Research Foundation of Korea (NRF) grant funded by the Korea government (MSIT) (2020R1A2C1007219 and RS-2025-23323627). This research was also supported in part by grant no. NSF PHY-2309135 to the Kavli Institute for Theoretical Physics (KITP).

\clearpage

\appendix

\section{Decoupling}
\label{A:disk_models}

In the main text we employ \citet{SS73} and \citet{Sirko_Goodman:2003} disk models. For the latter we use the implementation of \cite{pagn+2024, pag_2024zndo}: \url{https://github.com/DariaGangardt/pAGN}.

Here we provide analytical solutions for decoupling radii in each region of the \citet{SS73} disks. These are: the Inner Region where radiation dominates the pressure and electron scattering dominates the opacity: The Middle Region where gas pressure and electron scattering dominate, and the Outer Region where gas pressure and free-free opacity dominates. An additional region arises from choice of viscosity prescription in the Inner Region: proportional to total pressure ($b=0$) or gas pressure only ($b=1$). See \citep[\eg,][]{HKM09} for a review of these regions.

For each of the four regions, the disk- and migrator-decoupling separations are found by solving Eqs. (\ref{Eq:adec_tII}) or (\ref{Eq:adec_TI}) for $a$ with the appropriate scaling of dimensionless disk aspect ratio $h(N_{\cav} a )$ and surface density $\Sigma(N_{\cav} a )$ in each disk region. The GW inspiral rate is given by $\dot{a}_{\GW} = \beta_{\GW} a^{-3}$, with $\beta_{\GW} \equiv \left(64/5\right) (G^3/c^5) M^3 q/(1+q)^2$. We write the surface density and aspect ratio in each region as: $\Sigma = K_\Sigma(\alpha, \Mdot, M) a^{\gamma_{\Sigma}}$ and $h = K_h (\alpha, \Mdot, M) a^{\gamma_h}$, noting the implicit dependencies of each prefactor on disk properties. With these conventions one can insert the appropriate disk solutions into the following formulae for the binary decoupling separations:
\begin{enumerate}
    \item Inner region ($b=0$, $\gamma_{\Sigma} = 3/2$, $\gamma_h = -1$):
    \begin{eqnarray}
        a_{\dec D} = \left[\frac{2}{3}  \frac{\AGW}{\sqrt{GM}} \frac{N^{5/2}_{\cav}}{\alpha K^2_h} \right]^{2} ;  \qquad
        a_{\dec M} = \left[ \frac{\AGW}{2 f_{\mathrm{mig}}  N^5_{\cav}} \frac{M^2}{m \sqrt{G M}} \frac{K^2_h}{K_\Sigma} \right]^{1/8}
    \end{eqnarray}
    \item Inner region ($b=1$, $\gamma_{\Sigma} = -3/5$, $\gamma_h = -1$) $a_{\dec D}$ is unchanged and:
    \begin{eqnarray}
        a_{\dec M} &=& \left[ \frac{\AGW}{2 f_{\mathrm{mig}}  N^{29/10}_{\cav}} \frac{M^2}{m \sqrt{G M}} \frac{K^2_h}{K_\Sigma} \right]^{10/59}
    \end{eqnarray}
    \item Middle region ( $\gamma_{\Sigma} = -3/5$, $\gamma_h = 1/20$):
    \begin{eqnarray}
        a_{\dec D} = \left[\frac{2}{3}  \frac{\AGW}{\sqrt{GM}} \frac{N^{2/5}_{\cav}}{\alpha K^2_h} \right]^{5/13} ;  \qquad
        a_{\dec M} = \left[ \frac{\AGW}{2 f_{\mathrm{mig}}  N^{4/5}_{\cav}} \frac{M^2}{m \sqrt{G M}} \frac{K^2_h}{K_\Sigma} \right]^{5/19}
    \end{eqnarray}
    \item Outer region ( $\gamma_{\Sigma} = -3/4$, $\gamma_h = 1/8$):
    \begin{eqnarray}
        a_{\dec D} = \left[\frac{2}{3}  \frac{\AGW}{\sqrt{GM}} \frac{N^{4/110}_{\cav}}{\alpha K^2_h} \right]^{4/11};  \qquad
        a_{\dec M} = \left[ \frac{\AGW}{2 f_{\mathrm{mig}}  N^{1/2}_{\cav}} \frac{M^2}{m \sqrt{G M}} \frac{K^2_h}{K_\Sigma} \right]^{2/7}
    \end{eqnarray}
\end{enumerate}

\subsection{Insensitivity of decoupling scale to the alpha parameter}
The matching condition on the decoupling radii (see discussion around Eq. \ref{Eq:adec_TI}) dictates the tightest orbit
that the star can reach before it decouples from the binary. 
While the condition occurs for small values of $\alpha\sim 10^{-2}-10^{-3}$ for the typical system parameters chosen above, this largest-of-the-two decoupling separations is rather insensitive to changes in disk parameters. 
We find that decoupling always occurs
near the values found in Eqs. (\ref{Eq:adec_tII}) and (\ref{Eq:adec_TI}), even for large values of $\alpha$.
This is because, at near-Eddington accretion rates, and in the middle region of the SS disk (valid for $a \sim 500 r_G$), the migrator-decoupling separation increases very weakly as $a_{\dec M} \propto \alpha^{3/19}$. For low values of $\alpha$, disk-decoupling occurs in the middle region of the disk where it decreases weakly with $\alpha$, $a_{\dec D} \propto \alpha^{-4/13}$. However, for $\alpha \gtrsim 10^{-3}$, the location of disk-decoupling transitions to the inner disk region, where it plummets as $a_{\dec D} \propto \alpha^{-2}$ (for near Eddington accretion rates). 
Hence, for small $\alpha$, the decoupling scale is set by the weakly $\alpha$-dependent disk-decoupling, while for large $\alpha$, the weakly dependent migrator-decoupling separation sets the decoupling scale.

\section{Tidal heating during inspiral}
\label{A:Tidal_Heating}
Here we show that tidal heating by the binary for a star with radiative core and convective envelope (consistent with our $\gamma=5/3$ polytrope stripping model) can cause expansion of the star and overflow at larger separations, consistent with our choice of $R_* = 2 R_{\MS}(m_*)$ throughout. 

As the binary inspirals, the star slowly enters deeper into the tidal field of the SMBHs. During this process the tidal energy will be deposited into the star. Assuming that the star is tidally locked on a low eccentricity orbit to a central object with the binary total mass, then the rate of tidal energy dissipation into a Sun-like star is \citep[][see also Appendix A.2 of \citealt{Linial_circTDE:2024}]{GoodmanDickson:1998} 
\begin{equation}
    \dot{E}_{\mathrm{tide}} \approx 
    A \left(\frac{\rho_c}{m_*/R^3_*} \right) \frac{M^2m^{5/3}}{\left(M+m_*\right)^{8/3}}  \Omega^{-6}_* \Omega^{9} (N_{\cav} a )^2 e^2_*,
\end{equation}
where $A\approx 0.059$, $\rho_c$ is the density at the stellar core-envelope boundary, $\Omega_*=\sqrt{Gm_*/R^3_*}$, $\Omega=\sqrt{G(M+m_*)/(N_{\cav} a)^3}$, and in the single SMBH case, $e_*$ is the eccentricity of the stellar orbit. Because the star is not on a true elliptical orbit, we define an effective eccentricity by identifying the closest and furthest approach of the star to the primary (or secondary) SMBH with the pericenter and apocenter of an elliptical orbit, such that we replace $e_*$ above with $e_{\mathrm{eff,prm}} \equiv q/(1+q)/N_{\cav}$, for the primary, and $e_{\mathrm{eff,sec}} \equiv 1/(1+q)/N_{\cav}$, for the secondary. This is accurate as long as the star is fixed at the cavity edge with $r_{\cav}=N_{\cav}a$.  Combining primary and secondary SMBH contributions, we define,
\begin{equation}
e^2_{\mathrm{eff}} = e^2_{\mathrm{eff,prm}} + e^2_{\mathrm{eff,sec}} = \frac{1+q^2}{N^2_{\cav}(1+q)^2},
\end{equation}
to write the total dissipated energy as,
\begin{equation}
\dot{E}_{\mathrm{tide}, \BHcirc \BHcirc} \approx 
    4.3 \times 10^{38} \frac{\mathrm{erg}}{\mathrm{s}}\left(\frac{e_{\mathrm{eff}}}{0.125}\right)^2
    \left(\frac{a}{a_{\crit}}\right)^{-23/2}  = K_{\mathrm{tide}} a^{-23/2},
\end{equation}
which is independent of $M$ when evaluated at $a_{\crit}$, in the $m_*\ll M$ limit, and we evaluated at the Solar value, $\rho_c=0.2$g cm$^{-2}$.

We consider the maximum energy deposited into the star by assuming this energy accumulates over the entire migration of the star and SMBHB (most is deposited near the end of migration), and compare to the stellar energy $E_*=Gm^2_*/R_*$. We then solve for a characteristic binary separation $a_{\Heat}$ where tidal heating becomes important, 
\begin{equation}
    \frac{E^{\max}_{\mathrm{tide}}}{E_*} =-\frac{R_*}{Gm^2_*} \int^{a_{\Heat}}_{\infty} \frac{ \dot{E}_{\mathrm{tide}} }{ \dot{a} }  da = 1.
\end{equation}
Because the inspiral rate of the star and binary are locked, and GW inspiral dominates when at small separations where heating is relevant, we use $\dot{a} = \dot{a}_{\GW} = \beta_{\GW}a^{-3}$.
Integrating, 
\begin{eqnarray}
    a_{\Heat} = \left[\frac{2}{15} \frac{K_{\mathrm{tide}}}{E_{*} \beta_{\GW}} \right ]^{-2/15}.
\end{eqnarray}
For our fiducial system, this becomes $\sim 1.7 a_{\crit}(R_{\MS})$, and is just less than unity for a star with radius puffed up to twice its original size, $2 R_{\MS}$, consistent with our choice of stellar radius in the main text.
This ratio becomes larger with smaller binary mass ratio, larger cavity size and larger stellar mass and radius (for star with a radiative core and convective envelope). Note that in the case that $1/3$ of this heating rate goes into spin-synchronization \citep{Zahn:1977}, the above numbers decrease by a factor of three. It is unclear if this factor applies in the binary case.

\section{Accretion Model Derivation} 
\label{A:Acc_model}
To derive Eq. (\ref{Eq:Mring}) we assume that the rate at which mass in streams is pulled from the cavity-edge ring is proportional to the mass in the ring via $\dot{M}_{\CBDstm} = \chi M_{\ring}/P$, where $\chi$ represents the fraction of mass pulled out in a stream.
We model the ring mass by accounting for the rates at which mass enters and exits the co-orbital ring of the star,
\begin{equation}
    \dot{M}_{\ring} = \dot{M}_0 - \frac{\chi M_{\ring}}{P} + (1-f_{\acc})\frac{\chi M_{\ring}}{P} + \left<\dot{m}_*\right>_T + m_c \delta(t-t_c), 
    \label{Eq:ApxMring}
\end{equation}
with $\dot{M}_0$ the accretion rate entering the cavity ring, $\frac{\chi M_{\ring}}{P}$ the rate that streams are pulled from the ring, $(1-f_{\acc})\frac{\chi M_{\ring}}{P}$ the rate streams are expelled back to the ring, $\left<\dot{m}_*\right>_T$ the averaged stellar mass loss rate, and $m_c$ the core mass delivered at time $t_c>0$. 
Write $\chi$ from initial conditions $\dot{M}_{\ring} = \dot{m} = 0; M_{\ring}(t) = M_{\ring,0}$,
\begin{equation}
    \chi = \frac{\dot{M}_{0} P }{ f_{\acc} M_{\ring,0}},
\end{equation}
and substitute into Eq. (\ref{Eq:ApxMring}) to give Eq. (\ref{Eq:Mring}) of the text. 

The accretion rate onto the binary is set by assumption as,
\begin{eqnarray}
    \Mdot_{\bin} = f_{\acc}\frac{\chi M_{\ring}(t)}{P} = M_\ring \frac{\dot{M}_0}{M_{\ring0}} = \frac{M_{\ring}(t)}{\tdecay},
\end{eqnarray}
and it can be easily shown that, for a ring with width $\Delta r$,
\begin{eqnarray}
   \tdecay \equiv \frac{M_{\ring0}}{\dot{M}_0} = \frac{2 \pi \Sigma_0 r \Delta r}{3 \pi \Sigma_0 \nu} = \frac{2}{3}\frac{r \Delta r}{\nu} 
\end{eqnarray}
is the viscous time for material to cross the ring.

\section{Known QPE properties}
\label{A:QPE}

For convenience we summarize the properties of known QPEs and the comparative predictions for CBD-TDEs in Table \ref{Table:QPE}.

% --------------------------------------------------------------------
\begin{table}
 \begin{center}
 \caption{Summary of Observed QPE Sources and CBD-TDE properties. From left to right, columns list the source name or abbreviation, its redshift ($z$), the inferred central black hole mass ($M_\bullet$), the recurrence time ($P$), the eruption duration ($\delta t$), and the source X-ray luminosity ($L_X$).}
    \begin{tabular}{l c c c c c c c}
    \hline \hline
     Source & $z$ & $\log_{10}(M_{\bullet} / M_\odot )$  & $P \, (\unit{hr})$ & $\delta t \, (\unit{hr})$ & $L_{X} \, (\unit{erg} \, \unit{s}^{-1})$ \\
     $-$  & $\sim 40 - 220 \,\unit{Mpc}$ & $\sim 5 - 6.5$ & $\sim 2 - 19$ & $\sim 0.3 - 8$ & $\sim 10^{42} - 10^{44}$ \\
     %
     % source-name & redshift & mass & period & duration & x-ray-luminosity \\
     %
    \hline 
    GSN-069$^{1, \star}$ & $0.0181$ & $5.5 - 6.5$ & $8.5$ -- $9.5$ & $1.1$ & $1.3\times 10^{42}$  \\
    eRO-QPE1$^{2}$ & $0.0505$ & $5.8$ & 18.5 & 7.6 & $2.0 \times 10^{43}$ \\
    eRO-QPE2$^{2}$ & $0.0175$ & $5 - 6$ & $2.36 - 2.75$ & $0.18 - 0.44$ & $1.5\times 10^{42}$ \\
    RX J1301$^{3}$ & $0.02358$ & $5.9 - 6.5$ & $3.6 - 5.6$ & 0.33 & $1.2 \times 10^{42}$ \\
    \hline 
    XMM J0249$^{4, \ast}$ & $0.0186$ & $4.9$ & $2.5$ & $\sim 0.3$ & $3.4 \times 10^{41}$  \\
    eRO-QPE3$^{5, \dagger}$ & $0.024$ & $5.9 - 6.7$ & $20$ & $3.7$ & $4.2 \times 10^{41}$  \\
    eRO-QPE4$^{5}$ & $0.044$ & $7.2 - 7.8$ & $9.8-14.7$ & $0.46-0.58$ & $1.27 \times 10^{43}$  \\
    Swift J0230$^{6, \ddagger}$ & $0.037$ & $6.6$ & $528$ & $24 - 108$ & $5 \times 10^{42}$  \\
    AT2019qiz$^{7}$ & $0.0151$ & $5.4 - 6.3$ & $48.4$ & $8 - 10$ & $1.8 \times 10^{43}$  \\
    \hline 
    CBD-TDE & $z$ & $4 - 6$ & $ 13.8 (1+z) \left(\frac{\rhoavg}{\rho_{\mathrm{Fid}}}\right)^{-1/2}$ & $\sim 10 \% P$ & $\sim 10^{43} \left(\frac{M}{10^5 \Msun} \right)$  \\   
    \hline
    \multicolumn{6}{l}{$^\star$ Two apparent associated TDEs within the last 20 years, one of which occurred concurrent with QPE cessation in 2020.}\\
    \multicolumn{6}{l}{$^{\ast}$ Two observed QPE bursts followed by long term ($\sim 18\,\unit{yr}$) dimming consistent with a TDE.} \\
    \multicolumn{6}{l}{$^\dagger$ Decaying quiescent emission consistent with a prior TDE.}\\
    \multicolumn{6}{l}{$^\ddagger$ Estimated Eddington fraction for quiescent emission more consistent with a low-luminosity AGN than a thin-disk solution.}\\
    \multicolumn{6}{l}{$^1$\citet{Muniutti:GSN069review:2023};  $^2$\citet{Arcodia:Nature-QPE-eRO-12:2021}; $^3$\citet{Giustini:QPE-RXJ1301.9+2747:2020}; $^4$\citet{Chakraborty:QPE-XMMSL1:2021}; $^5$\citet{Arcodia:QPE-eRO-34:2024} }\\
    \multicolumn{6}{l}{$^6$\citet{Guolo:SwiftJ0230:2024}; $^7$\citet{Nicholl+2024}}\\
    \end{tabular}
     \label{Table:QPE}
 \end{center}
\end{table}
% --------------------------------------------------------------------

\bibliographystyle{apj} 
\bibliography{refs, qpe_refs}

\begin{thebibliography}{}
\expandafter\ifx\csname natexlab\endcsname\relax\def\natexlab#1{#1}\fi

\bibitem[{{Agazie} {et~al.}(2023){Agazie}, {Anumarlapudi}, {Archibald},
  {Arzoumanian}, {Baker}, {B{\'e}csy}, {Blecha}, {Brazier}, {Brook},
  {Burke-Spolaor}, {Burnette}, {Case}, {Charisi}, {Chatterjee},
  {Chatziioannou}, {Cheeseboro}, {Chen}, {Cohen}, {Cordes}, {Cornish},
  {Crawford}, {Cromartie}, {Crowter}, {Cutler}, {Decesar}, {Degan}, {Demorest},
  {Deng}, {Dolch}, {Drachler}, {Ellis}, {Ferrara}, {Fiore}, {Fonseca},
  {Freedman}, {Garver-Daniels}, {Gentile}, {Gersbach}, {Glaser}, {Good},
  {G{\"u}ltekin}, {Hazboun}, {Hourihane}, {Islo}, {Jennings}, {Johnson},
  {Jones}, {Kaiser}, {Kaplan}, {Kelley}, {Kerr}, {Key}, {Klein}, {Laal}, {Lam},
  {Lamb}, {Lazio}, {Lewandowska}, {Littenberg}, {Liu}, {Lommen}, {Lorimer},
  {Luo}, {Lynch}, {Ma}, {Madison}, {Mattson}, {McEwen}, {McKee}, {McLaughlin},
  {McMann}, {Meyers}, {Meyers}, {Mingarelli}, {Mitridate}, {Natarajan}, {Ng},
  {Nice}, {Ocker}, {Olum}, {Pennucci}, {Perera}, {Petrov}, {Pol}, {Radovan},
  {Ransom}, {Ray}, {Romano}, {Sardesai}, {Schmiedekamp}, {Schmiedekamp},
  {Schmitz}, {Schult}, {Shapiro-Albert}, {Siemens}, {Simon}, {Siwek}, {Stairs},
  {Stinebring}, {Stovall}, {Sun}, {Susobhanan}, {Swiggum}, {Taylor}, {Taylor},
  {Turner}, {Unal}, {Vallisneri}, {van Haasteren}, {Vigeland}, {Wahl}, {Wang},
  {Witt}, {Young}, \& {Nanograv Collaboration}}]{NANOG-GWB:2023}
{Agazie}, G., {Anumarlapudi}, A., {Archibald}, A.~M., {et~al.} 2023, \apjl,
  951, L8

\bibitem[{{Ajith} {et~al.}(2007){Ajith}, {Babak}, {Chen}, {Hewitson},
  {Krishnan}, {Whelan}, {Br{\"u}gmann}, {Diener}, {Gonzalez}, {Hannam}, {Husa},
  {Koppitz}, {Pollney}, {Rezzolla}, {Santamar{\'\i}a}, {Sintes}, {Sperhake}, \&
  {Thornburg}}]{Ajith_Phenom+2007}
{Ajith}, P., {Babak}, S., {Chen}, Y., {et~al.} 2007, Classical and Quantum
  Gravity, 24, S689

\bibitem[{{Amaro-Seoane} {et~al.}(2023){Amaro-Seoane}, {Andrews}, {Arca Sedda},
  {Askar}, {Baghi}, {Balasov}, {Bartos}, {Bavera}, {Bellovary}, {Berry},
  {Berti}, {Bianchi}, {Blecha}, {Blondin}, {Bogdanovi{\'c}}, {Boissier},
  {Bonetti}, {Bonoli}, {Bortolas}, {Breivik}, {Capelo}, {Caramete},
  {Cattorini}, {Charisi}, {Chaty}, {Chen}, {Chru{\'s}li{\'n}ska}, {Chua},
  {Church}, {Colpi}, {D'Orazio}, {Danielski}, {Davies}, {Dayal}, {De Rosa},
  {Derdzinski}, {Destounis}, {Dotti}, {Du{\c{t}}an}, {Dvorkin}, {Fabj},
  {Foglizzo}, {Ford}, {Fouvry}, {Franchini}, {Fragos}, {Fryer}, {Gaspari},
  {Gerosa}, {Graziani}, {Groot}, {Habouzit}, {Haggard}, {Haiman}, {Han},
  {Istrate}, {Johansson}, {Khan}, {Kimpson}, {Kokkotas}, {Kong}, {Korol},
  {Kremer}, {Kupfer}, {Lamberts}, {Larson}, {Lau}, {Liu}, {Lloyd-Ronning},
  {Lodato}, {Lupi}, {Ma}, {Maccarone}, {Mandel}, {Mangiagli}, {Mapelli},
  {Mathis}, {Mayer}, {McGee}, {McKernan}, {Miller}, {Mota}, {Mumpower},
  {Nasim}, {Nelemans}, {Noble}, {Pacucci}, {Panessa}, {Paschalidis}, {Pfister},
  {Porquet}, {Quenby}, {Ricarte}, {R{\"o}pke}, {Regan}, {Rosswog}, {Ruiter},
  {Ruiz}, {Runnoe}, {Schneider}, {Schnittman}, {Secunda}, {Sesana}, {Seto},
  {Shao}, {Shapiro}, {Sopuerta}, {Stone}, {Suvorov}, {Tamanini}, {Tamfal},
  {Tauris}, {Temmink}, {Tomsick}, {Toonen}, {Torres-Orjuela}, {Toscani},
  {Tsokaros}, {Unal}, {V{\'a}zquez-Aceves}, {Valiante}, {van Putten}, {van
  Roestel}, {Vignali}, {Volonteri}, {Wu}, {Younsi}, {Yu}, {Zane}, {Zwick},
  {Antonini}, {Baibhav}, {Barausse}, {Bonilla Rivera}, {Branchesi},
  {Branduardi-Raymont}, {Burdge}, {Chakraborty}, {Cuadra}, {Dage}, {Davis}, {de
  Mink}, {Decarli}, {Doneva}, {Escoffier}, {Gandhi}, {Haardt}, {Lousto},
  {Nissanke}, {Nordhaus}, {O'Shaughnessy}, {Portegies Zwart}, {Pound},
  {Schussler}, {Sergijenko}, {Spallicci}, {Vernieri}, \&
  {Vigna-G{\'o}mez}}]{LISA_Astro+2023}
{Amaro-Seoane}, P., {Andrews}, J., {Arca Sedda}, M., {et~al.} 2023, Living
  Reviews in Relativity, 26, 2

\bibitem[{{Andreoni} {et~al.}(2022){Andreoni}, {Margutti}, {Salafia},
  {Parazin}, {Villar}, {Coughlin}, {Yoachim}, {Mortensen}, {Brethauer},
  {Smartt}, {Kasliwal}, {Alexander}, {Anand}, {Berger}, {Bernardini}, {Bianco},
  {Blanchard}, {Bloom}, {Brocato}, {Bulla}, {Cartier}, {Cenko}, {Chornock},
  {Copperwheat}, {Corsi}, {D'Ammando}, {D'Avanzo}, {H{\'e}l{\`e}ne Datrier},
  {Foley}, {Ghirlanda}, {Goobar}, {Grindlay}, {Hajela}, {Holz}, {Karambelkar},
  {Kool}, {Lamb}, {Laskar}, {Levan}, {Maguire}, {May}, {Melandri},
  {Milisavljevic}, {Miller}, {Nicholl}, {Nissanke}, {Palmese}, {Piranomonte},
  {Rest}, {Sagu{\'e}s-Carracedo}, {Siellez}, {Singer}, {Smith}, {Steeghs}, \&
  {Tanvir}}]{Andreoni+2022}
{Andreoni}, I., {Margutti}, R., {Salafia}, O.~S., {et~al.} 2022, \apjs, 260, 18

\bibitem[{{Arcodia} {et~al.}(2021){Arcodia}, {Merloni}, {Nandra}, {Buchner},
  {Salvato}, {Pasham}, {Remillard}, {Comparat}, {Lamer}, {Ponti}, {Malyali},
  {Wolf}, {Arzoumanian}, {Bogensberger}, {Buckley}, {Gendreau}, {Gromadzki},
  {Kara}, {Krumpe}, {Markwardt}, {Ramos-Ceja}, {Rau}, {Schramm}, \&
  {Schwope}}]{Arcodia:Nature-QPE-eRO-12:2021}
{Arcodia}, R., {Merloni}, A., {Nandra}, K., {et~al.} 2021, \nat, 592, 704

\bibitem[{{Arcodia} {et~al.}(2022){Arcodia}, {Miniutti}, {Ponti}, {Buchner},
  {Giustini}, {Merloni}, {Nandra}, {Vincentelli}, {Kara}, {Salvato}, \&
  {Pasham}}]{Arcodia:eRO-QPE1:2022}
{Arcodia}, R., {Miniutti}, G., {Ponti}, G., {et~al.} 2022, \aap, 662, A49

\bibitem[{{Arcodia} {et~al.}(2024{\natexlab{a}}){Arcodia}, {Merloni},
  {Buchner}, {Baldini}, {Ponti}, {Rau}, {Liu}, {Nandra}, \&
  {Salvato}}]{Arcodia+2024}
{Arcodia}, R., {Merloni}, A., {Buchner}, J., {et~al.} 2024{\natexlab{a}}, \aap,
  684, L14

\bibitem[{{Arcodia} {et~al.}(2024{\natexlab{b}}){Arcodia}, {Liu}, {Merloni},
  {Malyali}, {Rau}, {Chakraborty}, {Goodwin}, {Buckley}, {Brink}, {Gromadzki},
  {Arzoumanian}, {Buchner}, {Kara}, {Nandra}, {Ponti}, {Salvato}, {Anderson},
  {Baldini}, {Grotova}, {Krumpe}, {Maitra}, {Miller-Jones}, \&
  {Ramos-Ceja}}]{Arcodia:QPE-eRO-34:2024}
{Arcodia}, R., {Liu}, Z., {Merloni}, A., {et~al.} 2024{\natexlab{b}}, \aap,
  684, A64

\bibitem[{{Armitage} \& {Natarajan}(2002)}]{ArmNat:2002:ApJL}
{Armitage}, P.~J., \& {Natarajan}, P. 2002, \apjl, 567, L9

\bibitem[{Artymowicz \& Lubow(1994)}]{AL94}
Artymowicz, P., \& Lubow, S.~H. 1994, \apj, 421, 651

\bibitem[{{Artymowicz} \& {Lubow}(1996)}]{ArtyLubow:1996}
{Artymowicz}, P., \& {Lubow}, S.~H. 1996, \apjl, 467, L77

\bibitem[{{Bartos} {et~al.}(2017){Bartos}, {Kocsis}, {Haiman}, \&
  {M{\'a}rka}}]{Bartos_AGNchan+2017}
{Bartos}, I., {Kocsis}, B., {Haiman}, Z., \& {M{\'a}rka}, S. 2017, \apj, 835,
  165

\bibitem[{{Batygin}(2015)}]{Batygin_MMRrev:2015}
{Batygin}, K. 2015, \mnras, 451, 2589

\bibitem[{{Begelman} {et~al.}(1980){Begelman}, {Blandford}, \&
  {Rees}}]{Begel:Blan:Rees:1980}
{Begelman}, M.~C., {Blandford}, R.~D., \& {Rees}, M.~J. 1980, \nat, 287, 307

\bibitem[{{Bellovary} {et~al.}(2016){Bellovary}, {Mac Low}, {McKernan}, \&
  {Ford}}]{Bellovary+2016}
{Bellovary}, J.~M., {Mac Low}, M.-M., {McKernan}, B., \& {Ford}, K.~E.~S. 2016,
  \apjl, 819, L17

\bibitem[{{Bowen} {et~al.}(2017){Bowen}, {Campanelli}, {Krolik}, {Mewes}, \&
  {Noble}}]{Bowen_slosh+2017}
{Bowen}, D.~B., {Campanelli}, M., {Krolik}, J.~H., {Mewes}, V., \& {Noble},
  S.~C. 2017, \apj, 838, 42

\bibitem[{{Chakraborty} {et~al.}(2021){Chakraborty}, {Kara}, {Masterson},
  {Giustini}, {Miniutti}, \& {Saxton}}]{Chakraborty:QPE-XMMSL1:2021}
{Chakraborty}, J., {Kara}, E., {Masterson}, M., {et~al.} 2021, \apjl, 921, L40

\bibitem[{{Chen} {et~al.}(2023){Chen}, {Jiang}, {Goodman}, \&
  {Ostriker}}]{ChenJiang+2023}
{Chen}, Y.-X., {Jiang}, Y.-F., {Goodman}, J., \& {Ostriker}, E.~C. 2023, \apj,
  948, 120

\bibitem[{{Christian} \& {Loeb}(2017)}]{ChristianLoeb:2017}
{Christian}, P., \& {Loeb}, A. 2017, \mnras, 469, 930

\bibitem[{{Cimerman} \& {Rafikov}(2024)}]{CimmermanRafi:2024}
{Cimerman}, N.~P., \& {Rafikov}, R.~R. 2024, \mnras, 528, 2358

\bibitem[{{Colpi} {et~al.}(2024){Colpi}, {Danzmann}, {Hewitson},
  {Holley-Bockelmann}, {Jetzer}, {Nelemans}, {Petiteau}, {Shoemaker},
  {Sopuerta}, {Stebbins}, {Tanvir}, {Ward}, {Weber}, {Thorpe}, {Daurskikh},
  {Deep}, {Fern{\'a}ndez N{\'u}{\~n}ez}, {Garc{\'\i}a Marirrodriga}, {Gehler},
  {Halain}, {Jennrich}, {Lammers}, {Larra{\~n}aga}, {Lieser},
  {L{\"u}tzgendorf}, {Martens}, {Mondin}, {Piris Ni{\~n}o}, {Amaro-Seoane},
  {Arca Sedda}, {Auclair}, {Babak}, {Baghi}, {Baibhav}, {Baker}, {Bayle},
  {Berry}, {Berti}, {Boileau}, {Bonetti}, {Brito}, {Buscicchio}, {Calcagni},
  {Capelo}, {Caprini}, {Caputo}, {Castelli}, {Chen}, {Chen}, {Chua}, {Davies},
  {Derdzinski}, {Domcke}, {Doneva}, {Dvorkin}, {Mar{\'\i}a Ezquiaga}, {Gair},
  {Haiman}, {Harry}, {Hartwig}, {Hees}, {Heffernan}, {Husa},
  {Izquierdo-Villalba}, {Karnesis}, {Klein}, {Korol}, {Korsakova}, {Kupfer},
  {Laghi}, {Lamberts}, {Larson}, {Le Jeune}, {Lewicki}, {Littenberg}, {Madge},
  {Mangiagli}, {Marsat}, {Vilchez}, {Maselli}, {Mathews}, {van de Meent},
  {Muratore}, {Nardini}, {Pani}, {Peloso}, {Pieroni}, {Pound},
  {Quelquejay-Leclere}, {Ricciardone}, {Rossi}, {Sartirana}, {Savalle},
  {Sberna}, {Sesana}, {Shoemaker}, {Slutsky}, {Sotiriou}, {Speri}, {Staab},
  {Steer}, {Tamanini}, {Tasinato}, {Torrado}, {Torres-Orjuela}, {Toubiana},
  {Vallisneri}, {Vecchio}, {Volonteri}, {Yagi}, \& {Zwick}}]{LISA_redbook:2024}
{Colpi}, M., {Danzmann}, K., {Hewitson}, M., {et~al.} 2024, arXiv e-prints,
  arXiv:2402.07571

\bibitem[{{Coughlin} {et~al.}(2017){Coughlin}, {Armitage}, {Nixon}, \&
  {Begelman}}]{Coughlin+2017}
{Coughlin}, E.~R., {Armitage}, P.~J., {Nixon}, C., \& {Begelman}, M.~C. 2017,
  \mnras, 465, 3840

\bibitem[{{Coughlin} \& {Nixon}(2022)}]{CoughlinNixon:2022}
{Coughlin}, E.~R., \& {Nixon}, C.~J. 2022, \mnras, 517, L26

\bibitem[{{Crida} {et~al.}(2006){Crida}, {Morbidelli}, \&
  {Masset}}]{Crida+2006}
{Crida}, A., {Morbidelli}, A., \& {Masset}, F. 2006, in AAS/Division for
  Planetary Sciences Meeting Abstracts, Vol.~38, AAS/Division for Planetary
  Sciences Meeting Abstracts \#38, 63.05

\bibitem[{{DariaGangardt} \& {Trani}(2024)}]{pag_2024zndo}
{DariaGangardt}, \& {Trani}, A.~A. 2024, {DariaGangardt/pagn: v1},
  doi:10.5281/zenodo.10723302

\bibitem[{{Derdzinski} \& {Mayer}(2023)}]{DerdzinskiMayer:2023}
{Derdzinski}, A., \& {Mayer}, L. 2023, \mnras, 521, 4522

\bibitem[{{Dittmann} \& {Cantiello}(2024)}]{DittmannCantiello:2024}
{Dittmann}, A.~J., \& {Cantiello}, M. 2024, arXiv e-prints, arXiv:2409.02981

\bibitem[{{Dittmann} {et~al.}(2023){Dittmann}, {Ryan}, \&
  {Miller}}]{Dittmann_decouple+2023}
{Dittmann}, A.~J., {Ryan}, G., \& {Miller}, M.~C. 2023, \apjl, 949, L30

\bibitem[{{D'Orazio} \& {Charisi}(2023)}]{DOrazioCharisi:2023}
{D'Orazio}, D.~J., \& {Charisi}, M. 2023, arXiv e-prints, arXiv:2310.16896

\bibitem[{{D'Orazio} {et~al.}(2024){D'Orazio}, {Duffell}, \&
  {Tiede}}]{binlite+2024}
{D'Orazio}, D.~J., {Duffell}, P.~C., \& {Tiede}, C. 2024, \apj, 977, 244

\bibitem[{D'Orazio {et~al.}(2016)D'Orazio, Haiman, Duffell, MacFadyen, \&
  Farris}]{D'Orazio:CBDTrans:2016}
D'Orazio, D.~J., Haiman, Z., Duffell, P., MacFadyen, A., \& Farris, B. 2016,
  Monthly Notices of the Royal Astronomical Society, 459, 2379

\bibitem[{{D'Orazio} {et~al.}(2013){D'Orazio}, {Haiman}, \&
  {MacFadyen}}]{DHM:2013:MNRAS}
{D'Orazio}, D.~J., {Haiman}, Z., \& {MacFadyen}, A. 2013, \mnras, 436, 2997

\bibitem[{{D'Orazio} {et~al.}(2015){D'Orazio}, {Haiman}, \&
  {Schiminovich}}]{PG1302Nature:2015b}
{D'Orazio}, D.~J., {Haiman}, Z., \& {Schiminovich}, D. 2015, Nature, 525, 351

\bibitem[{{Duffell} \& {MacFadyen}(2013)}]{DuffellMac:2013:smallqGapOpen}
{Duffell}, P.~C., \& {MacFadyen}, A.~I. 2013, \apj, 769, 41

\bibitem[{{Duffell} {et~al.}(2024){Duffell}, {Dittmann}, {D'Orazio},
  {Franchini}, {Kratter}, {Penzlin}, {Ragusa}, {Siwek}, {Tiede}, {Wang},
  {Zrake}, {Dempsey}, {Haiman}, {Lupi}, {Pirog}, \& {Ryan}}]{SBCC+2024}
{Duffell}, P.~C., {Dittmann}, A.~J., {D'Orazio}, D.~J., {et~al.} 2024, \apj,
  970, 156

\bibitem[{{Dunhill} {et~al.}(2015){Dunhill}, {Cuadra}, \&
  {Dougados}}]{Dunhill+2015}
{Dunhill}, A.~C., {Cuadra}, J., \& {Dougados}, C. 2015, \mnras, 448, 3545

\bibitem[{{EPTA Collaboration} {et~al.}(2023){EPTA Collaboration}, {InPTA
  Collaboration}, {Antoniadis}, {Arumugam}, {Arumugam}, {Babak}, {Bagchi}, {Bak
  Nielsen}, {Bassa}, {Bathula}, {Berthereau}, {Bonetti}, {Bortolas}, {Brook},
  {Burgay}, {Caballero}, {Chalumeau}, {Champion}, {Chanlaridis}, {Chen},
  {Cognard}, {Dandapat}, {Deb}, {Desai}, {Desvignes}, {Dhanda-Batra},
  {Dwivedi}, {Falxa}, {Ferdman}, {Franchini}, {Gair}, {Goncharov}, {Gopakumar},
  {Graikou}, {Grie{\ss}meier}, {Guillemot}, {Guo}, {Gupta}, {Hisano}, {Hu},
  {Iraci}, {Izquierdo-Villalba}, {Jang}, {Jawor}, {Janssen}, {Jessner},
  {Joshi}, {Kareem}, {Karuppusamy}, {Keane}, {Keith}, {Kharbanda}, {Kikunaga},
  {Kolhe}, {Kramer}, {Krishnakumar}, {Lackeos}, {Lee}, {Liu}, {Liu}, {Lyne},
  {McKee}, {Maan}, {Main}, {Mickaliger}, {Ni{\c{t}}u}, {Nobleson}, {Paladi},
  {Parthasarathy}, {Perera}, {Perrodin}, {Petiteau}, {Porayko}, {Possenti},
  {Prabu}, {Quelquejay Leclere}, {Rana}, {Samajdar}, {Sanidas}, {Sesana},
  {Shaifullah}, {Singha}, {Speri}, {Spiewak}, {Srivastava}, {Stappers},
  {Surnis}, {Susarla}, {Susobhanan}, {Takahashi}, {Tarafdar}, {Theureau},
  {Tiburzi}, {van der Wateren}, {Vecchio}, {Venkatraman Krishnan}, {Verbiest},
  {Wang}, {Wang}, \& {Wu}}]{EPTA_GWB_2023}
{EPTA Collaboration}, {InPTA Collaboration}, {Antoniadis}, J., {et~al.} 2023,
  \aap, 678, A50

\bibitem[{{Evans} \& {Kochanek}(1989)}]{EvansKochanek1989}
{Evans}, C.~R., \& {Kochanek}, C.~S. 1989, \apjl, 346, L13

\bibitem[{{Fabj} {et~al.}(2025){Fabj}, {Dittmann}, {Cantiello}, {Perna}, \&
  {Samsing}}]{Gaia_AGNstars+2024}
{Fabj}, G., {Dittmann}, A.~J., {Cantiello}, M., {Perna}, R., \& {Samsing}, J.
  2025, \apj, 981, 16

\bibitem[{{Fabj} {et~al.}(2020){Fabj}, {Nasim}, {Caban}, {Ford}, {McKernan}, \&
  {Bellovary}}]{Fabj+2020}
{Fabj}, G., {Nasim}, S.~S., {Caban}, F., {et~al.} 2020, \mnras, 499, 2608

\bibitem[{{Farr} \& {Mandel}(2018)}]{FarrMandel_comment:2018}
{Farr}, W.~M., \& {Mandel}, I. 2018, Science, 361, aat6506

\bibitem[{{Farris} {et~al.}(2015){Farris}, {Duffell}, {MacFadyen}, \&
  {Haiman}}]{Farris:2015:GW}
{Farris}, B.~D., {Duffell}, P., {MacFadyen}, A.~I., \& {Haiman}, Z. 2015,
  \mnras, 447, L80

\bibitem[{{Foster} \& {Backer}(1990)}]{FosterBacker:1990}
{Foster}, R.~S., \& {Backer}, D.~C. 1990, \apj, 361, 300

\bibitem[{{Franchini} {et~al.}(2024){Franchini}, {Bonetti}, {Lupi}, \&
  {Sesana}}]{Franchini_decouple+2024}
{Franchini}, A., {Bonetti}, M., {Lupi}, A., \& {Sesana}, A. 2024, \aap, 686,
  A288

\bibitem[{{French} {et~al.}(2020){French}, {Wevers}, {Law-Smith}, {Graur}, \&
  {Zabludoff}}]{French:QPE-galaxies:2020}
{French}, K.~D., {Wevers}, T., {Law-Smith}, J., {Graur}, O., \& {Zabludoff},
  A.~I. 2020, \ssr, 216, 32

\bibitem[{{Gangardt} {et~al.}(2024){Gangardt}, {Trani}, {Bonnerot}, \&
  {Gerosa}}]{pagn+2024}
{Gangardt}, D., {Trani}, A.~A., {Bonnerot}, C., \& {Gerosa}, D. 2024, \mnras,
  530, 3689

\bibitem[{{Gezari}(2021)}]{Gezari2021}
{Gezari}, S. 2021, \araa, 59, 21

\bibitem[{{Giustini} {et~al.}(2020){Giustini}, {Miniutti}, \&
  {Saxton}}]{Giustini:QPE-RXJ1301.9+2747:2020}
{Giustini}, M., {Miniutti}, G., \& {Saxton}, R.~D. 2020, in XIV.0 Scientific
  Meeting (virtual) of the Spanish Astronomical Society, 40

\bibitem[{{Gladman}(1993)}]{Gladman:1993}
{Gladman}, B. 1993, \icarus, 106, 247

\bibitem[{{Goodman} \& {Dickson}(1998)}]{GoodmanDickson:1998}
{Goodman}, J., \& {Dickson}, E.~S. 1998, \apj, 507, 938

\bibitem[{{Grishin} {et~al.}(2024){Grishin}, {Gilbaum}, \&
  {Stone}}]{Grishin_Therm+2023}
{Grishin}, E., {Gilbaum}, S., \& {Stone}, N.~C. 2024, \mnras, 530, 2114

\bibitem[{{Guolo} {et~al.}(2024){Guolo}, {Pasham}, {Zaja{\v{c}}ek}, {Coughlin},
  {Gezari}, {Sukov{\'a}}, {Wevers}, {Witzany}, {Tombesi}, {van Velzen},
  {Alexander}, {Yao}, {Arcodia}, {Karas}, {Miller-Jones}, {Remillard},
  {Gendreau}, \& {Ferrara}}]{Guolo:SwiftJ0230:2024}
{Guolo}, M., {Pasham}, D.~R., {Zaja{\v{c}}ek}, M., {et~al.} 2024, Nature
  Astronomy, 8, 347

\bibitem[{Haiman {et~al.}(2009)Haiman, Kocsis, \& Menou}]{HKM09}
Haiman, Z., Kocsis, B., \& Menou, K. 2009, \apj, 700, 1952

\bibitem[{{Hayasaki} \& {Loeb}(2016)}]{HayasakiLoeb:2016}
{Hayasaki}, K., \& {Loeb}, A. 2016, Scientific Reports, 6, 35629

\bibitem[{Hayasaki {et~al.}(2007)Hayasaki, Mineshige, \& Sudou}]{Hayasaki:2007}
Hayasaki, K., Mineshige, S., \& Sudou, H. 2007, Publications of the
  Astronomical Society of Japan, 59, 427

\bibitem[{{Hayasaki} {et~al.}(2013){Hayasaki}, {Stone}, \&
  {Loeb}}]{Hayasaki+2013}
{Hayasaki}, K., {Stone}, N., \& {Loeb}, A. 2013, \mnras, 434, 909

\bibitem[{{Hayasaki} {et~al.}(2018){Hayasaki}, {Zhong}, {Li}, {Berczik}, \&
  {Spurzem}}]{Hayasaki+2018}
{Hayasaki}, K., {Zhong}, S., {Li}, S., {Berczik}, P., \& {Spurzem}, R. 2018,
  \apj, 855, 129

\bibitem[{{Hills}(1975)}]{Hills1975}
{Hills}, J.~G. 1975, \nat, 254, 295

\bibitem[{{Hopkins} {et~al.}(2024){Hopkins}, {Squire}, {Quataert}, {Murray},
  {Su}, {Steinwandel}, {Kremer}, {Faucher-Giguere}, \&
  {Wellons}}]{Hopkins_MADisk+2024}
{Hopkins}, P.~F., {Squire}, J., {Quataert}, E., {et~al.} 2024, The Open Journal
  of Astrophysics, 7, 20

\bibitem[{{Hu{\v{s}}ko} {et~al.}(2022){Hu{\v{s}}ko}, {Lacey}, \&
  {Baugh}}]{Husko+2022}
{Hu{\v{s}}ko}, F., {Lacey}, C.~G., \& {Baugh}, C.~M. 2022, \mnras, 509, 5918

\bibitem[{{Jiang} {et~al.}(2019){Jiang}, {Stone}, \& {Davis}}]{Jiang+2019}
{Jiang}, Y.-F., {Stone}, J.~M., \& {Davis}, S.~W. 2019, \apj, 880, 67

\bibitem[{{Kanagawa} {et~al.}(2018){Kanagawa}, {Tanaka}, \&
  {Szuszkiewicz}}]{Kanagawa+2018}
{Kanagawa}, K.~D., {Tanaka}, H., \& {Szuszkiewicz}, E. 2018, \apj, 861, 140

\bibitem[{{Kley} {et~al.}(2019){Kley}, {Thun}, \&
  {Penzlin}}]{Kley_cavtraprad:2019}
{Kley}, W., {Thun}, D., \& {Penzlin}, A. B.~T. 2019, \aap, 627, A91

\bibitem[{{Komossa}(2015)}]{Komossa2015}
{Komossa}, S. 2015, Journal of High Energy Astrophysics, 7, 148

\bibitem[{{Kratter} \& {Shannon}(2014)}]{Kratter_CBDplanetstability:2014}
{Kratter}, K.~M., \& {Shannon}, A. 2014, \mnras, 437, 3727

\bibitem[{{Krauth} {et~al.}(2023){Krauth}, {Davelaar}, {Haiman},
  {Westernacher-Schneider}, {Zrake}, \& {MacFadyen}}]{Krauth:2023}
{Krauth}, L.~M., {Davelaar}, J., {Haiman}, Z., {et~al.} 2023, \mnras, 526, 5441

\bibitem[{{Linial} \& {Quataert}(2024)}]{Linial_circTDE:2024}
{Linial}, I., \& {Quataert}, E. 2024, \apj, 974, 67

\bibitem[{{Linial} \& {Sari}(2017)}]{LinialSari:2017}
{Linial}, I., \& {Sari}, R. 2017, \mnras, 469, 2441

\bibitem[{{Liu} {et~al.}(2009){Liu}, {Li}, \& {Chen}}]{LiuChen:2009}
{Liu}, F.~K., {Li}, S., \& {Chen}, X. 2009, \apjl, 706, L133

\bibitem[{{Liu} {et~al.}(2014){Liu}, {Li}, \& {Komossa}}]{Liu+2014}
{Liu}, F.~K., {Li}, S., \& {Komossa}, S. 2014, \apj, 786, 103

\bibitem[{{Lubow} \& {Artymowicz}(1997)}]{LubowArty:1997}
{Lubow}, S.~H., \& {Artymowicz}, P. 1997, in Astronomical Society of the
  Pacific Conference Series, Vol. 121, IAU Colloq. 163: Accretion Phenomena and
  Related Outflows, ed. D.~T. {Wickramasinghe}, G.~V. {Bicknell}, \&
  L.~{Ferrario}, 505

\bibitem[{{MacLeod} {et~al.}(2012){MacLeod}, {Guillochon}, \&
  {Ramirez-Ruiz}}]{MacLeod+2012}
{MacLeod}, M., {Guillochon}, J., \& {Ramirez-Ruiz}, E. 2012, \apj, 757, 134

\bibitem[{{MacLeod} \& {Lin}(2020)}]{MacLeodLin:2020}
{MacLeod}, M., \& {Lin}, D. N.~C. 2020, \apj, 889, 94

\bibitem[{{Malik} {et~al.}(2015){Malik}, {Meru}, {Mayer}, \&
  {Meyer}}]{MalikMMM:2015}
{Malik}, M., {Meru}, F., {Mayer}, L., \& {Meyer}, M. 2015, \apj, 802, 56

\bibitem[{{Martini}(2004)}]{PMartini:2004}
{Martini}, P. 2004, Coevolution of Black Holes and Galaxies, 169

\bibitem[{{Masset} {et~al.}(2006){Masset}, {Morbidelli}, {Crida}, \&
  {Ferreira}}]{Masset_traps+2006}
{Masset}, F.~S., {Morbidelli}, A., {Crida}, A., \& {Ferreira}, J. 2006, \apj,
  642, 478

\bibitem[{{McKernan} {et~al.}(2012){McKernan}, {Ford}, {Lyra}, \&
  {Perets}}]{McKernan+2012}
{McKernan}, B., {Ford}, K.~E.~S., {Lyra}, W., \& {Perets}, H.~B. 2012, \mnras,
  425, 460

\bibitem[{{Mei} {et~al.}(2021){Mei}, {Bai}, {Bao}, {Barausse}, {Cai}, {Canuto},
  {Cao}, {Chen}, {Chen}, {Ding}, {Duan}, {Fan}, {Feng}, {Fu}, {Gao}, {Gao},
  {Gong}, {Gou}, {Gu}, {Gu}, {He}, {Hendry}, {Hong}, {Hu}, {Hu}, {Hu}, {Huang},
  {Huang}, {Jiang}, {Jiang}, {Jiang}, {Jiang}, {Jin}, {Korol}, {Li}, {Li},
  {Li}, {Li}, {Li}, {Li}, {Li}, {Li}, {Li}, {Liang}, {Liang}, {Liao}, {Liu},
  {Liu}, {Liu}, {Liu}, {Liu}, {Liu}, {Liu}, {Lu}, {Lu}, {Lu}, {Luo}, {Luo},
  {Milyukov}, {Ming}, {Pi}, {Qin}, {Qu}, {Sesana}, {Shao}, {Shi}, {Su}, {Tan},
  {Tan}, {Tan}, {Tu}, {Wang}, {Wang}, {Wang}, {Wang}, {Wang}, {Wang}, {Wang},
  {Wang}, {Wang}, {Wang}, {Wang}, {Wei}, {Wu}, {Xiao}, {Xu}, {Xue}, {Yang},
  {Yang}, {Yang}, {Yang}, {Ye}, {Yeh}, {Yu}, {Zhai}, {Zhang}, {Zhang}, {Zhang},
  {Zhang}, {Zhang}, {Zhang}, {Zhang}, {Zhou}, {Zhou}, {Zhou}, {Zhu}, {Zi}, \&
  {Luo}}]{TianQin+2021}
{Mei}, J., {Bai}, Y.-Z., {Bao}, J., {et~al.} 2021, Progress of Theoretical and
  Experimental Physics, 2021, 05A107

\bibitem[{Merritt \& Milosavljevi{\'c}(2005)}]{MerrittMilos:2005:LRR}
Merritt, D., \& Milosavljevi{\'c}, M. 2005, Living Reviews in Relativity, 8

\bibitem[{{Miniutti} {et~al.}(2023){Miniutti}, {Giustini}, {Arcodia}, {Saxton},
  {Read}, {Bianchi}, \& {Alexander}}]{Muniutti:GSN069review:2023}
{Miniutti}, G., {Giustini}, M., {Arcodia}, R., {et~al.} 2023, \aap, 670, A93

\bibitem[{{Miniutti} {et~al.}(2019){Miniutti}, {Saxton}, {Giustini},
  {Alexander}, {Fender}, {Heywood}, {Monageng}, {Coriat}, {Tzioumis}, {Read},
  {Knigge}, {Gandhi}, {Pretorius}, \&
  {Ag{\'\i}s-Gonz{\'a}lez}}]{Miniutti:QPE-Nature:2019}
{Miniutti}, G., {Saxton}, R.~D., {Giustini}, M., {et~al.} 2019, \nat, 573, 381

\bibitem[{{Nicholl} {et~al.}(2024){Nicholl}, {Pasham}, {Mummery}, {Guolo},
  {Gendreau}, {Dewangan}, {Ferrara}, {Remillard}, {Bonnerot}, {Chakraborty},
  {Hajela}, {Dhillon}, {Gillan}, {Greenwood}, {Huber}, {Janiuk}, {Salvesen},
  {van Velzen}, {Aamer}, {Alexander}, {Angus}, {Arzoumanian}, {Auchettl},
  {Berger}, {de Boer}, {Cendes}, {Chambers}, {Chen}, {Chornock}, {Fulton},
  {Gao}, {Gillanders}, {Gomez}, {Gompertz}, {Fabian}, {Herman}, {Ingram},
  {Kara}, {Laskar}, {Lawrence}, {Lin}, {Lowe}, {Magnier}, {Margutti}, {McGee},
  {Minguez}, {Moore}, {Nathan}, {Oates}, {Patra}, {Ramsden}, {Ravi}, {Ridley},
  {Sheng}, {Smartt}, {Smith}, {Srivastav}, {Stein}, {Stevance}, {Turner},
  {Wainscoat}, {Weston}, {Wevers}, \& {Young}}]{Nicholl+2024}
{Nicholl}, M., {Pasham}, D.~R., {Mummery}, A., {et~al.} 2024, \nat, 634, 804

\bibitem[{{Park} \& {Hayasaki}(2020)}]{Park2020}
{Park}, G., \& {Hayasaki}, K. 2020, \apj, 900, 3

\bibitem[{{Peters}(1964)}]{Peters64}
{Peters}, P.~C. 1964, Physical Review, 136, 1224

\bibitem[{Quinlan(1996)}]{Quinlan:1996}
Quinlan, G.~D. 1996, New Astronomy, 1, 35

\bibitem[{Rafikov(2013)}]{Rafikov:2013}
Rafikov, R.~R. 2013, The Astrophysical Journal, 774, 144

\bibitem[{{Rafikov}(2016)}]{Rafikov:2016}
{Rafikov}, R.~R. 2016, \apj, 827, 111

\bibitem[{{Rees}(1988)}]{Rees1988}
{Rees}, M.~J. 1988, \nat, 333, 523

\bibitem[{{Reved} {et~al.}(2025){Reved}, {Friedland}, \&
  {Stone}}]{RevedStone_StarBinResCap:2024}
{Reved}, O., {Friedland}, L., \& {Stone}, N.~C. 2025, \mnras, 537, 661

\bibitem[{{Ricarte} {et~al.}(2016){Ricarte}, {Natarajan}, {Dai}, \&
  {Coppi}}]{Ricarte+2016}
{Ricarte}, A., {Natarajan}, P., {Dai}, L., \& {Coppi}, P. 2016, \mnras, 458,
  1712

\bibitem[{{Robson} {et~al.}(2019){Robson}, {Cornish}, \&
  {Liu}}]{RobsonCornsih+2019}
{Robson}, T., {Cornish}, N.~J., \& {Liu}, C. 2019, Classical and Quantum
  Gravity, 36, 105011

\bibitem[{{Roedig} {et~al.}(2014){Roedig}, {Krolik}, \&
  {Miller}}]{Roedig_SEDsigs+2014}
{Roedig}, C., {Krolik}, J.~H., \& {Miller}, M.~C. 2014, \apj, 785, 115

\bibitem[{{Secunda} {et~al.}(2019){Secunda}, {Bellovary}, {Mac Low}, {Ford},
  {McKernan}, {Leigh}, {Lyra}, \& {S{\'a}ndor}}]{Secunda+2019}
{Secunda}, A., {Bellovary}, J., {Mac Low}, M.-M., {et~al.} 2019, \apj, 878, 85

\bibitem[{{Shakura} \& {Sunyaev}(1973)}]{SS73}
{Shakura}, N.~I., \& {Sunyaev}, R.~A. 1973, \aap, 24, 337

\bibitem[{{Shi} {et~al.}(2012){Shi}, {Krolik}, {Lubow}, \&
  {Hawley}}]{ShiKrolik:2012:ApJ}
{Shi}, J.-M., {Krolik}, J.~H., {Lubow}, S.~H., \& {Hawley}, J.~F. 2012, \apj,
  749, 118

\bibitem[{{Sirko} \& {Goodman}(2003)}]{Sirko_Goodman:2003}
{Sirko}, E., \& {Goodman}, J. 2003, \mnras, 341, 501

\bibitem[{{Soberman} {et~al.}(1997){Soberman}, {Phinney}, \& {van den
  Heuvel}}]{SobermanPhinney+1997}
{Soberman}, G.~E., {Phinney}, E.~S., \& {van den Heuvel}, E.~P.~J. 1997, \aap,
  327, 620

\bibitem[{{Stone} {et~al.}(2017){Stone}, {Metzger}, \&
  {Haiman}}]{Stone_AGNchan+2017}
{Stone}, N.~C., {Metzger}, B.~D., \& {Haiman}, Z. 2017, \mnras, 464, 946

\bibitem[{{Tagawa} {et~al.}(2020){Tagawa}, {Haiman}, \&
  {Kocsis}}]{TagawaAGN+2020}
{Tagawa}, H., {Haiman}, Z., \& {Kocsis}, B. 2020, \apj, 898, 25

\bibitem[{{Thun} \& {Kley}(2018)}]{ThunKley:2018}
{Thun}, D., \& {Kley}, W. 2018, \aap, 616, A47

\bibitem[{{Tiede} {et~al.}(2020){Tiede}, {Zrake}, {MacFadyen}, \&
  {Haiman}}]{Tiede:2020}
{Tiede}, C., {Zrake}, J., {MacFadyen}, A., \& {Haiman}, Z. 2020, \apj, 900, 43

\bibitem[{{Tiede} {et~al.}(2022){Tiede}, {Zrake}, {MacFadyen}, \&
  {Haiman}}]{Tiede:2022}
---. 2022, \apj, 932, 24

\bibitem[{{Tiede} {et~al.}(2025){Tiede}, {Zrake}, {MacFadyen}, \&
  {Haiman}}]{Tiede+2024}
---. 2025, \apj, 984, 144

\bibitem[{{Ward}(1997)}]{Ward:1997}
{Ward}, W.~R. 1997, \icarus, 126, 261

\bibitem[{{Wevers} {et~al.}(2022){Wevers}, {Pasham}, {Jalan}, {Rakshit}, \&
  {Arcodia}}]{Wevers:QPE-galaxies:2022}
{Wevers}, T., {Pasham}, D.~R., {Jalan}, P., {Rakshit}, S., \& {Arcodia}, R.
  2022, \aap, 659, L2

\bibitem[{{Xin} {et~al.}(2024){Xin}, {Haiman}, {Perna}, {Wang}, \&
  {Ryu}}]{ChengCheng_PTDE+23}
{Xin}, C., {Haiman}, Z., {Perna}, R., {Wang}, Y., \& {Ryu}, T. 2024, \apj, 961,
  149

\bibitem[{{Zahn}(1977)}]{Zahn:1977}
{Zahn}, J.~P. 1977, \aap, 57, 383

\bibitem[{{Zalamea} {et~al.}(2010){Zalamea}, {Menou}, \&
  {Beloborodov}}]{Zalamea:2010}
{Zalamea}, I., {Menou}, K., \& {Beloborodov}, A.~M. 2010, \mnras, 409, L25

\bibitem[{{Zrake} {et~al.}(2024){Zrake}, {Clyburn}, \& {Feyan}}]{Zrake+2024}
{Zrake}, J., {Clyburn}, M., \& {Feyan}, S. 2024, arXiv e-prints,
  arXiv:2410.04961

\end{thebibliography}

\end{document}